\documentclass[english]{thesis}
\begin{document}
    \frontmatter


	\depto{Departamento de Física}

	\author{Vicente Kabeer Pedreros Aballay}

	\title{\textbf{Horndeski Gravity as an Effective Dark Energy Fluid in Cosmological N-Body Simulations}}

	\titulo{\textbf{Gravedad de Horndeski como un fluido efectivo de energía oscura en simulaciones cosmológicas de $N$-cuerpos}}

	\tesis{Magíster en Ciencias, Mención Física}

	\guia{Domenico Sapone}

	\comision{Cristián Barrera-Hinojosa,Claudia Scóccola,Spyros Sypsas}

	\auspicio{Fondecyt Regular 1200171\\ Fondecyt Regular 1251339\\
	Infraestructura de supercómputo del NLHPC (CCSS210001)}

	\anho{2026} 

\maketitle

\begin{abstract}
	Unravelling the nature of the cosmic expansion, driven by Dark Energy (DE), remains one of the key challenges in modern cosmology. Modified Gravity (MG) models provide an alternative to the standard cosmological constant paradigm. These models predict distinct signatures in the growth of structure, yet analytical predictions are challenging due to the highly non--linear equations involved. 
    
    To overcome this problem, we study MG as an effective DE fluid. To this end, we have incorporated linear DE perturbations into the N--body code \concept, which relies on the Einstein--Boltzmann solver \texttt{CLASS}. We replaced the \texttt{CLASS} version used by \concept\ with \texttt{mochi\_CLASS}, a patched version that includes Horndeski family models. By writing linear DE perturbations ($\delta\rho_{\rm DE}$, $\delta P_{\rm DE}$, $\theta_{\rm DE}$, and $\sigma_{\rm DE}$) in terms of metric perturbations in the synchronous gauge, the N--body code is able to evolve the effective fluid using the Boltzmann equation. Our current simulations encompass a range of cosmological scenarios ---$w_0w_a$CDM, Hu--Sawicki $f(R)$ and cubic Galileon--- each extended to include massive neutrinos.

    For each of the models, we have found deviations from General Relativity close to what is predicted in the linear regime: a sub-$0.2\%$ relative difference for \lcdm, up to $40\%$ for $f(R)$ and a sub-$25\%$ for cubic Galileon. Since our code is the first to implement Horndeski gravity into an N--body code, its validation is a non--trivial computational task. Our novel code opens the door to explore a vast number of modified gravity models in the context of current and future cosmological surveys.
\end{abstract}

\begin{resumen}
	Dilucidar la naturaleza de la expansión cósmica del universo, como consecuencia de la energía oscura, sigue siendo uno de los mayores desafíos de la cosmología moderna. Teorías de gravedad modificada ofrecen una alternativa al paradigma estándar de la constante cosmológica. Estos modelos predicen señales particulares en el crecimiento de estructuras, pero las predicciones analíticas son complejas debido a la alta no linealidad de las ecuaciones involucradas.
    
    Para saltar esta barrera, se estudian las modificaciones de gravedad como un fluido efectivo de energía oscura. Para ello, se incorporan perturbaciones lineales de energía oscura en el código de N cuerpos \concept, que depende del resolutor Einstein-Boltzmann \class. Se reemplazó la versión de \class\ utilizada por \concept\ por \mochi, un parche que incluye modelos de gravedad de la familia de Horndeski. Al escribir las perturbaciones lineales de energía oscura ($\delta\rho_{\rm DE}$, $\delta P _{\rm DE}$, $\theta_{\rm DE}$, y $\sigma_{\rm DE}$) en términos de perturbaciones a la métrica en el gauge sincrónico, el código de N cuerpos es capaz de evolucionar el fluido efectivo usando la ecuación de Boltzmann. Nuestras simulaciones abarcan un rango de escenarios cosmológicos –$w_0w_a$CDM, $f(R)$ Hu-Sawicki y Galileón cúbico–, cada uno incluyendo neutrinos masivos.

    Para cada uno de los modelos, se han encontrado desviaciones de Relatividad General acordes a predicciones lineales: una diferencia relativa bajo el $0.2\%$ para \lcdm\, hasta $40\%$ para $f(R)$ y bajo el $25\%$ para Galileón cúbico. Al ser el primer código en de N-cuerpos en incluir la familia de Horndeski, su validación ha presentado una tarea computacional no trivial. Este código abre la puerta a explorar una amplia de variedad de modelos de gravedad modificada en el contexto de los futuros y actuales \textit{surveys} cosmológicos.
\end{resumen}

\begin{dedication}
	A Euskadi y Frida
\end{dedication}

\begin{acknowledgements}
\vspace*{-1cm}
\begin{flushright}
\textit{You gave me more to live for}\\
\textit{more than you'll ever know.}\\
Jeff Buckley.
\end{flushright}
	Bueno, es difícil enumerar a toda la gente que ha sido parte de este proceso, pero aquí va un humilde intento. 

    Naturalmente, quiero agradecerle a mi mayor guía en este proceso, Domenico Sapone. Desde nuestras primeras interacciones, siento que ha depositado mucha confianza en mi, lo que me ha ayudado increíblemente a desarrollarme no sólo como un estudiante de postgrado, sino también personalmente. Luego, estoy infinitamente agradecido por el grupo de cosmología de la universidad, donde mi profesor guía, junto a Gonzalo Palma y Claudia Scóccola, siempre han tenido una amabilidad tremenda conmigo. Las dos iteraciones del grupo que me tocó ver fueron de gran apoyo emocional e intelectual. Entre otros académicos, me gustaría agradecerles a Spyros Sypsas, Daniela Grandón y Cristian Barrera, por todo lo que me han ayudado y sus sabios consejos. 

    También quiero agradecerles a las grandes amistades que hice durante estos años. A Cony, persona que me vio crecer muchísimo y me acompañó durante mucho tiempo. Cata, mi grandísima amiga de más de una década, quien ha sido crucial en muchas etapas. A PP, Nuco y Fran, Chaivo, Huenupi, Javier, Jordan, Charlie y muchas más personas, que hicieron de los cafés y las sacadas de vuelta, mi paso del tercero algo que siempre recordaré con mucho cariño. A mis grandes amistades, Donko, Xano, Lari, Primo y Pili, por siempre retarme, escucharme y reírse de mí. Más recientemente, a Vania, por el tremendo cariño, risas, conversaciones deep y los cafés cumas.

    Por último, pero para nada menos relevante a mi familia. Gracias a mi madre, por siempre estar disponible para escucharme y brindarme todas las facilidades que tiene a su alcance, no sabes cuánto te amo. A mi papá y hermanos, que pese a la distancia, está este amor constante. A mis abuelos, Coyo, Kena y Pancho, por esencialmente enseñarme qué es el amor.

    Muchas gracias a quienes no mencioné, pero que de alguna forma influyeron en la persona que soy hoy.

    Esta tesis no pudo llevarse a cabo sin el apoyo financiero de Fondecyt Regular 1200171, y Fondecyt Regular 1251339. Esta tesis fue parcialmente apoyada por la infraestructura de supercómputo del NLHPC (CCSS210001).
\end{acknowledgements}

{\setlength{\parskip}{0pt}
\tableofcontents
}

{\setlength{\parskip}{0pt}
\listoftables 
}

{\setlength{\parskip}{0pt}
\listoffigures 
}

	\mainmatter
		\chapter*{Introduction}
\addcontentsline{toc}{chapter}{Introduction}
\label{intro}
Throughout human history, key discoveries have revolutionised society, shaping the way we live and understand the world. Findings such as the mastery of fire, the invention of the wheel, the discovery of DNA, advancements in medicine, the theory of evolution, and the laws of motion have had a huge impact once they were understood and implemented on a large scale. One could argue that for every major finding, some understanding of physics is needed to fully exploit it: thermodynamics to light a fire, kinematics to travel efficiently by wheel, or quantum mechanics to predict the structure of DNA, as Schrödinger did.

Two of the most celebrated scientists in history, Sir Isaac Newton and Albert Einstein, completely revolutionised our understanding of the universe. Each developed groundbreaking models of gravity that required profound intuition and abstraction. Newton’s laws of motion and universal gravitation provided a framework that accurately described celestial and terrestrial phenomena for centuries. Later, Einstein reframed this understanding with his theory of General Relativity (GR) \cite{einstein_feldgleichungen_1915}, introducing a geometric view of gravity that explained phenomena Newtonian mechanics could not, such as the bending of light by massive objects and Mercury's orbit. Both theories not only aligned with existing observations but also made predictions that were experimentally confirmed, placing them as pillars of modern science.

While GR paved the way to understand the cosmos from a geometric viewpoint, it was Edwin Hubble who first suggested that the universe is undergoing an expansion, supported by observations of receding galaxies \cite{hubble_relation_1929}. This was key to the birth of \textit{Cosmology}, the study of the Universe on the largest scales, where, for many purposes, a galaxy can be treated as a point. Questions that cosmologists seek to address concern the origin, evolution, composition, and structure of the universe. This branch of physics emerged as a direct application of Einstein's theory of General Relativity and remains a young field of research. Unlike scientists in a laboratory, cosmologists have access to only one realisation of their experiment, the universe itself. Additionally, observations of the oldest light in the universe and the overall distribution of galaxies on the largest scales suggest that approximately 95\% of the universe's content is not directly detectable by telescopes, as it does not interact with light. These are predictions of the $\Lambda$CDM model, which is the current standard paradigm of cosmology. $\Lambda$ makes reference to the \textit{cosmological constant}, responsible for the accelerated expansion of the universe, and CDM stands for Cold Dark Matter, an exotic type of matter that only interacts with gravity. 

It is remarkable how cosmologists have developed robust theories supported by observations, despite the aforementioned challenges. Inflation is a mechanism aiming to explain the very first moments of our universe, while Dark Matter and Dark Energy are exotic components that explain late-time observations. This presents a unique opportunity for particle physicists, astrophysicists and cosmologists to collaborate in understanding the nature of Dark Matter, potentially leading to groundbreaking discoveries in fundamental physics.

While the success of the $\Lambda$CDM model is undeniable, significant challenges remain. These include the considerable discrepancy between the predicted and observed energy density of the cosmological constant. Also, there is tension between early-time and late-time observations, leading to statistically different values of the current expansion rate of the universe, known as the $H_0$ tension. The first open problem has motivated the idea of extending the model by replacing the cosmological constant with a Dark Energy fluid, modifying Einstein's theory of gravity, and including extra fields and interactions, among other approaches.

In this thesis, we specifically examine the scenario in which Einstein's theory of General Relativity is extended. We examine how these modifications affect a key observable: the matter power spectrum, which quantifies the variance of the universe's matter distribution. To achieve this, we use numerical $N$-body simulations capable of reproducing non-linear dynamics on small scales. These simulations evolve Dark Matter particles in a cosmological volume under the influence of gravity. To accurately reproduce results at the largest scales, relativistic corrections must be accounted for, and several schemes exist to do so. We incorporate models beyond GR through the effective fluid approach. In essence, it amounts to treating the extensions to GR as an effective Dark Energy fluid, so that the equations of motion are identical in both approaches. Hence, implementation in code simplifies considerably.

This work is outlined as follows:

\textbf{Chapter 1} introduces key aspects of the fiducial cosmological model at the background level. We explore the Cosmological Principle, the geometric foundations of spacetime, the Friedmann equations that govern the universe's expansion, and the universe's epochs.

\textbf{Chapter 2} presents an overview of Perturbation Theory. Following this, we stress the importance of gauge transformations and present the overall structure for analytically computing these perturbations. A brief discussion of the Boltzmann equation is included. We continue by connecting these fluctuations to key observables and discussing how they constrain cosmological parameters. The chapter concludes with the concordance cosmological model, its key ingredients and shortcomings.

\textbf{Chapter 3} is concerned with Modified Gravity, specifically on the most general scalar-tensor theory of gravity in four dimensions that preserves Lorentz invariance, Horndeski gravity. We briefly discuss the different types of instabilities that must be taken into account in both analytical and numerical computations. Then we move on to describe some of the most popular models within this family. Finally, we show how the effective fluid approach applies to Horndeski.

\textbf{Chapter 4} focuses on the numerical tools used, starting with Einstein-Boltzmann solvers. Then we discuss the functioning of Cosmological N-body codes, providing a brief overview of the COsmological N-body CodE in PyThon (\concept). We detail its tight integration with the code \texttt{CLASS} and the methodology for extending both codes to include new species.

\textbf{Chapter 5} describes the methodology for including Modified Gravity models into \concept. First, we discuss how it was previously implemented in \class, and then how to interface a new version of \class\ with the N-body code. The chapter ends by explaining how \mochi\ was extended to use the effective fluid approach and how this new species must be included in \concept.

\textbf{Chapter 6} shows the main results of this work and our analysis. We start by presenting the results of our custom solver for linear cosmological perturbations, followed by those of our new version of \mochi, which is compatible with \concept. We finalise by presenting the simulation results of our code for three theories of gravity.

\textbf{Chapter 7} presents the concluding remarks and future prospects derived from this work.

Finally, at the end of the manuscript, we include appendices complementing the technical aspects of this work.

		\chapter{The Standard Cosmological Model: Homogeneous Universe}
\label{sec:the standard cosmological model}
Our current standard paradigm for understanding our universe is summarised in the standard cosmological model, often referred to as the $\Lambda$CDM model. It is a direct consequence of Einstein's theory of General Relativity (GR), and it has been successful in fitting most observations. However, it relies on three unknown ingredients: Inflation, Cold Dark Matter (CDM) and Dark Energy (DE). 

Given that modern cosmology is a relatively new branch of physics, there are various conventions regarding the mathematical tools that are used. To describe the dynamics of the universe, throughout this thesis, we adopt the convention of setting the speed of light, reduced Planck's constant and Boltzmann's constant to unity. Furthermore, working within the framework of GR, we will denote a point in space-time as $x^\mu\equiv(x^0,x^1,x^2,x^3)$, with $x^0=t$ representing cosmic time. Greek indices will run from 0 to 3, and Latin indices from 1 to 3, so we can write $x^\mu\equiv(t,x^i)$, where $i=1,2,3$. Three-vectors will be written as $\vb x$. We denote the Minkowski metric describing flat spacetime as $\tensor{\eta}{_\mu_\nu}$ and a general metric as $\tensor{g}{_\mu_\nu}$, and we follow the ``mostly plus convention’’, where the Minkowski metric reads $\eta = \mathrm{diag}(-1,+1,+1,+1)$. A simple argument for preferring this sign convention is that, should we find ourselves in a universe with extra dimensions, the determinant of the metric would not change sign. Unless otherwise stated, derivatives with respect to cosmic time $t$ will be denoted with a dot, and derivatives with respect to conformal time $\tau$\footnote{The definition of these times comes later in this chapter} will be denoted by a prime:
\begin{equation}
    \dot{\O} \equiv \dv{\O}{t} \quad \O'\equiv \dv{\O}{\tau}.
\end{equation}
\section{Gravitation}
\subsection{Geometric Foundations}
General Relativity is a central pillar of modern cosmology, providing the relationship between the curvature of spacetime and the energy sources present in the universe. The most basic object that we need to understand is the \textit{metric}, which appropriately defines a line element $\dd s$ in any space-time, through the relation:
\begin{equation}
    \dd s^2 = \sum_{\mu,\nu=0}^3g_{\mu\nu} \,\dd x^\mu \dd x^\nu = g_{\mu\nu} \,\dd x^\mu \dd x^\nu,
\end{equation}
where, in the last equality, we adopted the Einstein summation convention, which omits the sum symbol when two indices are repeated.

When working in curved spacetime, concepts such as differentiation and the trajectory of a freely falling particle must be reconsidered to obtain physical conclusions independent of the chosen coordinates. For instance, derivatives change because coordinates vary from point to point. To preserve the invariance of differentiation, we introduce the covariant derivative:
\begin{align}
    \nabla_\mu A^\nu &= \partial_\mu A^\nu + \tensor{\Gamma}{^\nu_\mu_\alpha}A^\alpha, \\
    \nabla_\mu A_\nu &= \partial_\mu A_\nu - \tensor{\Gamma}{^\alpha_\mu_\nu}A_\alpha, \\
    \nabla_\mu\tensor{T}{^\lambda_\nu} &= \partial_\mu\tensor{T}{^\lambda_\nu} - \tensor{\Gamma}{^\alpha_\mu_\nu}\tensor{T}{^\lambda_\alpha} + \tensor{\Gamma}{^\lambda_\mu_\alpha}\tensor{T}{^\alpha_\nu},
    \label{covariant derivatives}
\end{align}
where $\partial_\mu\equiv\partial/\partial x^\mu$ is the usual derivative, and $\tensor{\Gamma}{^\nu_\mu_\alpha}$ are Christoffel symbols, defined from the metric as\footnote{This holds for a symmetric metric, ie: $g_{\mu\nu}=g_{\nu\mu}$.}:
\begin{equation}
    \tensor{\Gamma}{^\lambda_\mu_\nu} \equiv \frac{1}{2} g^{\lambda\rho} \qty(\partial_\nu g_{\mu\rho} + \partial_\mu g_{\nu\rho} - \partial_\rho g_{\mu\nu}),
    \label{Christoffel def}
\end{equation}
where $g^{\lambda\rho}$ is the inverse metric, satisfying $g^{\lambda\rho}g_{\rho\alpha}=\tensor{\delta}{^\lambda_\alpha}$, with $\tensor{\delta}{^\lambda_\alpha}$ the Kronecker delta. 

With the covariant derivative, it is possible to derive the \textit{geodesic equation}, which describes the paths that minimise the distance between two points, defined from an affine parameter along the path $x^\mu(\lambda)$:
\begin{equation}
    \dv[2]{x^\mu}{\lambda} + \tensor{\Gamma}{^\mu_\alpha_\beta}\dv{x^\alpha}{\lambda}\dv{x^\beta}{\lambda}=0.
    \label{geodesic}
\end{equation}

The fact that, in general, covariant derivatives do not commute is a direct manifestation of curvature in the specific space-time we are currently working in. This can be thought of as displacing arrows between two points of the surface of a sphere through different paths, since in general the final orientation of the arrows will not be the same. This difference is encoded in what is known as the \textit{Riemann tensor}, which can be defined in terms of the metric by:
\begin{equation}
    \tensor{R}{^\alpha_\mu_\beta_\nu} = \partial_\beta \tensor{\Gamma}{^\alpha_\mu_\nu} - \partial_\nu\tensor{\Gamma}{^\alpha_\mu_\beta} + \tensor{\Gamma}{^\alpha_\beta_\rho}\tensor{\Gamma}{^\rho_\mu_\nu}-\tensor{\Gamma}{^\alpha_\nu_\rho}\tensor{\Gamma}{^\rho_\mu_\beta},
\end{equation}
or as the information encoded in how covariant derivatives commute, acting on a generic 4-vector $u$:
\begin{equation}
    \comm{\nabla_\mu}{\nabla_\nu}u^\alpha = \tensor{R}{^\alpha_\beta_\mu_\nu}u^\beta.
    \label{riemann_com}
\end{equation}
Two useful contractions of this 4-rank tensor are the \textit{Ricci tensor} $R_{\mu\nu} \equiv \tensor{R}{^\rho_\mu_\rho_\nu}$ and the \textit{Ricci scalar} $R\equiv \tensor{R}{^\mu_\mu}$.
\subsection{Gravitational Field Equations}
With the principle of least action, we can derive the equations of motion for the metric (which is the dynamic object) from the Einstein-Hilbert action, which reads:
\begin{equation}
    S_{\mathrm{EH}} = \int\dd[4]{x} \sqrt{-g}\qty[\frac{R}{16\pi G_N} + \L_{\mathrm{m}}],
\end{equation}
where $R$ is the aforementioned Ricci scalar (which contains information about the curvature of space-time), $g$ is the determinant of the metric, $G_N$ is Newton's gravitational constant,  and $\L_{\mathrm{m}}$ is the Lagrangian for a matter field minimally coupled to the metric. The result of minimising the action is Einstein's field equations:
\begin{equation}
    G_{\mu\nu}\equiv R_{\mu\nu}-\frac{1}{2}g_{\mu\nu}R = 8\pi G_N \,T_{\mu\nu},
    \label{efe}
\end{equation}
where $G_{\mu\nu}$ is called the \textit{Einstein tensor} and $T_{\mu\nu}$ is the \textit{energy momentum tensor}, defined through:
\begin{equation}
    \delta S_{\mathrm{m}} = \int\dd[4]x\delta\qty(\sqrt{-g}\L_{\mathrm{m}}) = -\frac{1}{2}\int\dd[4]x\sqrt{-g}\,T_{\mu\nu}^{(\mathrm{m})}\delta g^{\mu\nu}.
    \label{def_emt}
\end{equation}

\cref{efe} can be interpreted analogously to Maxwell's equations, as both describe the relationship between a field and its sources. In electrodynamics, the electromagnetic field is generated by charges and currents, whereas in GR the gravitational field is represented by the curvature of spacetime, which is determined by matter. In fact, any energy source will bend space-time, since we know from special relativity that there is a relation between mass and energy.

\section{The Homogeneous Universe}
The cosmological principle is a fundamental pillar of modern cosmology. It states that the energy content of the universe is distributed homogeneously and isotropically on large scales ($\gtrsim 100$ Mpc; currently debated \cite{gaite_fractal_1999,fanha_robustness_2026}). Meaning that if we look on sufficiently large scales, we could not tell where we are (homogeneity), nor which direction we are looking (isotropy). This principle assumes there is no privileged point in space, which explains the observed uniformity in the distribution of galaxies and their similar separations regardless of the observer's point of view. Essentially, it reflects the same idea as Nicolaus Copernicus' heliocentric model, since we can not be audacious enough to think that we are at the centre of the universe.
\subsection{Expansion of the Universe}
Based on measurements of the distance and radial velocities of galaxies, Georges Lemaître and Edwin Hubble proposed that the universe is expanding following a linear relation between the distance and recession velocity of a galaxy \cite{lemaitre_homogeneous_1931, hubble_relation_1929}. In 1998, two independent collaborators went beyond to suggest that this expansion is accelerated, based on newer observations \cite{riess_observational_1998, perlmutter_measurements_1999}.
The fact that the universe is expanding forces us to reconsider some basic concepts. One of the first questions that one can ask themselves is how we should measure distances. Since galaxies far away from us are distancing themselves apart, the distance between these galaxies is not trivial to define. It may be reasonable to think that the distance between two objects will grow over time, since the space-time is expanding. This is usually referred to as the \textit{physical distance}. In contrast, someone could say that, since every object is in the same space-time, taking expansion into account in the definition of distance is sensible. This distance is known as the \textit{comoving distance}. 

It is possible to characterise the expansion of the universe through the \textit{scale factor} $a(t)$, which is a function of time only; otherwise, homogeneity and isotropy would break down. The scale factor can then be used to relate the distances that were previously mentioned, such that if $r$ is the physical distance between two objects, and $x$ is the comoving distance between the same objects, we have:
\begin{equation}
    r(t)=a(t)x(t).
\end{equation}

If we take the time derivative of the physical distance, by applying the chain rule on the right-hand side, we get:
\begin{align}
    \dv{r}{t} &= \dv{(a x)}{t}, \nonumber\\
    & = \dot{a} x + a \dot{x}, \nonumber \\
    & = \dot{a} \frac{r}{a} + a v, \nonumber \\
    & = H(t) r(t) + v_p(t),
\end{align}
where on the last line we defined the Hubble parameter $H(t)\equiv \dot{a}(t)/a(t)$ and the peculiar velocity $v_p$ of the object with respect to the Hubble flow as $v_p \equiv a(t) v(t)$. In simple terms, the change of the physical distance with time is given by the expansion of the background (referred to as the Hubble flow) and the peculiar movement of the object with respect to the background. It is worth noting that for an object with no peculiar velocity (or negligible compared to the Hubble flow), we see that the change in time of the physical distance between two points is proportional to the physical distance, which is known as the Hubble-Lemaître law\footnote{It is commonly known as the Hubble law, but in 2018 the International Astronomical Union voted to recommend renaming it as the Hubble-Lemaître law, to give Georges Lemaître its due credit.}.

In the same spirit as we defined two distances, two definitions of time can be made. These were mentioned in the introduction to this chapter, where the usual time will be called \textit{cosmic time} and denoted by $t$, while the \textit{conformal time} will be denoted by $\tau$ and is defined by:
\begin{equation}
    \dd\tau \equiv \frac{\dd t}{a(t)}.
\end{equation}

A neat result from this is that for particles travelling at the speed of light (recalling $c=1$), the comoving distance that they travel in a time interval $\Delta\tau$ is $\Delta\tau$. Additionally, the \textit{conformal Hubble parameter} is defined as 
\begin{equation}
    \H(t) \equiv a'(t)/a(t) = a(t)H(t).
    \label{hubble law}
\end{equation}

The space-time coordinate system that describes a homogeneous and isotropic universe is the \textit{Friedmann-Lemaître-Robertson-Walker} (FLRW) metric. It was first worked out by Friedmann as a solution of Einstein's field equations \cite{friedman_uber_1922}, then independently from results that depended on the metric \cite{lemaitre_homogeneous_1931}, and later by only assuming homogeneity and isotropy by Robertson \cite{robertson_kinematics_1935} and Walker \cite{walker_milnes_1937}. The line element of this metric is depicted by:
\begin{equation}
    \dd s^2 = -\dd t^2 + a^2(t)\qty[\frac{\dd r^2}{1-Kr^2} + r^2\dd \Omega^2],
    \label{FLRW metric}
\end{equation}
where $\dd \Omega$ is the differential of solid angle and $K$ is the 4-dimensional curvature of space-time:
\begin{equation}
    K = \begin{cases}
        + 1 & \mathrm{closed},\\
        0 & \mathrm{flat},\\
        -1 & \mathrm{open}.
    \end{cases}
    \label{curvature values}
\end{equation}
An intuitive way to understand curvature is by considering the motion of freely falling particles in the absence of local energy sources. Suppose two particles are initially separated and undergo free fall. In a flat universe, the distance between particles remains constant, whereas in an open universe it increases over time, unlike a closed universe, where it decreases.

A remarkable result from plugging in the FLRW metric \cref{FLRW metric} into the geodesic equation in \cref{geodesic}, is that the energy of massless particles in an expanding universe, decays with the expansion of the universe $E=E_0/a$. From another perspective, we can think of the wavelength of photons as being stretched by the expansion, so their energy must decrease proportionally to the expansion of the universe. This ``stretching'' of the wavelength is commonly referred to as \textit{redshift} and denoted by $z$, such that:
\begin{equation}
    1+z \equiv \frac{\lambda_{\mathrm{obs}}}{\lambda_{\mathrm{emit}}} = \frac{a_{\mathrm{obs}}}{a_{\mathrm{emit}}},
\end{equation}
where $\lambda$ is the wavelength of the photon, and naturally the subscripts stand for observed (obs) and emitted (emit). It is customary to set the observed scale factor to unity today.

A common way to write the same metric in \cref{FLRW metric} is to use the metric $\gamma_{ij}$ describing a 3-dimensional space with constant curvature, such that the FLRW metric reads as:
\begin{equation}
    \dd s^2 = -c^2 \dd t^2 + a^2(t)\gamma_{ij}\dd x^i\dd x^j,
    \label{FLRW metric gamma}
\end{equation}
where just for a moment we have restituted the $c$ factors, and:
\begin{equation}
    \gamma_{ij}\dd x^i\dd x^j=\dd \chi^2 + f_K(\chi)^2\dd \Omega^2,
\end{equation}
where $f_\chi$ is given by:
\begin{equation}
    f_K(\chi) = \begin{cases}
        \sinh(\chi), & K=-1,\\
        \chi,        & K=0,\\
        \sin(\chi),  & K=+1,
    \end{cases}
\end{equation}
and it essentially encodes the change of variables $\dd r=\sqrt{1-Kr^2}\dd \chi$. 

If we are interested in measuring the distance between two galaxies, and we do so by starting at time $t_A$ from galaxy $A$, and moving at the speed of light to galaxy $B$, arriving at time $t_B$, we must set $\dd s=0$ since photons move on null geodesics. If we also consider that these points are in the same direction ($\dd\Omega=0$), then \cref{FLRW metric gamma} is left as:   
\begin{equation}
    \dd s = -c^2\dd t^2 +a^2 \dd \chi^2 = 0,
\end{equation}
By integrating over the radial coordinate, we find:
\begin{equation}
    \chi = \int_{\chi_A}^{\chi_B} \dd \chi = \int_{t_A}^{t_B}\frac{c}{a(t)}\dd t = c\int_{a_A}^{a_B}\frac{\dd a}{a^2H(a)} = -c\int_{z_A}^{z_B}\frac{\dd z}{H(z)},
    \label{comoving distance}
\end{equation}
where we have used the relations $\dd t=\dd a/(aH)$ and $\dd a=-\dd z/(1+z)^2$ to have the integral in terms of redshift. If we want to compute this distance between us and an object to an arbitrary redshift $z$, we have:
\begin{equation}
    \chi(z) = c \int_0^z\frac{\dd z'}{H(z')}.
\end{equation}
which is known as the \textit{comoving distance}. If we consider the distance that light travels since some initial time $t=0$ without any interactions, we obtain the \textit{comoving horizon}:
\begin{equation}
    \chi_\text{h} = \int_0^t \frac{\dd t'}{a(t')},
\end{equation}
which is the maximum distance with causal connection, since information cannot travel faster than light.

\subsection{Friedmann Equations}
Up until now, the discussion has focused on the space-time geometry side. We have to include species of energy that bend space-time. The energy contents of the universe are usually described as a perfect fluid, for which the energy-momentum tensor takes the form:
\begin{equation}
    T_{\mu\nu} = P g_{\mu\nu} + (\rho + P) U_\mu U_\nu,
    \label{background emt}
\end{equation}
where $U^\mu\equiv\partial x^\mu/\partial\tau$ is the 4-velocity of the fluid, $\rho$ its energy density and $P$ its pressure. By taking the $0-0$ and $i-j$ components of \cref{efe}, we find the Friedmann equations for a universe with curvature $K$:
\begin{align}
    \qty(\frac{\dot{a}}{a})^2 &= \frac{8\pi G_N}{3} \rho - \frac{K}{a^2},\label{friedmann constraint}\\
    \frac{\ddot{a}}{a} &= -\frac{4\pi G_N}{3}\qty(\rho+3P) \label{friedmann evolution},
\end{align}
It is also possible to derive both Friedmann equations using Newtonian gravity in an expanding background, but we prefer this approach because it is straightforward to include more species than just matter.
                                
From \cref{friedmann constraint} it is possible to find a total critical energy density, such that if the real energy density of the universe is exactly equal to the critical density, the universe is Euclidean; otherwise, it exhibits a curvature contribution that scales as $1/a^2$. This critical density is defined as:
\begin{equation}
    \rho_\text{cr} \equiv \frac{3H^2}{8\pi G_N}.
\end{equation}
It is helpful to rewrite \cref{friedmann constraint} as follows:
\begin{equation}
    H^2=H_0^2 \qty(\Omega_\text{r,0}a^{-4}+\Omega_\text{m,0}a^{-3} + \Omega_\Lambda),
\end{equation}
where we assumed a flat universe filled with radiation, matter and a cosmological constant. Additionally, we introduced density parameters, defined as:
\begin{equation}
    \Omega_\text{s} \equiv \frac{\rho_{\mathrm{s}}}{\rho_{\mathrm{cr}}},
\end{equation}
such that $\Omega_\text{m,0}$ would correspond to the density parameter of matter, evaluated at the present time.

It is of special interest to include all species in the energy-momentum tensor and study their dynamics, which are given by contracted Bianchi identities satisfied by Einstein's tensor:
\begin{equation}
    \nabla_\mu G^{\mu\nu}=0,
\end{equation}
which implies that:
\begin{equation}
    \nabla_\mu T^{\mu\nu}=0.
    \label{covt}
\end{equation}

By considering the $\nu=0$ component of \cref{covt}, we get a continuity equation for the energy content of the universe:
\begin{equation}
    \dot{\rho} + 3H(\rho+P) = 0.
    \label{continuitybk}
\end{equation}
By assuming that a species is a barotropic fluid, the equation of state reads:
\begin{equation}
    P(\rho)=w\rho,
    \label{equation of state}
\end{equation}
and $w$ is the equation of state parameter, which is 1/3 for radiation, 0 for pressurless matter, and -1 for the cosmological constant. Dropping the assumption that $w$ is a constant of time, we define:
\begin{equation}
    \Tilde{w}_\text{s}(z) \equiv \frac{1}{\ln(1+z)}\int_0^z\frac{w_\text{s}(z')}{1+z'}\dd z'.
\end{equation}
Then the evolution of the energy content of the universe for a single species is:
\begin{equation}
    \rho_{\mathrm{s}}(z) = \rho_{\mathrm{s,0}}(1+z)^{3(1+\Tilde{w}_\text{s}(z))},
\end{equation}
and for the case of a constant equation of state parameter simplifies to:
\begin{equation}
    \rho_{\mathrm{s}}(z) = \rho_{\mathrm{s,0}}(1+z)^{3(1+w_\text{s})}.
    \label{evo_rho}
\end{equation}

\section{Eras of the Universe}
From \cref{evo_rho} we can deduce that at different stages of the history of the universe, different components will dominate over the rest, which is clearer in Fig.~\ref{omega_evol}. There are three distinct epochs:
\begin{figure}[ht]
    \centering
    \includegraphics{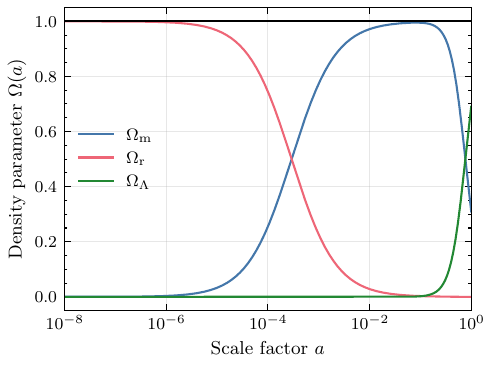}
    \caption{Density parameters as a function of the scale factor, obtained with \class.}
    \label{omega_evol}
\end{figure}
\begin{itemize}
    \item \textbf{Radiation Dominated Era} (RDE): if we found ourselves in a fluid characterised by relativistic particles, such as photons, massless neutrinos and gravitons, the appropriate equation of state parameter corresponds to $w_\text{r}=1/3$. Hence, the time dependence of its energy content is:
    \begin{equation}
        \rho_\text{r}\propto a^{-4}.
    \end{equation}
    Three of the four powers in the scale factor come from the fact that the physical volume decays as the cube of the scale factor (one per dimension), and the extra power comes from the stretching that the wavelength of these particles experiences due to the expansion.
    \item \textbf{Matter Dominated Era} (MDE): considering a fluid made up of non-relativistic particles only, then its pressure is exponentially suppressed when compared to its energy density, implying that to consider the case of $w_\text{m}=0$ is very well justified. This is the case for baryonic and dark matter, which are often referred to as dust since they constitute a gas of weakly interacting particles. In this epoch of the universe, the matter component evolves with time as:
    \begin{equation}
        \rho_\text{m} \propto a^{-3},
    \end{equation}
    and it is simple to see that, since these particles do not experience a redshift in their wavelength, we have only 3 powers of the scale factor.
    \item \textbf{Cosmological Constant Dominated Era} ($\Lambda$DE): When the universe is experiencing an accelerated expansion, we think that the cosmological constant is responsible for this. Since, by definition, this energy content is a constant of time, we must take the equation of state parameter $w_\Lambda=-1$. Recent measurements of Supernovae with Baryonic Acoustic Oscillations (BAO) suggest that this might not be the case, and this parameter might actually be a function of time \cite{desi_collaboration_desi_2025}. From \cref{friedmann evolution} it can be deduced that in order to have an accelerating universe, $w_{\rm DE}<-1/3$.
\end{itemize}

It is of special interest to identify the times when the universe transitions from one era to another. In particular, the time when the radiation and matter content of the universe were roughly the same is called \textit{equality} time. In terms of the scale factor, this time corresponds to
\begin{equation}
    a_{\rm eq} = \frac{\Omega_{\rm r,0}}{\Omega_{\rm m,0}} \sim 2.9\cdot 10^{-4}.
\end{equation}

As a first approximation, the size of the observable universe, understood as the distance that light has travelled since the Big Bang, is roughly given by the Hubble radius:
\begin{equation}
    R_H\equiv\frac{1}{H}.
\end{equation}
Together with the wavelength of a given perturbation $\lambda \sim a/k$, the two scales define two distinct dynamical regimes, which are: superhorizon ($\lambda > R_H$ or $k<aH$) and subhorizon ($\lambda < R_H$ or $k>aH$) evolution. This distinction is crucial for analysing the evolution of perturbations, since the wavelength of a perturbation grows at a rate that differs from the Hubble radius's scaling. 

\begin{table}[h!]
    \centering
    \begin{tabular}{|l|c|c|c|c|}\hline
        \textbf{Era} & $a(t)$ & $\rho(t)$ & $\lambda(t)$ & $R_H(t) $ \\\hline
        RDE & $t^{1/2}$     & $a^{-4}$& $t^{1/2}$& $t$\\\hline
        MDE & $t^{2/3}$     & $a^{-3}$& $t^{2/3}$& $t$\\\hline
        $\Lambda$DE & $e^{Ht}$ & \text{const.}& $e^{Ht}$ & \text{const.}\\\hline
    \end{tabular}
    \caption{Summary for time dependence of scale factor, energy density, wavelength and Hubble radius for a single-component universe. This is a first approximation only, and for a more precise treatment, it is preferred to solve the system numerically.}
    \label{background summary table}
\end{table}

From Table \ref{background summary table}, it is clear that if a particular perturbation begins in the superhorizon regime at early times, it will eventually enter the horizon. The time $t_*$ when a mode satisfies $k = a(t_*)H(t_*)$, is known as the \textit{horizon crossing}. When perturbations have a particular wavelength larger than the horizon, we cannot observe them evolve.

		\chapter{The Standard Cosmological Model: Inhomogeneous Universe}
\label{sec:PT}
As discussed in the previous chapter, the cosmological principle states that on large scales the universe is statistically homogeneous and isotropic. Below these scales, we observe that the universe has structure, and this principle breaks down. Observations have shown that the distribution of galaxies is not random, but they form filaments and clusters, and in their absence there are voids, forming the \textit{cosmic web} \cite{york_sloan_2000, colless_2df_2001, laureijs_euclid_2011, weinberg_observational_2013}. This must be a consequence of initial conditions that were not perfectly homogeneous, which evolved into the rich structure we observe. To systematically study these inhomogeneities, we consider departures from the homogeneous universe using perturbation theory. Of course, this is only valid up to a certain scale at which non-linearities become important, and perturbation theory is no longer adequate.

\section{The Perturbed Metric}
To study the dynamics of these small perturbations, we are going to expand the metric, as to consider an ``averaged'' metric $\bar{g}_{\mu\nu}$ (FLRW metric, also called the background metric) and a ``perturbed'' metric $\delta g_{\mu\nu}$, such that the metric now reads:
\begin{equation}
    g_{\mu\nu}(t,\vb{x}) = \bar{g}_{\mu\nu}(t) + \delta g_{\mu\nu}(t,\vb{x}),
    \label{perturbed metric}
\end{equation}
and the inverse of the perturbed metric is:
\begin{equation}
    \delta g^{\mu\nu} = -\bar{g}^{\mu\lambda}\bar{g}^{\nu\sigma}\delta g_{\lambda\sigma}.
\end{equation} 
Essentially, we are replicating what has historically been done in Physics: a Taylor expansion. Cosmology is no exception to this, and we must consider that the perturbed metric $\delta g_{\mu\nu}$ must have entries much smaller than the zeroth-order part, meaning that we are assuming that up to first order, the universe looks like FLRW with small perturbations on top.

For the rest of this work, we restrict ourselves to the flat universe, i.e., we fix $K=0$. Thus, the background metric is:
\begin{equation}
   \bar{g}_{\mu\nu} = a^2(\tau)\qty(-\dd\tau^2 + \delta_{ij}\dd x^i\dd x^j) = a^2(\tau)\eta_{\mu\nu},
\end{equation}
such that if we write the perturbed metric as $\delta g_{\mu\nu} = a^2 h_{\mu\nu}$, we can write the metric up to first order as:
\begin{equation}
    g_{\mu\nu} = a^2(\tau) \qty(\eta_{\mu\nu} + h_{\mu\nu}).
\end{equation}

The advantage is that these modes do not mix in linear order, so that we can treat them with separate equations. Additionally, all the equations for these fluctuations are linear, so Fourier analysis is apt, and we can deal with each mode $k$ of the perturbations independently. However, it is important to highlight that the linear treatment of these perturbations is only valid up to a scale $k_\text{NL}\sim 0.1 \,\mathrm{Mpc}^{-1}$ \cite{dodelson_modern_2003}, where non-linearities dominate structure formation. At these smaller scales, there have been some analytic treatments such as the Effective Field Theory of Large Scale Structure (EFTofLSS) \cite{baumann_cosmological_2012, carrasco_effective_2012}, and N-body simulation codes.

\subsection{SVT Decomposition}
This perturbed metric, being symmetric, has 10 independent components. These degrees of freedom can be decomposed into four scalar, two divergenceless 3-vectors and one traceless divergenceless rank-2 tensor. This decomposition is known as the SVT decomposition, which stands for Scalar-Vector-Tensor decomposition. Nonetheless, since GR is a covariant theory under general coordinate transformations, we must subtract 4 degrees of freedom, corresponding to the reparametrisation of each coordinate.

In the most general case, the line element reads \cite{weinberg_cosmology_2008}:
\begin{equation}
    \dd s^2 = a^2\qty[-(1+2A)\dd\tau^2 - 2B_i\dd x^i\dd\tau + \qty(\delta_{ij}+ h_{ij})\dd x^i\dd x^j],
    \label{perturbed line element}
\end{equation}
where the perturbations $A,B_i$ and $h_{ij}$ are functions of time and space. Up until now, we have one scalar degree of freedom $A$, a 3-vector $B_i$ and the symmetric spatial tensor $h_{ij}$. By virtue of Helmholtz's theorem, we can split any 3-vector into the gradient of a scalar and a divergenceless vector $\hat{B}_i$:
\begin{equation}
    B_i = \partial_iB + \hat{B}_i.
\end{equation}
Furthermore, if we expand the spatial perturbation as:
\begin{equation}
    h_{ij} = 2H_L\delta_{ij} + 2H_{Tij} + 2H_{ij},
\end{equation}
where $H_L$ is a scalar, $H_{ij}$ is a traceless divergenceless rank-2 tensor, and\footnote{$\partial_{(i}A_{j)} \equiv \frac{1}{2}\qty(\partial_{i}A_{j} + \partial_{j}A_{i})$}:
\begin{equation}
    H_{Tij} = \qty(\partial_i\partial_j - \frac{1}{3}\delta_{ij}\laplacian)H_T + \partial_{(i}\hat{E}_{j)}.
\end{equation}
Now we can clearly see the four scalars $A, B, H_L, H_T$, the two divergenceless 3-vectors $\hat{B}_i,\hat{E}_i$, and the traceless divergenceless tensor $H_{ij}$. These add up to ten degrees of freedom: four from the scalars, four from the two 3-vectors (they are divergenceless, so one is subtracted for each vector), and two from the tensor (symmetry removes three, traceless removes one, and the divergenceless removes three). For completeness, the perturbed metric now reads:
\begin{align}
    h_{00}(\tau, \vb{x}) &= -2A,\label{metric perturbations1}\\
    h_{0i}(\tau, \vb{x}) &= -\partial_iB - \hat{B}_i,\label{metric perturbations2}\\
    h_{ij}(\tau, \vb{x}) &= 2\qty[H_L\delta_{ij} + \qty(\partial_i\partial_j - \frac{1}{3}\delta_{ij}\laplacian)H_T + \partial_{(i}\hat{E}_{j)} + H_{ij}].\label{metric perturbations3}
\end{align}

It is possible to show that this perturbative expansion does not mix the different types of degrees of freedom. Meaning that we could focus on only one type of mode, and safely neglect the other two modes. This expansion, although advantageous, has one clear problem. We know from GR that the field equations are invariant under coordinate changes, but at this stage, we can see that the perturbed metric is gauge-dependent. To avoid this, we can take two approaches. One is to create gauge invariant combinations of the perturbed quantities, such as Bardeen potentials \cite{bardeen_gauge-invariant_1980}, but that approach is beyond this work. The other is to take advantage of this gauge freedom, such that we can fix up to four degrees of freedom in the equations. Later, we will discuss the most common gauges in detail.

Some exotic inflationary models predict the generation of vector modes, but these modes rapidly decay. One case where these perturbations are relevant is related to astrophysical processes, where rotating galaxies might induce vorticity that, in principle, could be detected. A similar phenomenon called \textit{shell crossing} occurs in N-body simulations, where particles crossing each other introduces local rotations in the cosmic fluid elements. \cite{barrera-hinojosa_vector_2021, barrera-hinojosa_looking_2022}.

On the other hand, tensor modes have attracted interest because they may offer a way to probe inflation by detecting primordial gravitational waves. Experimentally, the cosmological community is actively searching for the distinct imprint these tensor perturbations would leave on the Cosmic Microwave Background in the form of B-mode polarisation. Detection of this signal would be definitive observational evidence for inflation, but it would also allow us to measure the scalar-to-tensor ratio. This would give us insight into the energy scale at which inflation occurred, which is orders of magnitude higher than we can achieve with particle accelerators on Earth.

Scalar perturbations are present in many of the observables, and particularly in the matter power spectrum. Thus, we focus our attention on the scalar sector. Before jumping into observables, we should discuss the most commonly used gauges.
\subsection{The Gauges}
The two most widely used gauges in cosmology are the synchronous and Newtonian gauges. However, in this work, we are also interested in the implications of the N-body gauge.
\subsubsection{Synchronous Gauge} 
The synchronous gauge is the original gauge used by Lifshitz \cite{lifshitz_republication_2017} and the first cosmologists who worked on cosmological perturbation theory. It is the preferred gauge of \texttt{CLASS}\ \cite{lesgourgues_cosmic_2011, diego_blas_cosmic_2011, lesgourgues_classIII_2011} and \texttt{CAMB} \cite{lewis_efficient_2000}, two of the most widely used Einstein-Boltzmann solvers in Cosmology, which solve the system of equations within this framework. The primary drawback of this approach is that metric perturbations are not uniquely defined by this gauge condition, leaving some freedom to make a transformation. From another viewpoint, the picture is unclear as to which perturbations are physical and which are merely coordinate artefacts. This ambiguity historically motivated the development of gauge-invariant perturbation theory.

By construction, this gauge has its time-time and time-space components unperturbed, meaning that the line element from \cref{perturbed line element} reduces to:
\begin{equation}
    \dd s^2 = a^2(\tau)\qty[-\dd \tau^2+\qty(\delta_{ij}+h_{ij})\dd x^i\dd x^j],
    \label{ds synchronous}
\end{equation}
where the metric perturbation corresponds to $h_{ij}$, and it was characterized in \cref{metric perturbations1,metric perturbations2,metric perturbations3}.

Later, it will be proven useful to work in Fourier space, so we write the scalar mode of the metric perturbation in Fourier space, in terms of two scalar fields $h(\vb{k},\tau)$ and $\eta(\vb{k},\tau)$:
\begin{equation}
    h_{ij}(\tau,\vb{x})=\int\frac{\dd[3]\vb k}{(2\pi)^3}\vb{k}e^{i\vb{k}\vdot\vb{x}}\qty[\vu{k}_i\vu{k}_jh(\tau,\vb{k}) + \qty(\vu{k}_i\vu{k}_j-\frac{1}{3}\delta_{ij})6\eta(\vb{k},\tau)].
\end{equation}

\subsubsection{Conformal Newtonian Gauge} 
It is also known as the longitudinal gauge, and it is defined by only having two scalar perturbations $\Psi$ and $\Phi$, which are related to the Bardeen potentials in \cite{bardeen_gauge-invariant_1980}, through $\Phi_\text{A} = \Psi$ and $\Phi_\text{H}=-\Phi$ \cite{ma_cosmological_1995}. Thus, we have to impose $A=\Psi$, $H_L=-\Phi$ and everything else is set to zero, but it is possible to generalise this gauge to include vector and tensor modes. This gauge is chosen such that the coordinate system is fixed to the background of the universe, such that peculiar velocities and other effects caused by inhomogeneities are manifest. This gauge is usually the first one studied in cosmology, since the metric is diagonal and it has a clear Newtonian limit. For most of this manuscript, we will restrict ourselves to this gauge.

Thus, the line element that we are interested in is:
\begin{equation}
    \dd s^2 = a^2(\tau)\qty{-\qty[1+2\Psi(\vb{x},\tau)]\dd\tau^2 + \qty[1-2\Phi(\vb{x},\tau)]\delta_{ij}\dd x^i\dd x^j}.
    \label{cng metric}
\end{equation}
\subsubsection{N-body Gauge}
This particular gauge was motivated by N-body codes, hence the name. Since simulations aim to reproduce our universe, we must account for relativistic effects. Developing a fully relativistic code is a huge task in its own right, and several researchers are working on numerical relativity. One more amenable approach is to take advantage of Newtonian codes that have been extensively tested and optimised, and to apply relativistic corrections to them. However, from the start, there is a clear problem: transfer functions are initialised in a particular gauge, and then a Newtonian code uses them while completely forgetting the gauge. Then, to apply the linear relativistic potential to the particles from the simulation throughout time, we have to choose a gauge in which the time evolution is performed. This gauge enables the N-body code to have a dynamic space-time, effectively perturbing the Newtonian physics upon which the code was built. These ideas will become clearer in \cref{sec:numerics}.

The N-body gauge was introduced by Fidler \cite{fidler_general_2015}, who demonstrated that, at linear order, the Newtonian evolution of the system is unperturbed in the matter-only case. Thus, relativistic effects are absorbed by the gauge, and there is no need to modify the Newtonian code. If we wish to express final results in a more conventional gauge, naturally, a gauge transformation suffices.

The primary requirement of this gauge is the elimination of the local spatial volume deformation by enforcing $H_L=0$. To fully fix the scalar coordinates, the temporal-spatial shift is set to $B = v$, where $v$ is the total velocity potential of the combined cosmic fluid. The line element for scalar perturbations in the N-body gauge is:
\begin{equation}
    \dd s^2 = a^2 \qty[ -e^{2\xi} \dd\tau^2 - 2\partial_i B \, \dd x^i \dd\tau + \qty( \delta_{ij} - 2D_{ij}H^\mathrm{Nb}_T) \dd x^i \dd x^j],
\end{equation}
where $\xi$ is a linear metric perturbation sourced by radiation pressure and anisotropic stress, which by a first-order Taylor expansion reads $e^{2\xi}\simeq 1+2\xi$. Additionally, the operator $D_{ij}$ corresponds to:
\begin{equation}
    D_{ij} \equiv \partial_i\partial_j - \frac{1}{3}\delta_{ij}\laplacian.
\end{equation}

Because $H_L$ is absent, the N-body gauge density perturbation $\delta_\mathrm{Nb}$ directly matches the ``counting density'' constructed from dark matter particles in an N-body simulation. This makes the N-body gauge exceptionally powerful: the leading-order mass conservation equation for non-relativistic species becomes identical to the Newtonian continuity equation.

To initialise a simulation in this framework, linear outputs from an Einstein-Boltzmann solver computed in the synchronous (or conformal Newtonian) gauge are transformed directly into the N-body gauge using straightforward algebraic gauge transformations.

\section{The Perturbed Energy Momentum Tensor}
To obtain the sources that curve space-time, we must characterise the energy-momentum tensor. One simple way to do it is to focus on perfect fluids:
\begin{equation}
    \tensor{T}{_\mu_\nu} = P\tensor{g}{_\mu_\nu} + (\rho + P)\tensor{U}{_\mu} \tensor{U}{_\nu},
\end{equation}
where $\rho$ and $P$ are the fluid density and pressure, $U^\mu$ is the 4-velocity of the fluid given at first order by $U^\mu=(1-\Psi,\vb{u})/a(\tau)$. This allows us to consider perturbations on the sources, such as dark matter and dark energy, by characterising the tensor as:
\begin{equation}
    T_{\mu\nu} = \bar{T}_{\mu\nu} + \delta T_{\mu\nu},
\end{equation}
where the components of the background tensor are:
\begin{align}
    \bar{T}_{00} &= -\bar{\rho}(t),\\
    \bar{T}_{0i} &= \bar{T}_{i0} = 0,\\
    \bar{T}_{ij} &= a^2(t) \bar{p}(t)\delta_{ij}.
\end{align}
As a first-order approximation, the form of the perturbed energy-momentum tensor $\delta T_{\mu\nu}$ may be deduced by perturbing the perfect fluid:
\begin{align}
    \delta T_{\mu\nu} &= \delta\qty(P\tensor{g}{_\mu_\nu} + (\rho + P)\tensor{U}{_\mu} \tensor{U}{_\nu}),\nonumber
    \\ &= \bar{P}\delta g_{\mu\nu} + \qty(\bar{\rho} + \bar{P})\qty(\delta U_\mu \bar{U}_\nu + \bar{U}_\mu \delta U_\nu) + \delta P \bar{g}_{\mu\nu} + \qty(\delta\rho + \delta P) \bar{U}_\mu\bar{U_\nu},
\end{align}
such that we can write the components of the energy-momentum tensor as:
\begin{align}
    \delta\tensor{T}{_0_0} &= -\bar{\rho}\delta g_{00} + \delta\rho,\label{emt decomp1}\\
    \delta\tensor{T}{_0_i} &= \bar{P}\delta g_{i0} - (\bar{\rho}+\bar{P})\delta U_i,\label{emt decomp2}\\
    \delta\tensor{T}{_i_j} & = \bar{P}\delta g_{ij} + a^2\qty(\delta_{ij}\delta P +\Sigma_{ij}), \label{emt decomp3}
\end{align}
where the barred quantities are related to the background and only depend on time, due to homogeneity and isotropy, while the perturbations also depend on $\vb{x}$. The object $\tensor{\Sigma}{^i_j}$ is an anisotropic stress tensor, whose off-diagonal terms are the only non-zero components. This is one simple deviation from the perfect fluid case by modifying the isotropic pressure term through $\bar{P}\delta_{ij} \rightarrow \bar{P}\delta_{ij} + \Sigma_{ij} = \bar{P}\qty[\delta_{ij} + \Pi_{ij}]$.

Just like for the perturbed metric, the object $T_{\mu\nu}$ has ten degrees of freedom, of which six are physical, and four are gauge choices. Alternatively, we can split the degrees of freedom into five corresponding to the perfect fluid and five that deviate from this restriction. The perfect fluid degrees of freedom are the ones that keep the perfect fluid structure of the energy-momentum tensor, and they correspond to perturbations in density, pressure and velocity:
\begin{align}
    \rho &= \bar{\rho} + \delta\rho,\\
    P &= \bar{P} + \delta P,\\
    U^i &= \bar{U}^i + \delta U^i,
\end{align}
where $\delta U^0$ is not free, because of the constraint $U^\mu U_\mu=-1$. Additionally, we have set $\bar{U}^i = 0$ to only consider peculiar velocities on top of the background (fluid is at rest in the background). We now define a couple of common variables:
\begin{align}
    \delta &\equiv \frac{\delta\rho}{\bar{\rho}},\label{density contrast def}\\
    v^i &\equiv u^i = a \delta U^i \label{peculiar velocity def},
\end{align}
where $\delta$ is the fluctuation in the matter density with respect to the average, also called the overdensity field, and $v^i$ is the peculiar velocity represented by the velocity observed by a comoving observer, such that $v^i = \dd x^i/\dd\tau$.
\section{The Perturbed Einstein Equations}
Now, we have perturbed the Energy-Momentum tensor $\delta T_{\mu\nu}$, and we are also allowed to compute a perturbed Einstein tensor $\delta G_{\mu\nu}$ from the perturbed metric in \cref{perturbed metric}. To this end, we have to compute the perturbed Christoffel symbols, perturbed Riemann tensor and contract it into the perturbed Ricci tensor and scalar. In the conformal Newtonian gauge, components of the Einstein tensor read:
\begin{align}
    \delta \tensor{G}{_0_0} &= 2\laplacian \Phi - 6\H\dot{\Phi},\\
    \delta G_{i0} &= 2\partial_i\qty(\dot{\Phi} + \H \Psi),\\
    \delta G_{ij} &= \delta_{ij}\qty[\ddot{\Phi} + 2\H\qty(\dot{\Psi} + 2\dot{\Phi}) + \laplacian(\Psi-\Phi) + 2(2\dot{\H} + \H^2)(\Phi+\Phi)] + \partial_i\partial_i (\Phi - \Psi).
\end{align}
Thus, we are ready to write the Einstein field equations at linear order:
\begin{align}
    G_{\mu\nu} &= \bar{G}_{\mu\nu} + \delta G_{\mu\nu},\\
    T_{\mu\nu} &= \bar{T}_{\mu\nu} + \delta T_{\mu\nu},\\
    \implies \delta G_{\mu\nu} &= 8\pi G_N \delta T_{\mu\nu},
\end{align}
where we have cancelled out the background quantities, since $\bar{G}_{\mu\nu} = 8\pi G_N \bar{T}_{\mu\nu}$ holds exactly.
Depending on which gauge is chosen, these equations look different. If one is interested in working in the conformal Newtonian gauge, the evolution for the perturbations in Fourier space is:
\begin{align}
    k^2 \Phi + 3 aH \left( \Phi' + aH \Psi \right) &= -4\pi G_N a^2 \delta \rho, \label{00einstein}\\
    k^2 \left( \Phi' + aH \Psi \right) &= 4\pi G_N a^2 (\bar{\rho} + \bar{P}) \theta, \label{0ieinstein}\\
    \Phi'' + aH (\Psi' + 2 \Phi') + \qty(3a^2H^2+2aH') \Psi + \frac{k^2}{3} (\Phi - \Psi) &= 12\pi G_N a^2 \delta P,\label{ij traceless Einstein} \\
    k^2 (\Phi - \Psi) &= 12 \pi G_N a^2 (\bar{\rho} + \bar{P}) \sigma,\label{ij trace einstein}
\end{align}
where a prime denotes a derivative with respect to conformal time, $\theta=\partial^iu_i$ is the divergence of the velocity, $H$ is the Hubble parameter and $\sigma$ is related to the anisotropic stress, defined through $(\bar{\rho} +\bar{P})\sigma\equiv-(\vu{k}_i\vu{k}_j-\frac{1}{3}\delta_{ij})\Sigma^{ij}$ (recall that $\Sigma$ is traceless).

For the synchronous gauge Einstein field equations are \cite{ma_cosmological_1995}:
\begin{align}
    k^2\eta-\frac{1}{2}aHh'&=-4\pi G_Na^2\delta \rho,\\
    k^2\eta'&=4\pi G_Na^2(\bar{\rho}+\bar{P})\theta,\\
    h''+2aHh'-2k^2\eta&=-24\pi G_Na^2\delta P,\\
    h''+6\eta''+2aH\left(h'+6\eta'\right)-2k^2\eta&=-24\pi G_Na^2(\bar{\rho}+\bar{P})\sigma.
\end{align}

It is important to stress that these sets of equations concern only scalar perturbations of the metric at linear order. Additionally, this set of equations is not a complete system, since we have to further characterise the energy sources. In thermal equilibrium, they can be modelled as perfect fluids with no anisotropic stress, and through an equation of state such as \cref{equation of state}. However, if particle interactions cannot be neglected, the perturbed quantities $\delta\rho, \delta P$ and $\sigma_{ij}$ must be calculated from the distribution function for each particle species, which is governed by the Boltzmann equation. Providing these additional equations closes the system, allowing it to be solved with adequate initial conditions. This is why codes such as \texttt{CAMB} and \texttt{CLASS} are referenced as Einstein-Boltzmann solvers.

\section{Conservation of the Energy Momentum Tensor}
Given that we have fully characterised the scalar degrees of freedom of both the perturbed metric and the energy-momentum tensor, we may consider the equations arising from the conservation law $\nabla_\mu \tensor{T}{^\mu_\nu}=0$. At the background level, the $\nu=0$ component yields the continuity \cref{continuitybk}. Similarly, we can find a continuity equation for the perturbed density:
\begin{equation}
    {\delta\rho}' = -3\H(\delta P + \delta \rho) + (\bar{\rho} + \bar{P})(-\partial_iv^i + 3\Phi').
\end{equation}
The effects present in the equation are the background expansion, the velocity divergence, which is the local fluid expansion, and fluctuations in the scalar potential $\Phi$. Likewise, for the spatial component of the conservation equations, we have:
\begin{equation}
    (\bar{\rho} + \bar{P})v_i' = -(\bar{\rho}' + \bar{P}')v_i - \partial_i\qty[\delta P + (\bar{\rho} + \bar{P})(\Psi -\sigma)] - 4\H(\bar{\rho} + \bar{P})v_i,
\end{equation}
where we see the time evolution of the background inertia, the gradients of the pressure and gravitational potentials, and, lastly, the background expansion. These equations may be recast by using the continuity equation of the background \cref{continuitybk} together with the equation of state in \cref{equation of state}:
\begin{align}
    \delta' &= -(1+w)(\theta - 3\Phi') - 3\H\qty(\frac{\delta P }{\delta\rho} - w)\delta,\label{delta cng}\\
    \theta' &= -\H(1-3w)\theta - \frac{w'}{1+w}\theta + \frac{1}{1+w}\frac{\delta P}{\delta\rho}k^2\delta - k^2\sigma + k^2\Psi\label{theta cng}.
\end{align}

It is clear that these equations are written for the Newtonian gauge, but the same analysis can be done when working on the synchronous gauge, and the corresponding equations are:
\begin{align}
    \delta' &= -(1+w)\qty(\theta - \frac{h'}{2}) - 3\H\qty(\frac{\delta P }{\delta\rho} - w)\delta,\\
    \theta' &= -\H(1-3w)\theta - \frac{w'}{1+w}\theta + \frac{1}{1+w}\frac{\delta P}{\delta\rho}k^2\delta - k^2\sigma.
\end{align}

It is convenient to have the gauge transformations for the perturbed variables, which read:
\begin{align}
    \delta^\text{s} &= \delta^\text{c} - \alpha \frac{\bar{\rho}'}{\bar{\rho}},\\
    \theta^\text{s} &= \theta^\text{c} - \alpha k^2,\\
    \delta P^\text{s} &= \delta P^\text{c} - \alpha \bar{P}',\\
    \sigma^\text{s} &= \sigma^\text{c},
\end{align}
where $\alpha = (h' + 6\eta')/2k^2$, and the superscript make reference to either the synchronous or Newtonian Gauge. For the N-body gauge, the transformations are:
\begin{align}
    \delta^\text{Nb}_\text{a} &= \delta^\text{s/c}_\text{a} + \frac{3\H(1+w_\text{a})\theta_\text{tot}^\text{s/c}}{k^2},\label{eq:enbody transf1}\\
    \theta^\text{Nb}_\text{a} &= \theta^\text{s/c}_\text{a} -\frac{3}{k^2}\partial_\tau\qty[\H \theta_\text{tot}^\text{s/c}] + \begin{cases}
        \frac{h'}{2} & \text{Synchronous},\\
        -3\Phi' & \text{Newtonian},
    \end{cases}
    \label{eq:enbody transf2}
\end{align}
where the subscript ``a'' is for the species of type ``a''. These equations are valid for uncoupled fluids, such that each perturbation is concerned with a unique species. This is particularly neat, since if we are interested in studying late-time dynamics, these are the equations that we need to solve, together with the perturbed Einstein equations. However, the 6 equations are not independent because of the Bianchi identity. They are consistent, so we can use whichever subset best suits our particular case.

One important remark is that \cref{theta cng} is not defined for $ w =- 1$. This is the case for the cosmological constant, but by definition, this species has no perturbations, since it is a constant. Nevertheless, if we leave the parameter $w_\text{DE}$ as a free function that can depend on time, we might arrive at a case where this function crosses the value of -1. This particular behaviour is known as \textit{phantom crossing} \cite{kunz_crossing_2006}, and one way to treat this issue is through the Parametrised Post Friedmann (PPF) approach \cite{fang_crossing_2008, hu_parametrized_2008}. Consequently, the equation for the evolution of $\theta$ is recast into a new variable $V=(1+w)\theta$, which often works better for numerical solutions. Hence, the final version of the equations for the evolution of $\delta$ and $V$ is:
\begin{align}
    \delta' &= -V + 3(1+w)\Phi' - 3\H\qty(\frac{\delta P}{\delta \rho} - w)\delta,\label{delta cng2}\\
    V' &= -\H(1-3w)V +\frac{\delta P}{\delta \rho}k^2\delta + k^2(1+w)(\Psi-\sigma)\label{V cng2}.
\end{align}

At this stage, we have more variables than equations, since in principle we should find $\delta,\theta,\delta P, \Phi,\Psi$ and $\sigma$. To obtain a closed system, one of the common approaches is to consider perfect fluids only, such that $\sigma=0$ and the scalar metric perturbations are equal, through \cref{ij trace einstein}. Additionally, we can relate the pressure and density perturbation through the sound speed $c_s^2$, defined as:
\begin{equation}
    c_s^2 \equiv \frac{\dot{\bar{P}}}{\dot{\bar{\rho}}} = w - \frac{w'}{3\H(1+w)}.
    \label{sound speed}
\end{equation}

In the case of adiabatic perturbations\footnote{By adiabatic perturbations, we mean perturbations that were initially adiabatic, which does not imply that they stay adiabatic as the universe evolves.} we can relate the pressure perturbation to the density perturbation via:
\begin{equation}
    \delta P = c_s^2 \delta \rho.
\end{equation}

From \cref{sound speed}, it is clear that for radiation and matter, the sound speed is 1/3 and 0, respectively. A null sound speed physically means that the perturbations in the fluid do not propagate, allowing them to cluster. On the contrary, for values greater than zero, the fluid can support waves that propagate due to local gradients in pressure and density. This leads to the definition of the \textit{sound horizon}, which is a scale below which perturbations cannot grow due to the counteracting of pressure. The crossover scale is quantified by the Jeans scale:
\begin{equation}
    \lambda_J = c_s \sqrt{\frac{\pi}{G_N\bar\rho}}.
\end{equation}
Whenever a perturbation has a scale smaller than the Jeans scale, it starts oscillating, which is the case for baryonic acoustic oscillations.

\section{Solutions for Linear Perturbations}
In order to find the solution for the linear perturbations, we might notice that \cref{delta cng2,V cng2} may be combined into a single second-order differential equation. The resulting equation, without further assumptions on the fluid, is:
\begin{align}
    \delta'' &+ \H\qty(1-6w+3c_s^2)\delta' + \qty{3\qty(c_s^2-w)\qty[\H' + \H^2\qty(1-3w)] +3\H(c_s^2-w)'+k^2c_s^2}\delta \nonumber\\
    &= 3(1+w)\Phi'' + \qty[3\H\qty(1-2w-3w^2)+3w']\Phi' - k^2(1+w)(\Psi-\sigma).
\end{align}
There are different approximation schemes to solve this equation analytically, like assuming that we are deep in MDE and matter composes a perfect fluid, then the equation for matter perturbations simplifies to:
\begin{equation}
    \delta_{\rm m}'' + \H\delta_{\rm m}' = 3\Phi'' + 3\H\Phi'-k^2\Phi.
\end{equation}
If we also consider sub-horizon modes, such that time derivatives\footnote{Using Einstein equations, it is possible to show that the potential is almost constant during MDE} of the potential are suppressed compared to $k^2\Phi$, and use \cref{00einstein,0ieinstein}, we end up with:
\begin{equation}
    \delta_{\rm m}'' + \H\delta_{\rm m}'+\qty(c_{\rm s}^2k^2-\frac{3}{2}\H^2)\delta_{\rm m} = 0,
    \label{delta 2nd order ODE}
\end{equation}
which has analytic solutions. First, if we consider the regime where the Hubble flow is zero ($\H \to0$), we find the classical fluid wave equation $\delta '' + c_{\rm s}^2k^2\delta=0$. Additionally, here is where we find the restriction:
\begin{equation}
    c_{\rm s}^2k^2-\frac{3}{2}\H^2>0,
\end{equation}
for the perturbation to grow, which is closely related to the Jeans length previously introduced. When $c_{\rm s}k\ll\H$, we might find growing solutions of \cref{delta 2nd order ODE}, which are easier to treat when e-folds $N\equiv\ln{a}$ are used as the time variable. The solution is given by:
\begin{equation}
    \delta_{\rm m}(a) = A a + B a^{-3/2}, \quad \theta_{\rm m}(a) = -A a^{1/2} + Ca^{-1} - B a^{-2},
    \label{analytic delta solution}
\end{equation}
where $A, B, C$ are constants given by initial conditions.

For more realistic scenarios, such as a universe similar to what we observe today, with a cosmological constant and non-negligible matter component, we often turn to numerical solutions. While common choices among cosmologists include well-established codes such as \texttt{CAMB} and \texttt{CLASS}, they can obscure the dynamics and physics underlying the system of equations they aim to solve. Therefore, we have developed a custom solver written in \texttt{Python} that is publicly available\footnote{\url{https://github.com/vPedreros/perturbations}}. Rather than reducing the system to a second-order differential equation, our code directly integrates the first-order coupled system of equations. The main goal of this solver is to have full control over studying late-time dynamics, so we consider Dark Matter and Dark Energy perturbations (without safeguards against phantom crossing).

\begin{figure}[ht]
    \centering
    \includegraphics{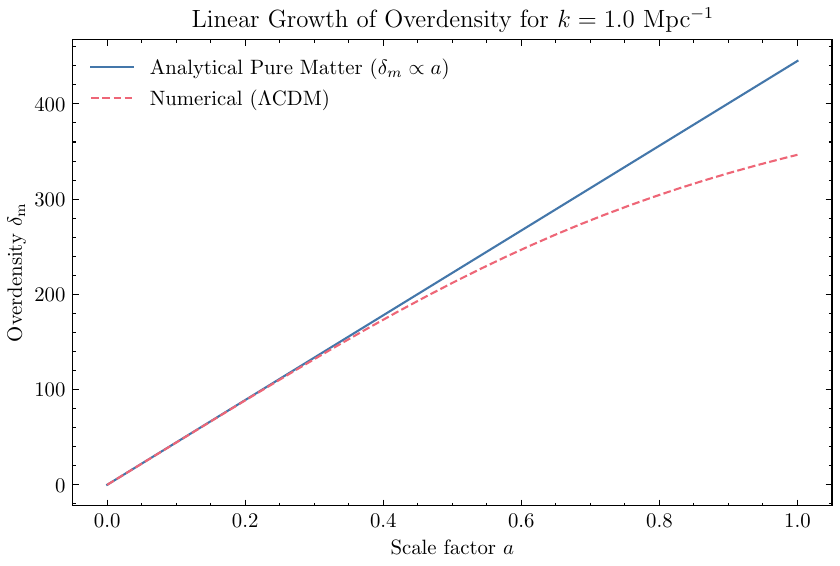}
    \caption{Comparison between analytic solution from a matter-only universe, with the numerical solution for a universe with DM and DE.}
    \label{anl vs numeric}
\end{figure}

In Figure \ref{anl vs numeric}, it is clear that the numerical solution that considers a Dark Energy component differs at late times from the analytic solution for a matter-only universe. Given that Dark Energy drives the accelerated expansion of the universe, there is a friction that counteracts gravity, implying that the growth of perturbations is suppressed. A careful reader might note that the values for $\delta$ are quite large, and orders of magnitude larger than unity, implying that linear theory should break. The reason for this is that the numeric solver considers an absolute amplitude that is arbitrary, so in general cosmologists prefer to work with a normalised version of the perturbation called the linear \textit{growth factor}.

The general solution for $\delta$ in a universe composed of multiple species includes a growing and a decaying solution, which reads:
\begin{equation}
    \delta(\vb x, t) = D_+(a)\delta_+(\vb x) + D_-(a)\delta_-(\vb x),
\end{equation}
where $D_+(a)$ is the scale-independent growth factor that determines how the density contrast depends on the scale factor. It mainly describes the growth of structures at late times, and as an example for a matter-dominated universe, it takes the simple form $D_+(a)=a$, as it can be deduced from \cref{analytic delta solution}. Decaying modes are not interesting, and we neglect them. The definition of the growth factor is\footnote{This is for the case of DE in the form of a cosmological constant.}:
\begin{equation}
    D_+(a) = \frac{5\Omega_{\text{m},0}}{2}\frac{H(a)}{H_0}\int_0^a\frac{\dd a'}{(a')^3}\qty(\frac{H_0}{H(a')})^3.
\end{equation}

So far, we have been able to explain late-time dynamics by making assumptions about the fluid that we are interested in. However, these assumptions do not capture the full thermodynamics of more complicated scenarios, such as the photon-baryon plasma in the early universe.
\section{The Boltzmann Equation}
So far, we have described the universe from a macroscopic perspective, treating species as perfect fluids. Here, we give a brief microscopic description of the same fluids. This is necessary because contemporary codes rely on connecting the two descriptions. Mathematically, we saw that it is necessary to make assumptions about the species to solve Einstein's field equations, but it is possible to go beyond these assumptions and actually solve for perturbed quantities using the Boltzmann equation.
\subsection{Boltzmann Equation in Minkowski}
Given that it is not feasible to describe every single particle and its interactions, we turn to a statistical description. The object that allows us to do this is the distribution function $f$, which can be defined by relating the number of particles inside a volume of phase space:
\begin{equation}
    \dd N(\vb{x},\vb{p},t) = f(\vb{x},\vb{p},t) \dd V_\text{phase space} = f(\vb{x},\vb{p},t) \dd[3]x \frac{\dd[3]p}{(2\pi\hbar)^3},
\end{equation}
where the factor $(2\pi\hbar)^3$ represents the fundamental volume of a single quantum state in the phase space, from Heisenberg's uncertainty principle. This ensures that the distribution function represents the average occupancy number per quantum state.

In the scenario where particles only interact gravitationally, as is the case for dark matter particles, one expects that the number of particles is conserved, which can be formalised by stating that the distribution function has its total time derivative equal to zero:
\begin{equation}
    \dv{f}{t} = 0 \iff \pdv{f}{t} + \dv{\vb x}{t}\cdot\pdv{f}{\vb x} + \dv{\vb p}{t}\cdot\dv{f}{\vb p}=0.
\end{equation}
In the classical picture, this equation can be rewritten as:
\begin{equation}
    \pdv{f}{t} + \frac{\vb p}{m}\cdot\pdv{f}{\vb x} - m \grad\Phi \cdot \pdv{f}{\vb p} = 0,
\end{equation}
where $\Phi$ is the Newtonian potential that satisfies the Poisson equation $\nabla^2\Phi = 4\pi G\rho$. We have found that the partial time derivative of the distribution function is determined by the number of particles entering and leaving the phase space we are considering.

Motivated by the statistical description, it is possible to define macroscopic quantities that characterise our fluid. In this case, we define the particle density $\rho_\text{m}$, momentum density $\pi^i = \rho_\text{m}v^i$ and anisotropic stress tensor $\kappa^{ij}$ as:
\begin{align}
    \rho_\text{m}(\vb{x},t) &\equiv m\int \dd[3]p f(\vb{x},\vb{p},t),\\
    \rho_\text{m} v^i(\vb{x},t) &\equiv \int \dd[3]p \,p^i f(\vb{x},\vb{p},t),\\
    \rho_\text{m} v^iv^j(\vb{x},t)+\kappa^{ij}(\vb{x},\vb{p},t) &\equiv \frac{1}{m}\int \dd[3]p \,p^ip^j f(\vb{x},\vb{p},t).
\end{align}

We find conservation equations by taking moments of the Boltzmann equation, which corresponds to integrating in momenta, the distribution function times the mass of the particle, multiplied by powers of the peculiar velocity of the particle:
\begin{equation}
    \int \dd[3]p\,m\frac{p^{i_1}}{m}\dots \frac{p^{i_n}}{m}\left[\dv{f}{t}\right]=0.
\end{equation}

Thus, the zeroth and first moments of the Boltzmann equation give the energy conservation and Euler equation:
\begin{align}
    \dot{\rho}_\text{m} + \partial_i(\rho_\text{m}v^i)&=0,\label{eq:continuity boltz mink}\\
    \dot{v}^i + v^j\partial_jv^i &= -\partial^i\Phi.\label{eq: euler boltz mink}
\end{align}

In principle, every moment of the Boltzmann equation should be taken into account, but it suffices to consider these two equations. For cold dark matter particles, we do not see the anisotropic stress tensor, since they have a zero velocity dispersion.

In general, to properly study the particles in our universe, it is necessary to consider more than just long-range field forces. For example, we know that photons interact with electrons, and this is quite important for understanding the CMB. To take into account such interactions, we must introduce the \textit{collision functional} term, such that the Boltzmann equation in general reads:
\begin{equation}
    \dv{f}{t} = C[f],
\end{equation}
and we refer to \cite{dodelson_modern_2003} for a proper derivation of such collision terms, which in general must be treated with Quantum Field Theory.
\subsection{Boltzmann Equation in FLRW}
Now we turn to our case of interest, which is the expanding universe. First, we must find an expression for the comoving momentum, which is defined as $P^\mu = \dv*{x^\mu}{\lambda}$, and it satisfies the \text{on-shell constraint} $P^2 = -m^2$, where $m$ corresponds to the particle rest mass. Given that our interest is in writing a Boltzmann equation for the distribution function $f(\vb x, \vb p, t)$, we have to transform our comoving momentum to a physical one. To this end, by exploiting the on-shell constraint, it is straightforward to find:
\begin{equation}
    P^\mu = \qty[E(1-\Psi), p^i\frac{1+\Phi}{a}].
\end{equation}
Thus, we can write the relativistic Boltzmann equation as\footnote{The terms proportional to $\pdv*{f}{x^i}$ and $\pdv*{\hat{p}^i}{t}$ can be dropped for the homogeneous universe.}:
\begin{equation}
    \dv{f}{t} = \pdv{f}{t} + \dv{x^i}{t}\pdv{f}{x^i} + \dv{p}{t}\pdv{f}{p} + \dv{\hat{p}^i}{t}\pdv{f}{\hat{p}^i} = C[f].
\end{equation}
We can go further by exploiting the geodesic equation in \cref{geodesic}, and the corresponding Christoffel symbols, arriving at:
\begin{align}
    \dv{f}{t} &= \pdv{f}{t} + \frac{p}{E}\frac{\hat{p}^i}{a}(1+\Phi+\Psi)\pdv{f}{x^i} - p\qty(H-\dot\Phi+ \frac{E}{ap}\hat{p}^i\partial_i\Psi)\pdv{f}{p}\nonumber\\
    &+\frac{E}{ap}\qty[-\partial_i\qty(\frac{p^2}{E^2}\Phi + \Psi) + \hat{p}^i\hat{p}^k\partial_k\qty(\frac{p^2}{E^2}\Phi+\Psi)]\pdv{f}{\hat{p}^i}=C[f].
\end{align}

Now we compare the equation in Minkowski space to this case when taking the Newtonian limit. For that, we consider non-relativistic particles $(E\simeq m)$, assume that the distribution function does not depend on $\hat{p}^i$, and linearise the equation. This results in the linear-order Boltzmann equation for non-relativistic particles:
\begin{equation}
    \pdv{f}{t} + \frac{p}{m}{\frac{\hat{p}^i}{a}}\pdv{f}{x^i} - \qty(Hp-\dot{\Phi} +\frac{m}{ap}\hat{p}^i\partial_i\Psi)\pdv{f}{p} = C[f].
    \label{non-relativistic expanding boltzmann}
\end{equation}
If we go beyond and recall that, during the matter-dominated epoch, the metric perturbations are the same and do not depend on time, we observe that the only difference from the Minkowski case is the appearance of scale factors and the Hubble flow term.

Interestingly, if we go to the homogeneous and isotropic universe by dropping the metric perturbations and the spatial derivative of the distribution function in \cref{non-relativistic expanding boltzmann}, integration over all momenta yields:
\begin{equation}
    \pdv{n(t)}{t} + 3 H n(t) = \int\frac{\dd[3]p}{(2\pi)^3}C[f],
\end{equation}
and it is possible to see that the number density scales as $a^{-3}$ when the collision term is absent, as we previously saw for non-relativistic particles.

To describe macroscopic properties of our collection of particles, we introduce the relativistic expression of the energy-momentum tensor given a distribution function:
\begin{equation}
    \tensor{T}{^\mu_\nu} = \frac{g_*}{\sqrt{-g}}\int\frac{\dd P_1 \dd P_2 \dd P_3}{(2\pi)^3} \frac{P^\mu P_\nu}{P^0}f(\vb{x},\vb{p},t),
\end{equation}
where $g_*$ is the degeneracy factor\footnote{It counts how many different particle states are described by our distribution function.} and $g$ is the determinant of the metric.

When taking moments of a Boltzmann equation, we notice a pattern. We arrive at a set of equations, where each moment gives us the evolution of an object $A$ that depends on a new object $B$, whose evolution $\dot{B}$ is determined by the next moment of the Boltzmann equation. Thus, the effective approach consists of adding an effective term as a source to the Euler equation, which accounts for the effects of higher-order unknown objects in the moments. In fact, this is important for studying non-linearities, as described in \cite{baumann_cosmological_2012,carrasco_effective_2012}.
\section{The Cosmic Microwave Background}
The discovery of the Cosmic Microwave Background (CMB) by Penzias and Wilson in the 1960s is a cornerstone of modern cosmology \cite{penzias_measurement_1965}. Observations of the CMB provide precise constraints on cosmological parameters today. The study of the CMB is so crucial in physics that it has led to Nobel Prizes in Physics in 1978, 2006, and 2019.

As the oldest light that we can observe from the universe, the CMB is arguably the most relevant source of information from the primordial universe. It contains information on how initial quantum fluctuations that evolved into the classical regime were imprinted on this snapshot, often called the \textit{last scattering surface}.

To understand the underlying physics, we have to go further back than the emission of the CMB, which occurred when the universe was approximately 380,000 years old. At this stage, the universe was filled with radiation, well described by a hot baryon-photon plasma. Thomson scattering made the mean free path of photons so short that they could not free-stream away from the plasma. As we discussed, the most interesting dynamics emerge when studying small deviations from homogeneity. In this case, anisotropies are separated into primary and secondary, where the first correspond to effects that happened up until the photon decoupling, and secondary anisotropies are influences on the photons while they travel from the last scattering surface towards us.

\subsection{The Temperature Angular Power Spectrum}
The main object of discussion in this section will be the temperature fluctuation $\Theta$, leaving out photon polarisation, which is a rich topic in its own right \cite{baumann_probing_2009}. The quantities of interest for the CMB are spanned across the sky, so they are apt for a description with angular coordinates on a sphere, parametrised by introducing a unit vector $\vu{n}$. They are naturally described by an expansion in spherical harmonics, for which the temperature fluctuation reads:
\begin{equation}
    \Theta(\vu{n}) \equiv \frac{\Delta T (\vu{n})}{\bar{T}} = \sum_{\ell=0}^{\infty} \sum_{m=-\ell}^\ell a_{\ell m} Y_{\ell m} (\vu{n}),
    \label{def temperature fluctuation}
\end{equation}
where $\bar{T}$ is the average temperature of $\sim 2.7$ K, and $\Delta T$ is a small deviation, such that $\Theta$ is of order $10^{-5}$. The relevant statistical information of the fluctuation is encoded in the $a_{\ell m}$ coefficients. Statistical isotropy tells us that the two-point correlation function of these coefficients must take the form:
\begin{equation}
    \expval{a_{\ell m} a_{\ell'm'}^*} = \delta_{\ell\ell'}\delta_{m m'} C_\ell,
    \label{def C_ell}
\end{equation}
where the $C_\ell$ are known as multipole moments. From this definition, it is possible to write the two-point correlation of the temperature fluctuation as:
\begin{equation}
    \expval{\Theta(\vu{n}_1)\Theta(\vu{n}_2)} = \frac{1}{4\pi}\sum_{\ell=0}^\infty(2\ell + 1) C_\ell \mathcal{P}_\ell(\vu{n}_1\cdot\vu{n}_2),
\end{equation}
where $\mathcal{P}_\ell(x)$ are Legendre polynomials. Here we identify $C_\ell$ as the angular power spectrum of temperature fluctuations, and it is possible to write it as:
\begin{equation}
    C_\ell = \frac{1}{2\ell +1} \sum_{m=-\ell}^\ell \expval{a_{\ell m}a_{\ell m}^*}.
\end{equation}
In an idealised case, we may estimate the angular power spectrum with two equations, based on projecting the temperature map in spherical harmonics and constructing the estimator for $C_\ell$:
\begin{align}
    \hat{C}_\ell = \frac{1}{2\ell +1}\sum_{m=-\ell}^\ell a_{\ell m} a_{\ell m}^*,\\
    a_{\ell m} = \int\dd[2]\vu{n} \Theta(\vu{n}) Y^*_{\ell m}(\vu{n}),
\end{align}
but reality demands a much more sophisticated treatment for estimating the angular power spectrum. Another important concept is that of \textit{cosmic variance}, which is given by:
\begin{equation}
    \Delta C_\ell = \sqrt{\frac{2}{2\ell+1}}C_\ell,
\end{equation}
which is larger for lower multipoles (larger scales), since there are fewer independent modes on these scales.

Now we turn to a non-exhaustive list of effects of the CMB.
\begin{figure}[ht]
    \centering
    \includegraphics[width=0.9\textwidth]{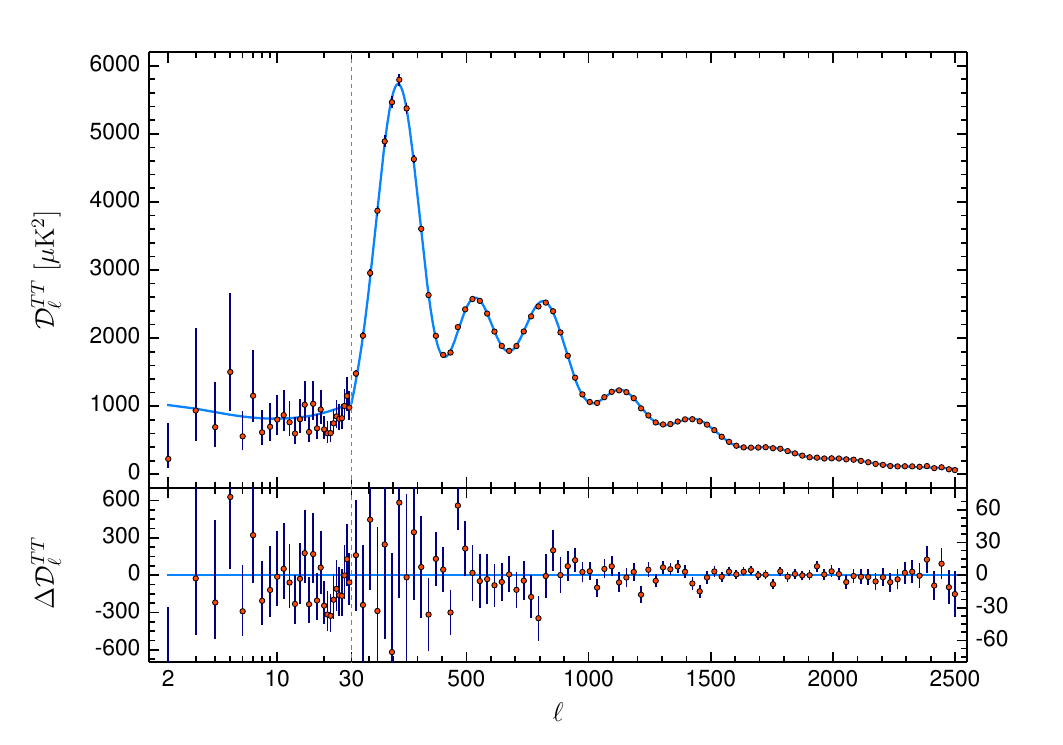}
    \caption{CMB temperature power spectrum as a function of multipoles, taken from \cite{planck_collaboration_planck_2020}. The blue line is the \lcdm\ model fitted to the data (red points). The error bars represent 1$\sigma$ Gaussian uncertainties. The lower panel shows residuals with respect to a base-\lcdm\ model. Scatter on the lower multipoles is due to cosmic variance.}
    \label{CMB temperature power spectrum}
\end{figure}
\subsection{Primary Anisotropies}
A phenomenon that causes fluctuations in temperature is the \textit{Sachs-Wolfe effect} (SW). Essentially, fluctuations in the gravitational potential well affect the temperature of photons at the last-scattering surface. Photons leaving an overdense region lose more energy than if they leave an underdense region, so this directly affects the frequency of the photons. This is the main contribution to the CMB temperature power spectrum on large scales ($\ell<100$).

One of the physical processes that is needed to explain the oscillatory behaviour in \cref{CMB temperature power spectrum} is the baryonic acoustic oscillations (BAO). The interplay between radiation pressure and gravity compressing the plasma gives rise to baryon-photon acoustic waves. The main contribution to the gravitational potential wells comes from dark matter, so we can neglect the baryonic contribution. These overdense regions (where the potential well is deeper) also have higher radiation pressure, which eventually repels the material, which then falls back into the potential wells. From this oscillating behaviour, we can define the \textit{sound horizon} as the maximum scale that sound waves could travel before photon decoupling. This scale is extremely important, since it defines a standard ruler for measuring distances at very large scales. This is why the BAO is a powerful cosmological probe.

\subsection{Secondary Anisotropies}
There is a similar effect to the one previously described, related to the potential. This is called the \textit{Integrated Sachs-Wolfe effect} (ISW), which accounts for the time evolution of the gravitational potential. The standard paradigm predicts that during MDE, the gravitational potentials are constant, so this effect is only relevant at very late times and serves as a probe for the growth of structure and potential modifications to GR.

Additionally, the CMB is sensitive to lensing caused by objects between the last-scattering surface and the observers. This is a direct consequence of GR, and the same physics can be used in the weak lensing (WL) observable to probe dark matter.
\subsection{Constraints on Cosmological Parameters}
From a CMB temperature power spectrum, it is possible to constrain several parameters. Particularly, the position of the first peak is sensitive to the spatial curvature, and the current constraint is consistent with a flat universe. Acoustic waves, as we explained, depend on both the cold dark matter and baryon concentrations, so they also constrain the density parameters $\Omega_\text{c}$ and $\Omega_\text{b}$. Hence, the CMB is a direct probe of the existence of dark matter, its concentration, and its role in the BAO. Additionally, the combination of CMB temperature and lensing power spectra gives tight constraints on the matter clustering $\sigma_8$ and the total matter content $\Omega_\text{m}$. Lastly, this probe becomes stronger when cross-correlated with supernova observations, providing constraints on dark energy.
\section{Large Scale Structure}
Another powerful cosmological probe is the Large Scale Structure (LSS), which provides direct evidence of inhomogeneities in the universe. Several experiments have confirmed that galaxies form large-scale structures, meaning that their distribution is not random. Therefore, information is encoded by the structures that we should study.

To characterise the structure, we will be using the density contrast $\delta$ defined in \cref{density contrast def}, where an overdense region will correspond to $\delta>0$, and an underdense region to $\delta<0$.

\subsection{The Matter Power Spectrum}
Given that the density contrast is a random variable, we must turn to a statistical description. The Ergodic hypothesis allows us to relate theoretical ensemble averages to spatial observations in our universe, assuming that spatial averages over sufficiently large volumes resemble the ensemble average. For a random Gaussian field, all the information is contained in the one and two-point correlation functions, which are related to the field's mean and variance. In our case, we know that on sufficiently small scales, non-linearities take over the description, making it highly non-Gaussian. However, in practice, it is challenging to compute higher-order correlation functions, so we restrict ourselves to the two-point correlation function in this section. Since by definition the mean of the field is zero, we have:
\begin{equation}
    \expval{\delta} = 0,
\end{equation}
allowing us to write the variance of the field ($\sigma^2$) as:
\begin{equation}
    \expval{\delta^2} = \sigma ^2.
\end{equation} 
The two-point correlation function in real space is defined as:
\begin{equation}
    \xi(\vb{x},\vb{y}) = \expval{\delta(\vb{x})\delta(\vb{y})},
\end{equation}
and it is clear that if we add more fields inside the expectation value, we would have higher-order correlation functions. If the field that we are interested in is statistically homogeneous and isotropic, then we can simplify the expression of the correlation function to $\xi(|\vb{x}-\vb{y}|)$, since it would not be sensitive to the exact point and direction that we are evaluating it, rather to only the distance between the points that we are computing the correlation. This motivates us to work in Fourier space, for which we adopt the convention\footnote{Here we introduce the notation $\int_{\vb{k}} \equiv \int \frac{\dd[3]\vb{k}}{(2\pi)^3}$}:
\begin{align}
    \delta(\vb{x}) &= \int_{\vb{k}}e^{i\vb{k}\cdot\vb{x}}\delta(\vb{k}),\\
    \delta(\vb{k}) &= \int\dd[3]\vb{x}\,e^{-i\vb{k}\cdot\vb{x}}\delta(\vb{x}).
\end{align}
Therefore, we can define the \textit{matter power spectrum} $P_\text{m}(k)$ via:
\begin{equation}
    \expval{\delta(\vb{k})\delta^*(\vb{k'})} = (2\pi)^3 \delta^{(3)}(\vb k - \vb k') P_\text{m}(k),
\end{equation}
where we have used the Fourier transforms, an intermediate variable and the fact that the overdensity field is statistically isotropic and homogeneous. In consequence, we clearly see that the matter power spectrum depends on the magnitude of the wave vector, and on time (we are momentarily omitting the label). Since this is connected to the two-point correlation function of the overdensity field, we usually think of the matter power spectrum as the Fourier transform of this two-point correlation function, hence containing information about the variance. Physically, regions with higher density values indicate strong matter clustering, reflected in an enhancement of the power spectrum amplitude on these scales.

In practice, the power spectrum is computed on smoothed fields using a filter kernel $W(\vb r)$, such that the smoothed field is a convolution of the field by the kernel:
\begin{equation}
    \delta(\vb x) = \int\dd[3] \vb r W (\vb r) \delta(\vb x + \vb r),
\end{equation}
and the variance of this field is given by:
\begin{equation}
    \sigma^2_R = \frac{1}{2\pi^2}\int\dd k k^2 P_\text{m}(k) |W(kR)|^2.
    \label{sigma_R}
\end{equation}
Conventionally, the spherical top-hat function is used as a filter, and in real space is depicted by:
\begin{equation}
    W_R(r) = \begin{cases}
        \frac{3}{4\pi R^3} & r<R,\\
        0 & r>R,
    \end{cases}
\end{equation}
while its Fourier transform corresponds to:
\begin{equation}
    W(kR) = \frac{3j_1(kR)}{kR} = \frac{3 \qty[\sin{(kR)}-kR\cos{(kR)}]}{(kR)^3},
\end{equation}
where $j_1(x)$ is the spherical Bessel function of the first kind and order one. While this definition holds for any value of $R$, it is customary to set $R=8\text{Mpc}/h$ and define the cosmological parameter $\sigma_8$. This corresponds to the root-mean-square of the amplitude of matter perturbations, smoothed over 8Mpc$/h$ and evaluated at the present time. This is also a normalisation of the matter power spectrum.

\begin{figure}[ht]
\centering
\includegraphics[width=0.8\textwidth]{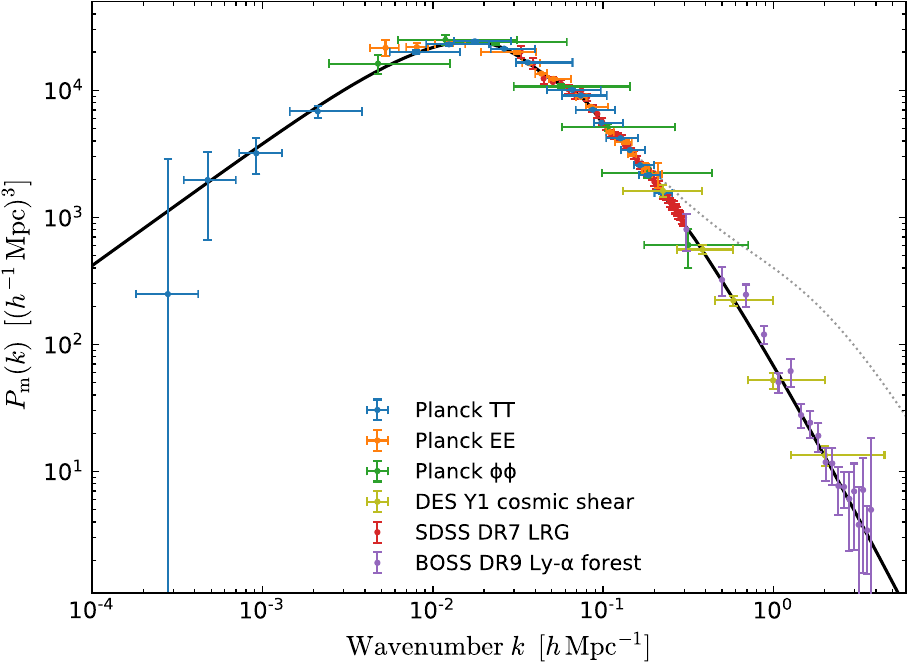}
\caption{Linear matter power spectrum at the present time. The solid black line corresponds to the theoretical prediction given by a Planck 2018 cosmology. Colour points are observations from different experiments. Taken from \cite{aghanim_planck_2020}.}
\label{fig: matter power spectrum}
\end{figure}

Measuring the matter power spectrum across different scales is one of the main goals of modern galaxy surveys. There is a consideration to be made, since the density contrast is not directly observable. Cosmological analysis must be performed using tracers, such as dark matter halos, galaxies and galaxy clusters. These objects tend to accumulate in regions of higher density, making them biased tracers of the underlying matter distribution. One simple way to model this is to think that the number density of galaxies is not proportional to the matter density, rather they are related by:
\begin{equation}
    n_g(\vb x) = \bar{n}_g\qty(1+b_1\delta_m(\vb x)),
\end{equation}
where $b_1$ is a linear galaxy bias and $\delta_\text{m}(\vb x)$ the underlying matter density contrast. In general, different tracers have different bias expansions and parameters. Figure \ref{fig: matter power spectrum} shows a compilation of observations, illustrating how the broad range of scales is probed by different experiments.

\subsection{Primordial Fluctuations and Linear Growth}
Just as with the CMB, the reason filaments, clusters, and voids are distributed non-randomly is related to the initial conditions. The origin of such structure and inhomogeneities is primordial. This is of particular interest since LSS serves as a probe for the statistics of the primordial curvature perturbation $\mathcal{R}_{\vb k}$. From linear theory, the density contrast is often characterised as \cite{dodelson_modern_2003}:
\begin{equation}
    \delta_{\vb k}(t) = \frac{2}{5} \frac{k^2}{\Omega_{\text{m},0} H_0^2} T_\text{m}(k)D_+(a)\mathcal{R}_{\vb k},
    \label{delta from primordial}
\end{equation}
where $T_\text{m}(k)$ is the matter transfer function and $D_+(a)$ is the growth factor. Physically, we have a primordial fluctuation of mode $\vb k$, that evolves into the density contrast $\delta_{\vb k}$ up until a precise time $t$. The time evolution is captured from the growing mode of the growth factor, meaning that it encodes the late time scale-independent evolution of perturbations. As for the transfer function, it contains information about evolution across the different eras of the universe, capturing the scale-dependent linear dynamics of modes inside the horizon. The 2/5 factor comes from the ratio $\Psi = \frac{3}{5} \mathcal{R}$, and the rest of the constants from the Poisson equation. Analytic expressions have been derived for the transfer function, such as Bardeen, Bond, Kaiser, and Szalay fitting function \cite{bardeen_statistics_1986}.

From \cref{delta from primordial}, we can derive an analytical expression for the linear matter power spectrum, which is valid for long wavelengths:
\begin{equation}
    P_\text{m}(k) = \frac{4}{25} \frac{k^4}{\Omega_{\text{m},0}^2 H_0^4} T^2_\text{m}(k)D^2_+(a)P_\mathcal{R}(k),
    \label{linear th matter power spectrum}
\end{equation}
where $P_\mathcal{R}(k)$ is the primordial power spectrum. Observations of the statistics of the primordial curvature, probed on scales $0.005 \text{Mpc}^{-1}\lesssim k \lesssim0.2\text{Mpc}^{-1}$, are consistent with an adiabatic, Gaussian, nearly scale-invariant primordial power spectrum \cite{akrami_planck_2020}:
\begin{equation}
    \Delta_\mathcal{R}(k) \equiv \frac{k^3P_\mathcal{R}(k)}{2\pi^2} = A_s \qty(\frac{k}{k_*})^{n_s-1},
    \label{primordial powerspec}
\end{equation}
where $A_s$ and $n_s$ are cosmological parameters, corresponding to the primordial power spectrum amplitude and the spectral index, respectively. The scale $k_*=0.05\text{Mpc}^{-1}$ is an arbitrary pivot scale. We also defined the dimensionless primordial power spectrum $\Delta_\mathcal{R}(k)$. For other power spectra, their corresponding dimensionless version is similarly defined.

By combining \cref{linear th matter power spectrum,primordial powerspec}, we get a successful description of the large-scale structure in the linear regime. This is most evident in Figure \ref{fig: matter power spectrum}, where we highlight that the power spectrum grows as $\sim k^{n_s}$, and then it decays as $\sim k^{n_s-3}$, and the turnover occurs at a scale $k_\text{eq}$. This particular scale corresponds to the scale of the Hubble horizon at the time of equality (where the universe transitioned from RDE to MDE), and is given by:
\begin{equation}
    k_\text{eq} \sim 0.02 \text{Mpc}^{-1}.
\end{equation}

\subsection{Constraints on Cosmological Parameters}
After discussing the key features of the matter power spectrum, we now detail how this observable depends on different cosmological parameters.

Two key parameters arise from the primordial origin of the matter density contrast and are clearly seen in \cref{primordial powerspec}. Naturally, if the amplitude of primordial fluctuations is stronger, we expect to have an enhanced amplitude for the matter power spectrum. On the other hand, if the primordial power spectrum turns out to have a stronger scale dependence, the overall scale dependence of the matter power spectrum will be modified.

We can also clearly see that the Hubble constant $H_0$ and the matter density parameter $\Omega_{\text{m},0}$ will change the observed matter power spectrum. Even though they appear explicitly in \cref{linear th matter power spectrum}, their impact is not that trivial. The equality time must change for a different value of $\Omega_{\text{m},0}$, so naturally the equality scale $k_\text{eq}$ is different, thus changing the position of the turnover point. Additionally, if the matter content is higher, it is natural to expect an enhanced power spectrum. Given that $H_0$ contains information about the expansion history of the universe, it also dictates when equality takes place in time, so it also shifts the peak horizontally.

Since matter is composed of cold dark matter and baryons, we should disentangle the effect. Baryon content plays a major role in the BAO features, so the wiggles seen in the matter power spectrum will be affected by the baryon density parameter, suppressing power on smaller scales when compared to a pure dark matter universe.

Additionally, recent particle experiments have established that neutrinos must have a small mass \cite{valle_neutrino_2005, schwetz_three-flavour_2008}. This is not the standard paradigm in cosmology, so we should consider how this massive particle might affect the power spectrum. Essentially, given that they are a very light particle, they have a non-negligible pressure\footnote{A formal treatment with the Boltzmann equation shows that the pressure of a non-relativistic species with a null chemical potential is exponentially suppressed by its mass: $P\sim \exp(-m/T)$.}. Thus, at small scales, they counteract matter clustering, resulting in a suppression of power on small scales.

It is of particular interest for this work to see how this observable depends on DE parameters. The effect of DE on the clustering of matter is carried by the growth factor, since it depends on the Hubble parameter $H(a)$. For different DE density and equation-of-state parameters, the entire expansion history is affected. Lastly, it also must depend on the $\sigma_8$ parameter, but this suffers from a well-known degeneracy with the matter density parameter.

\section{The \lcdm{} Model}
So far, we have discussed in detail the theoretical foundations necessary to discuss the cosmological concordance model. The \lcdm\ model has excellent statistical agreement between the theoretical predictions and observations \cite{aghanim_planck_2020}. It has been continuously tested, and now there is not much tinkering allowed since it is well established. It is difficult to explore alternatives without significantly deviating from it, and in the following chapter, we explore one of these alternatives. This success is despite three exotic ingredients whose underlying physics are still unknown: inflation, dark matter, and dark energy.

\subsection{Inflation}
Cosmic inflation came to be as an elegant and simple solution for both the \textit{flatness} and \textit{horizon} problems \cite{starobinsky_new_1980, guth_inflationary_1981}. The first corresponds to the incredibly low values measured for the curvature of the universe, which cannot be explained in simple terms\footnote{Otherwise, it is a fine-tuning problem.}. On the other hand, the \textit{horizon problem} is related to the fact that measurements of the CMB reveal that the universe at the time of last scattering was highly homogeneous and isotropic. The issue is that the size of the horizon at this instant in time corresponds to a subtended angle of around 1 degree of the whole map, implying that distant regions were causally disconnected, so there is no way for them to be so highly homogeneous.

Even though these problems might not appear to be big issues at first glance, as scientists, we would rather have a mechanism that explains these observations. Cosmic inflation solves them by introducing an early epoch of violent expansion before the Hot Big Bang. In fact, this expansion was so violent that the size of the universe was exponentially increasing with time. By imposing that the universe is undergoing an accelerated expansion, it is possible to show that:
\begin{equation}
    \ddot{a}>0 \iff \dv{t}\qty(\frac{1}{aH}) < 0,
\end{equation}
meaning that an accelerated expansion is equivalent to the comoving Hubble horizon decreasing. The inflationary mechanism has two distinct features, where the first shows that $\Omega_\text{K}$ decreases over time, explaining the observed flat universe. The second solves the horizon problem, since causally connected regions at the beginning of inflation evolved into disconnected regions by the end of it.

We must impose that this period comes to an end, which is measured in e-folds of expansion $N=\log(a_\text{ini}/a_\text{end})$. Since comoving distances decrease during inflation, there is a time for every scale at which it leaves the horizon. Given that we can only measure distances up to the observable universe, we require that inflation must last a certain number of e-folds, around 60. In other words, if we take the observable universe to have scale $\lambda_0$, then this scale must have left the horizon during inflation.

According to \cref{friedmann evolution}, the mechanism that drives the accelerated expansion should satisfy:
\begin{equation}
    \bar{\rho} + 3\bar{P} < 0.
    \label{restriction for acc universe}
\end{equation}
which, in the simplest case, is achieved by what is known as single-field slow-roll inflation. In summary, it relies on a scalar field, the \textit{inflaton}, as the main component of the universe. The inflationary mechanism is driven by the rolling of this field down the potential, ending when the scalar field reaches its minimum. For a detailed discussion, refer to \cite{achucarro_inflation_2022}.

Inflation imprints its features into the LSS, as the matter power spectrum is directly connected to the primordial power spectrum. Additionally, it provides an explanation for the high homogeneity and isotropy of the CMB, while accounting for the small deviations from it. It also predicts the generation of primordial gravitational waves from tensor perturbations of the metric. Furthermore, primordial tensor fluctuations would induce B-modes in the CMB polarisation, which are not produced by scalar perturbations, and would be strong evidence that inflation took place in the past.
\subsection{Cold Dark Matter}
Dark matter is a non-baryonic form of matter that does not interact with photons, and in our concordance cosmological model, it must be cold, meaning it should only clump efficiently in the early universe. In contrast, a hot dark matter would imply that structure would not form according to observations, and thus excludes neutrinos as a candidate for dark matter. One of the most popular ideas is that dark matter consists of elementary particles that were produced during inflation.

We have compelling evidence that, in the total matter content of the universe, about 85\% must be Cold Dark Matter (CDM), corresponding to about 25\% of the total energy content of the universe. Probes of CDM range from small to large scales in the universe. The first suggestion of dark matter came from measuring the rotational velocities of galaxies, where Vera C. Rubin showed that the velocities of stars at the outskirts of spiral galaxies deviated from Newtonian predictions. If we do not throw away Newtonian gravity, we must include extra matter that is not seen (since it does not interact with photons), that matches the rotation curves. Observations of galaxy clusters and gravitational lensing also point to a mass deficit. We have discussed how the CMB is strong evidence for the presence of CDM, but the observed LSS also needs the inclusion of this type of matter to explain the structure formation.

The characterisation of the dark matter mechanism is at the interface of different disciplines. Particle physicists, astrophysicists and cosmologists have placed constraints through experiments at particle accelerators, Milky Way astronomical observations, and galaxy surveys. For example, cosmological observations support the idea that dark matter should have a negligible equation of state parameter and sound speed, whereas particle experiments have shown that dark matter is weakly interacting with standard model particles.
\subsection{Dark Energy}
The cosmic expansion of the universe, discovered in 1998 \cite{perlmutter_measurements_1999,riess_observational_1998}, motivated the exploration of extensions to General Relativity (GR), since GR, as formulated by the standard model, cannot account for this expansion. The simplest way to do it is to include what is known as the \textit{cosmological constant}, commonly denoted by $\Lambda$. This component has an equation of state parameter $w_\Lambda = -1$, which states that the pressure it exerts equals minus its energy density, both of which are constants of time. The latter can be directly inferred by the continuity equation in \cref{continuitybk}.

As we saw in our brief discussion of inflation, in order to have an accelerated expansion, we need to have the second time derivative of the scale factor to be positive, which is the same as imposing $w_\text{DE}<-1/3$. Thus, one common way to extend the DE model is to treat its equation-of-state parameter as a free function of time. This family of models is referred to as $w$CDM, and the famous CPL parametrisation \cite{chevallier_accelerating_2001, linder_exploring_2003} can be thought of as a first-order Taylor expansion of the equation of state parameter, such that:
\begin{equation}
    w(a) = w_0 + w_a (1-a).
    \label{CPL parameterization}
\end{equation}

The dynamical approach is, in some ways, similar to the inflationary mechanism. We may introduce a dynamical scalar field that satisfies a Klein-Gordon equation in an expanding universe:
\begin{equation}
    \ddot{\phi} + 3H\dot{\phi} + \dv{V(\phi)}{\phi}=0.
    \label{klein-gordon}
\end{equation}
To obtain an expanding universe, it is required that $\dot{\phi}^2 < V(\phi)$, meaning that the potential must be shallow in order for the scalar field to slowly roll along the potential. These models receive the name of \textit{Quintessence} models, and cosmologists often propose different forms of the potential that satisfy the slow-roll condition.

\subsection{Problems and Tensions}
Despite the incredible success of the \lcdm\ model, it has its shortcomings.

The vacuum energy of quantum fields is an excellent candidate to the origin of the cosmological constant, which contributes a constant energy density of the order $\Lambda_{\rm{th}} \sim 10^{-60} \text{ to } 1\, M^4_{\rm{Pl}}$, depending on the UV cutoff of the theory \cite{weinberg_cosmological_1989, carroll_cosmological_2001, joyce_beyond_2015}. However, current measurements give $\Lambda \sim 10^{-120}M^4_{\rm{Pl}}$ \cite{aghanim_planck_2020}, which is 15 to 30 orders of magnitude smaller (in mass scale). This is known as the \textit{cosmological constant problem}, which requires a fine-tuning of the bare constant in the action of the theory to perfectly cancel the energy contribution from the quantum fields, to leave a small contribution that matches measurements. This disagreement has motivated the exploration of dark energy (DE) fields and modifications of gravity (MG) that could explain the universe's expansion.

On a similar note, several candidates for explaining the nature of CDM have been ruled out \cite{bozorgnia_dark_2025}, and the B-modes in the CMB polarisation are yet to be detected \cite{aghanim_planck_2020}. Hence, the nature of the three non-standard ingredients that the \lcdm\ model relies on remains an open question.

The $H_0$ tension is a famous issue arising from different inferred values of the Hubble constant from early- and late-time observations. The first measurements made by Hubble hinted at a value of $H_0\sim 500$ km Mpc$^{-1} s^{-1}$ \cite{hubble_relation_1929}. Then, improved measurements yielded values ranging between $50 \text{ to } 199$ km Mpc$^{-1} s^{-1}$, where variations came from statistical errors and calibrations mainly. In recent years, with precise measurements, the discrepancy has been established as one of the major puzzles of modern cosmology \cite{di_valentino_challenges_2022}. Observations of the CMB power spectrum point to a value of $H_0$ that is in a $\sim4\sigma$ tension with local observations employing the distance ladder \cite{aghanim_planck_2020, brout_pantheon_2022}, and several low-redshift probes also show disagreement with the CMB constraints \cite{di_valentino_snowmass2021_2021, perivolaropoulos_challenges_2022, liu_cosmological-model-independent_2023}. Several ideas have been proposed to solve this issue, such as early dark energy \cite{kamionkowski_hubble_2022} or interactions between dark components \cite{yao_can_2024}. Nowadays, there is a strong interest in using GWs as a probe for measuring distances, providing an independent way to infer $H_0$. In summary, there is no consensus on whether this is due to statistical analysis or a hint of new physics we are not considering \cite{di_valentino_snowmass2021_2021-1}.

Although not as significant as the $H_0$ tension, there is a considerable statistical discrepancy at the perturbation level. The $S_8$ tension emerged when recent WL surveys \cite{dark_energy_survey_collaboration_dark_2016, des_collaboration_dark_2022, kuijken_gravitational_2015, dalal_hyper_2023, aihara_hyper_2018} and CMB measurements led to different values for the growth parameter combination:
\begin{equation}
    S_8\equiv \sigma_8 \sqrt{\frac{\Omega_\text{m}}{0.3}}.
\end{equation}
In other words, the amplitude of the fluctuations inferred through low-redshift exhibits a $\sim2.5\sigma$ tension with Planck 2018 results. People actively working on solving this tension have proposed several methods, such as properly assessing non-linear effects and baryonic feedback \cite{amon_non-linear_2022, di_valentino_reducing_2018, di_valentino_exploring_2018, preston_non-linear_2023}.

This does not, in any way, mean that the \lcdm\ model needs to be disregarded. In fact, in this work we focus on tackling the cosmological problem and show that deviations from the concordance model are tightly constrained.
		 \chapter{Modified Gravity}
\label{sec:MG}
Through laboratory measurements, Solar System tests, direct detection of gravitational waves, and imaging of black holes, GR has been confirmed to great accuracy. However, these tests are not probes of gravity at cosmological scales, where the Hubble expansion becomes important. This new regime is characterised by new length scales, curvatures and a diffuse presence of matter, requiring the same level of precision and scrutiny as local tests. Upcoming Stage IV Large Scale Structure (LSS) Surveys (such as Euclid and Vera C. Rubin Observatory’s Legacy Survey of Space and Time) will enable us to perform precise observational tests of gravity at these scales. The cosmological constant problem has further motivated this exploration, since it may be that we are applying the wrong theory of gravity in our analysis.

We are allowed to tinker with the theory of gravity, but not in an arbitrary way. Lovelock's theorem states that \textit{the only second order, local gravity field equations derivable from an action containing only the 4-dimensional spacetime metric, are Einstein Field Equations} \cite{lovelock_four-dimensionality_1972}. From another perspective, to escape General Relativity (GR), we must break one of these assumptions, meaning that we could:
\begin{itemize}
    \item Add fundamental fields to our action.
    \item Go to higher dimensions.
    \item Consider higher order derivatives.
    \item Drop the idea of locality.
    \item Give up on deriving gravity from an action.
\end{itemize}
To a reader, the last possibility may seem egregious, but it could be that gravity emerges from another theory, such as thermodynamics. In our case, we go with the most popular idea: adding fundamental fields to our action.
\section{Horndeski gravity}
Gregor Horndeski presented in 1974 the most general Lorentz-invariant four-dimensional scalar-tensor theory of gravity \cite{horndeski_second-order_1974}. For about 40 years, the theory was mostly undiscussed, until in 2011 it was revisited in \cite{charmousis_general_2012}, aiming to study it on a FLRW background, as an approach to the cosmological constant problem. It came back to light since it was noted that there is an equivalence between Generalised Galileons\footnote{Which were studied in cosmological contexts at the time.} and Horndeski \cite{kobayashi_generalized_2011, deffayet_k-essence_2011}.

Some of the theories that are within this theory are $f(R)$, Brans-Dicke, kinetic braiding and non-minimal coupling models. It also encompasses most of the models explored in the context of cosmic inflation and cosmic acceleration, as subcases \cite{kobayashi_generalized_2011}. Its action is depicted by\footnote{This is not the original way in which Horndeski presented his action in 1974, rather a modern version.}:
\begin{equation}
    S\qty[g_{\mu\nu},\phi] = \int \dd[4]x\sqrt{-g}\qty[\mpls\sum_{i=2}^5\mathcal{L}_i\qty(g_{\mu\nu},\phi) + \mathcal{L}_{{\rm{m}}}\qty(g_{\mu\nu},\psi_{\rm{m}})],
    \label{Horndeski action}
\end{equation}
where $\mpls=(8\pi G)^{-1}$ is the reduced Planck mass, $\mathcal{L}_{\rm{m}}$ is the Lagrangian density of a cold dark matter component (since we are interested in late time dynamics), while the rest are:
\begin{align}
    \mathcal{L}_2 &= G_2(\phi, X) = K(\phi, X),\\
    \mathcal{L}_3 &= -G_3(\phi, X)\Box\phi,\\
    \mathcal{L}_4 &= G_4(\phi, X) R + G_{4X}(\phi,X)\left[(\Box\phi)^2-\nabla_\mu\nabla_\nu\phi\nabla^\mu\nabla^\nu\phi\right]\\
    \mathcal{L}_5 &= G_5(\phi,X)G_{\mu\nu}\nabla^\mu\nabla^\nu\phi \nonumber \\
    &-\frac{1}{6}G_{5X}(\phi,X) \left[(\Box\phi)^3+2\nabla_\mu\nabla^\nu\phi\nabla_\nu\nabla^\alpha\phi\nabla_\alpha\nabla^\mu\phi - 3\nabla_\mu\nabla_\nu\phi\nabla^\mu\nabla^\nu\phi\Box\phi\right],
\end{align}
where $\phi$ is a scalar field, $X\equiv -(\partial_\mu\phi\,\partial^\mu\phi)/2$ is the usual kinetic term of the scalar field, $\Box\phi\equiv \nabla_\mu\nabla^\mu\phi$ is the d'Alembertian of $\phi$, and $\{K,G_3,G_4,G_5\}$ are free functions of $\phi$ and $X$. To have some intuition of the free functions, we can think of $K$ as a generalization of the kinetic term of the Lagrangian, $G_3$ as a term that provides a screening mechanism, $G_4$ as a generalization of the dependence on the Ricci scalar (plus some extra exotic derivatives of the scalar field), and $G_5$ as a term that contains derivatives of the scalar field and a dependence on the Einstein Tensor. It is worth stressing that the equations of motion are still second order, since the commutation of higher order derivatives can always be rewritten in terms of the Ricci tensor through \cref{riemann_com}. As an example, it is straightforward to prove that:
\begin{equation}
    \nabla_\mu\qty(\Box{\phi}\nabla^\mu\phi+\nabla^\mu X)=\qty(\Box{\phi})^2-R_{\mu\nu}\nabla^\mu\phi\nabla^\nu\phi - \qty(\nabla_\mu\nabla_\nu\phi)\qty(\nabla^\mu\nabla^\nu\phi).
    \label{high_dervs}
\end{equation}

The model was considerably reduced since the gravitational wave event GW170817 and its electromagnetic counterpart GRB170817A constrained the difference between the propagation speed of gravitational waves and the speed of light \cite{creminelli_dark_2017, ezquiaga_dark_2017}. It was found to be:
\begin{equation}
    -3 \cdot10^{-15} \leq \frac{c_g}{c} - 1 \leq 7\cdot 10^{-16}.
\end{equation}
Since within the Horndeski family are models that allow for superluminal propagation of gravitational waves, this small difference had a major impact on two of the $G$. Explicitly, this can be seen from the sound speed for tensor perturbation given in \cite{kobayashi_generalized_2011}:
\begin{equation}
    c^2_{\rm{T}} = \frac{G_4 - X G_{5\phi} - X G_{5X}\Ddot{\phi}}{G_4 - 2XG_{4X}-X\qty(G_{5X}\dot{\phi}H-G_{5\phi})},
\end{equation}
which implies that:
\begin{equation}
    G_{4X} \approx 0,\,\,\,\,G_{5}\approx \text{const.}
\end{equation}
Note that we adopted the notation:
\begin{equation}
    G_i\equiv G_i(\phi,X),\,\,\,\, G_{i\phi} \equiv \frac{\partial G_i}{\partial \phi},\,\,\,\, G_{iX} \equiv \frac{\partial G_i}{\partial X},
\end{equation}
and it will be used for the remainder of this thesis.

In summary, we end up with what is commonly referred to as the \textit{reduced} Horndeski theories of gravity, and their Lagrangians reduce to:
\begin{align}
    \mathcal{L}_2 &= G_2(\phi, X) = K(\phi, X),\\
    \mathcal{L}_3 &= -G_3(\phi, X)\Box\phi,\\
    \mathcal{L}_4 &= G_4(\phi) R,\\
    \mathcal{L}_5 &= 0.
    \label{red_lagr}
\end{align}
For this work, we will stick to this reduced version of Horndeski gravity. 

Before moving to finding relevant equations of motion in this, it is worth noting that one recovers GR by setting $K=G_3=0$ and $G_4=1/2$, which gives the Einstein-Hilbert action:
\begin{equation}
    S[g_{\mu\nu},\phi]=\int\dd[4]x \sqrt{-g}\qty[\mpls\frac{R}{2}+\L_\text{m}(g_{\mu\nu},\psi_\text{m})].
\end{equation}
Eventually, we will be interested in interpreting an MG model as an effective DE fluid. Thus, for our purposes, we should include a cosmological constant in the Lagrangian such that we get the \lcdm\ model. For that purpose, we should set $K=-\Lambda, G_3=0$ and $G_4=1/2$, giving the action:
\begin{equation}
    S[g_{\mu\nu},\phi]=\int\dd[4]x \sqrt{-g}\qty[\mpls\qty(\frac{R}{2}-\Lambda)+\L_\text{m}(g_{\mu\nu},\psi_\text{m})].
\end{equation}
\subsection{Gravitational field equations}
We write the action for the reduced Horndeski as the purely geometric part, plus the matter field contributions as:
\begin{equation}
    S = \int\dd[4]x \sqrt{-g} \qty(\mpls\L_H + \L_{\rm{m}}) = S_H + S_{\rm{m}},
\end{equation}
where $\L_H = \sum_{i=2}^4\L_i$, $S_H = \int\dd[4]x\sqrt{-g}\mpls \L_H$, and $S_{\rm{m}} = \int\dd[4]x\sqrt{-g} \L_{\rm{m}}$.
What follows is to vary the action $S$ with respect to the inverse of the metric, for which we use the definition of the energy-momentum tensor \cref{def_emt} and set $\delta\phi=0$:       
\begin{align}
    \delta{S} &= \delta S_H + \delta S_{\rm{m}},\nonumber\\
    &= \int\dd[4]x\mpls\,\delta\qty(\sqrt{-g}\L_H) - \frac{1}{2}\int\dd[4]x\sqrt{-g}\,T_{\mu\nu}^{({\rm{m}})}\delta g^{\mu\nu},\nonumber\\
    &= \int\dd[4]x \frac{\sqrt{-g}\mpls}{2}\qty[\frac{2}{\sqrt{-g}}\delta(\sqrt{-g}\L_H) - \frac{1}{\mpls}T_{\mu\nu}^{({\rm{m}})}\delta g^{\mu\nu}],\nonumber\\
    &= \int\dd[4]x \frac{\sqrt{-g}\mpls}{2}\qty[-g_{\mu\nu}\L_H  + 2\pdv{\L_H}{X}\pdv{X}{g^{\mu\nu}}- \frac{1}{\mpls}T_{\mu\nu}^{({\rm{m}})}]\delta g^{\mu\nu},\nonumber\\
    &= \int\dd[4]x \frac{\sqrt{-g}\mpls}{2}\qty[-g_{\mu\nu}\L_H  - \pdv{\L_H}{X}\nabla_\mu\phi\nabla_\nu\phi- \frac{1}{\mpls}T_{\mu\nu}^{({\rm{m}})}]\delta g^{\mu\nu}.
\end{align}
Imposing the principle of least action, the gravitational field equations satisfied by any model of gravity within the reduced Horndeski family are:
\begin{equation}
    -g_{\mu\nu}\L_H  - \pdv{\L_H}{X}\nabla_\mu\phi\nabla_\nu\phi = \frac{1}{\mpls}T_{\mu\nu}^{({\rm{m}})},
\end{equation}
and their explicit form is\footnote{For a proper derivation, see Annex \ref{annex: Horndeski}.}:
\begin{align}
    \frac{1}{\mpls}T_{\mu\nu}^{({\rm{m}})} &= -K_X\nabla_\mu\phi\nabla_\nu\phi - Kg_{\mu\nu} + G_{3X}\Box\phi\nabla_\mu\phi\nabla_\nu\phi + 2\nabla_{(\mu}G_3\nabla_{\nu)}\phi -g_{\mu\nu}\nabla_\lambda G_3\nabla^\lambda\phi \nonumber \\
    &\quad+2G_4G_{\mu\nu}+2g_{\mu\nu}(G_{4\phi}\Box\phi-2XG_{4\phi\phi})-2G_{4\phi}\nabla_\mu\nabla_\nu\phi - 2G_{4\phi\phi}\nabla_\mu\phi\nabla_\nu\phi.
    \label{grav_field_eqs}
\end{align}
From the gravitational field equations, we can derive the Friedmann equations by assuming a homogeneous and isotropic Universe described by the FLRW metric. Substituting this metric into \cref{grav_field_eqs}, and considering a perfect fluid as the matter content, yields the two Friedmann equations governing the dynamics of the scale factor:
\begingroup
    \sloppy
\begin{align}
    \label{background1}
    -\frac{\bar{\rho}_{\rm{m}}}{\mpls} &= 2XK_X - K + 6X\dot{\phi}HG_{3X} - 2XG_{3\phi} -6H^2G_4-6H\dot{\phi}G_{4\phi} ,\\
    -\frac{\bar{P}_{\rm{m}}}{\mpls} &= K -2X\left(G_{3\phi}+\Ddot{\phi}G_{3X}\right) + 2\left(3H^2+2\dot{H}\right)G_4 + 2\left(\Ddot{\phi}+2H\dot{\phi}\right)G_{4\phi}\nonumber\\
    &\quad + 4XG_{4\phi\phi}.
    \label{background2} 
\end{align}
\endgroup

\subsection{Scalar field equation}
Given that our Lagrangian $\L_H$ is a function of the scalar field and its first and second derivatives, the scalar field equations are derived by varying the action with respect to the scalar field. This gives the familiar Euler-Lagrange equations from classical field theory:
\begin{equation}
    \pdv{\L_H}{\phi}-\nabla_\mu\pdv{\L_H}{(\nabla_\mu\phi)}+ \nabla_\mu\nabla_\nu\pdv{\L_H}{(\nabla_\mu\nabla_\nu\phi)} = 0.
\end{equation}
By using \cref{high_dervs} to get rid of higher order derivatives, we can write the scalar field equation as \footnote{Again, for a proper derivation see Annex \ref{annex: Horndeski}.}:
\begin{align}
    0 &= -K_\phi-\nabla_\mu K_X\nabla^\mu\phi-K_X\Box\phi+2G_{3\phi}\Box\phi+\nabla_\mu G_{3\phi}\nabla^\mu\phi+\nabla_\mu G_{3X}\nabla^\mu X\nonumber\\ 
    &\quad+ G_{3X}\left[(\Box\phi)^2-(\nabla_\mu\nabla_\nu\phi)^2-R_{\mu\nu}\nabla^\mu\phi\nabla^\nu\phi\right] - G_{4\phi}R.
\end{align}

\subsection{Linear perturbations}
So far, we have found everything needed to study the background evolution, but most of our interest lies in understanding how the whole system evolves. Perturbation theory allows us to systematically study how things deviate from the background, as we explored in Chapter \ref{sec:PT}.

As expected, the equations become cumbersome when we use the $G$ functions from the Horndeski action. For this section, we adopt the $\alpha$ functions\footnote{It is very important to note that we have a different effective Planck mass than in \cite{bellini_maximal_2014}. This is a consequence of factoring out the Planck mass in the action \cref{Horndeski action} from the Lagrangians $\L_i$. This decision is to have our equations as friendly as possible to implement them in \class, and for that we are following \cite{zumalacarregui_hi_class_2017}.} that were first introduced in \cite{bellini_maximal_2014}:
\begin{align}
    M^2_* & \equiv 2G_4,\\
    \alpha_{\rm{M}} & \equiv \dv{\ln M^2_*}{\ln a},\\
    H^2M^2_*\alpha_{\rm{K}} & \equiv 2X\qty(K_{X}+2XK_{XX}-2G_{3\phi}-2XG_{3\phi X}) +\frac{12H\phi'X}{a}\qty(G_{3X}+XG_{3XX}),\\
    HM^2_*\alpha_{\rm{B}} & \equiv \frac{2\phi'}{a}\qty(XG_{3X}-G_{4\phi}).
    \label{alpha_funcs}
\end{align}
This choice became popular since each of these functions has a physical interpretation:
\begin{itemize}
    \item Planck mass run rate $\alpha_{\rm{M}}$: gives the evolution of the dimensionless effective Planck mass $M^2_*$. This is the correct thing to study, since a constant redefinition of the Planck mass does not affect physics.
    \item Kineticity $\alpha_{\rm{K}}$: corresponds to the kinetic energy of scalar perturbations, coming directly from the action. Large values act to suppress the sound speed of scalar perturbations.
    \item Braiding $\alpha_{\rm{B}}$: mixing of the kinetic terms of the scalar and metric. It also contributes to the kinetic energy of scalar perturbations by backreacting through gravity. Causes clustering of Dark Energy.
    \item Tensor speed excess $\alpha_{\rm{T}}$: deviation of the speed of gravitational waves from the speed of light. Both $G_4$ and $G_5$ affect this function, but in our reduced Horndeski action, it vanishes.
\end{itemize}
By perturbing the metric, we find the linear perturbations for the gravitational field equations. Following \cite{bellini_maximal_2014}, rather than having $\delta\phi$ in the equations it is convenient to define:
\begin{equation}
    V_X \equiv -\frac{\delta\phi}{\phi'},
\end{equation}
Allowing us to find the time-time equation:
\begin{align}
    2k^2\eta &=-\frac{a^2 \delta\rho }{M^2_*\mpls}+\frac{aH}{2}(2-\alpha_{\rm{B}})h'+a^2H^2(\alpha_{\rm{K}}+3\alpha_{\rm{B}})V_X' \nonumber\\
    &\quad+\qty[\alpha_{\rm{B}}k^2+3aH'(2-\alpha_{\rm{B}})+a^2H^2(\alpha_{\rm{K}}+3\alpha_{\rm{B}})+3a^2\frac{\rho+P}{M^2_*\mpls}]aHV_X,
\end{align}
the time-space equation:
\begin{equation}
    2k^2\eta' =\frac{a^2}{M^2_*\mpls}(\rho+P)\theta+\alpha_{\rm{B}}aHk^2V_X'+k^2\qty[2aH'+a^2H^2\alpha_{\rm{B}} + \frac{a^2}{M^2_*\mpls}(\rho+P)]V_X,
\end{equation}
The space-space traceless equation:
\begin{align}
    h''&=-\frac{3a^2}{M^2_*\mpls}\delta P -aH(2+\alpha_{\rm{M}})h'+2k^2\eta-3aH\alpha_{\rm{B}}V_X'' + 2aH k^2\alpha_{\rm{M}}V_X\nonumber\\
    &\quad-3\qty[-2aH'+aH\alpha_{\rm{B}}'+\alpha_{\rm{B}}(a^2H^2(4+\alpham)+aH')]V_X'\nonumber\\
    &\quad-3\qty[2aH''+2a^2HH'(3+\alpha_{\rm{B}}+\alpha_{\rm{M}}) +a^3H^3\alpha_{\rm{M}}(\alpha_{\rm{B}}-4)+a^2H^2\alpha_{\rm{B}}']V_X,\nonumber\\
    &\quad-\frac{3a^2}{M^2_*\mpls}\qty[(\rho+P) (V_X'+aHV_X) +P'\,V_X],
\end{align}
and the space-space trace equation:
\begin{align}
    6\eta''&+h''+aH(2+\alpha_{\rm{M}})\qty(6\eta'+h')-2k^2\eta= 2aH k^2\alpha_{\rm{M}}V_X-\frac{3a^2}{M^2_*\mpls}(\rho+P)\sigma.
\end{align}
Additionally, the scalar field equation of motion is:
\begin{align}
    a^2 H^2\qty(\alpha_{\rm{K}} + \frac{3}{2}\alpha_{\rm{B}}^2)V_X'' + A V_X'+ \qty(3B + Ck^2)V_X &= \frac{C}{2}h'-\frac{3\alpha_{\rm{B}}}{2M^2_*\mpls}a^2\delta P \nonumber\\ &\quad - \qty(2\alpha_{\rm{M}}+\alpha_{\rm{B}})\frac{a^3H\delta\rho}{2M^2_*\mpls},
    \label{scalar 1st order}
\end{align}
where $A,B$ and $C$ are given in \cref{eqAannex,eqBannex,eqCannex}. Note that we dropped the subscript m, so that later we can consider multiple species rather than pressureless cold dark matter only.

\subsection{Instabilities}
It is important to highlight that, given the generality of this family of models, it is natural to expect problems for some particular choice of theory. Before we attempt to make a theory consistent with observations or numerically stable, we must avoid physical instabilities.

The simplest instability that might arise is related to a wrong sign in the kinetic term. These are known as \textit{ghost instabilities}, and they arise when the kinetic term is negative, meaning that the kinetic energy is not bounded from below. Thus, we would find ourselves in a universe in which the vacuum could decay into an infinite number of particles with positive and negative energy. Mathematically, it can be seen from \cref{scalar 1st order} that the combination
\begin{equation}
    D_{\rm kin} \equiv \alphak + \frac{3}{2}\alphab^2,
\end{equation}
must be positive, together with the effective Planck mass squared in order to avoid ghost instabilities.

Additionally, we could find an imaginary sound speed for the scalar field, defined as:
\begin{equation}
    c_{\mathrm{s}}^{2} \equiv \frac{1}{D_\text{kin}} \frac{C}{a^2H^2},
\end{equation}
which must be strictly positive, otherwise it would imply that short-wavelength modes would grow violently rather than oscillate. They are known as \textit{gradient instabilities}, and they typically manifest on time scales shorter than a Hubble time, rendering the cosmological background unviable. This is because the friction term driven by the Hubble expansion, which accompanies the $V_X'$ term, effectively counteracts the tachyon's effect.

Finally, \textit{tachyonic instabilities} are present when the effective mass of the scalar field $m_{\rm eff}^2 = 3B$ is negative. Although in principle these instabilities are not instantly fatal, their growth rate determines whether they can still allow a stable background.

\section{Effective fluid approach}
In \cite{kunz_dark_2007} it was found that at the linear level, there is no way of distinguishing a model of modified gravity from a dynamical dark energy one. This is seen when rewriting the gravitational field equations from an MG gravity model, as the Einstein field equations plus an effective dark energy fluid:
\begin{equation}
    \frac{1}{\mpls} T^{\rm{(DE)}}_{\mu\nu} = G_{\mu\nu} - \frac{1}{\mpls} T_{\mu\nu},
\end{equation}
where $G_{\mu\nu}$ is the usual Einstein tensor found in GR, $T^{\rm{(DE)}}_{\mu\nu}$ contains all the extra terms that come from the MG model, and $T_{\mu\nu}$ contains every energy source that we consider, except for DE. Thus, every quantity related to $T_{\mu\nu}$, will have an implicit sum running over every species (cold dark matter, baryons, neutrinos, radiation, and other exotic species can be considered). This motivates us to rewrite the background \cref{background1,background2} in terms of an effective dark energy density and pressure, both at background:
\begin{equation}
    \frac{\bar{\rho}_{\rm{DE}}}{3\mpls} = H^2[1-(1-\alpha_{\rm{B}})M^2_*]-\frac{K}{3}+\frac{2X}{3}(K_X-G_{3\phi})
    \label{eff de density}
\end{equation}
\begin{equation}
    \frac{\bar{P}_{\rm{DE}}}{3\mpls}=\frac{K}{3}-\frac{HM^2_*\alpha_{\rm{B}}}{3}\frac{\ddot{\phi}}{\dot{\phi}}-\frac{2X}{3}(G_{3\phi}-2G_{4\phi\phi})+\frac{3H^2+2\dot{H}}{3}(M^2_*-1)+\frac{4H\dot{\phi}}{3}G_{4\phi}
    \label{eff de pressure}
\end{equation}
which allows us to define an effective dark energy equation of state parameter as:
\begin{equation}
    w_{\rm{DE}} = \frac{K-HM^2_*\alpha_{\rm{B}}(\ddot{\phi}/\dot{\phi}) -2X(G_{3\phi}-2G_{4\phi\phi})+3H^2+2\dot{H}(M^2_*-1)+4H\dot{\phi}G_{4\phi}}{3H^2[1-(1-\alpha_{\rm{B}})M^2_*]-K+2X(K_X-G_{3\phi})}.
    \label{eff de w}
\end{equation}
Hence the modified Friedmann equations take the following familiar form:
\begin{align}
    3H^2&=\frac{1}{\mpls}\qty(\bar{\rho}+\bar{\rho}_{\rm{DE}}),\\
    -\left(2\dot{H} +3H^2\right) &= \frac{1}{\mpls} \qty(\bar{P} + \bar{P}_{\rm{DE}}),
\end{align}
where the equation of state for the effective DE fluid holds by definition:
\begin{equation}
     \bar{P}_{\rm{DE}} = w_{\rm{DE}}\,\bar{\rho}_{\rm{DE}}.
\end{equation}
Importantly, if we take the limit back to \lcdm, we correctly have:
\begin{equation}
    \bar{\rho}_{\rm{DE}} = \mpls\Lambda,\quad \bar{P}_{\rm{DE}} = -\mpls\Lambda, \quad w_{\rm{DE}}=-1.
\end{equation}

Similarly, the perturbed gravitational field equations can be rewritten in terms perturbations of an effective DE fluid:
\begin{align}
    k^2\eta-\frac{1}{2}aHh'&=-4\pi Ga^2\qty(\delta \rho+\delta \rho_{\rm{DE}}),\\
    k^2\eta'&=4\pi Ga^2\qty[(\bar{\rho}+\bar{P})\theta + (\bar{\rho}_{\rm{DE}}+\bar{P}_{\rm{DE}})\theta_{\rm{DE}}],\\
    h''+2aHh'-2k^2\eta&=-24\pi Ga^2\qty(\delta P + \delta P_{\rm{DE}}),\\
    h''+6\eta''+2aH\left(h'+6\eta'\right)-2k^2\eta&=-24\pi Ga^2\qty[(\bar{\rho}+\bar{P})\sigma + (\bar{\rho}_{\rm{DE}}+\bar{P}_{\rm{DE}})\sigma_{\rm{DE}}],
\end{align}
where $\delta \rho_{\rm{DE}},\delta P_{\rm{DE}}, \theta_{\rm{DE}}$ and $\sigma_{\rm{DE}}$ contain all deviations from GR and are given by:
\begin{align}
    &\frac{\delta\rho_{\rm{DE}}}{\mpls} = \frac{\delta\rho}{\mpls}\frac{(1-M^2_*)}{M^2_*}+\frac{H}{2a}\alpha_{\rm{B}}h' - H^2(\alpha_{\rm{K}} + 3\alpha_{\rm{B}})V_X'\nonumber\\ 
    &\quad\quad\quad\quad- \qty[\alpha_{\rm{B}}k^2+3aH'(2-\alpha_{\rm{B}})+a^2H^2(\alpha_{\rm{K}}+3\alpha_{\rm{B}})+3a^2\frac{\rho+P}{M^2_*\mpls}]\frac{H}{a}V_X,\label{eq:delta rho de}\\
    &\frac{\delta P_{\rm{DE}}}{\mpls} = \frac{\delta P }{\mpls}\frac{(1-M_*^2)}{M_*^2}+ \frac{H}{3a}\alpha_{\rm{M}}h'+\frac{H}{a}\alpha_{\rm{B}}V_X''-\frac{2H}{3a}k^2\alpha_{\rm{M}} V_X \nonumber\\
    &\quad\quad\quad\quad +\qty[\frac{2H'}{a}+\frac{H}{a}\alpha_{\rm{B}}'+4H^2\alpha_{\rm{B}}+\frac{H'}{a}\alpha_{\rm{B}} + H^2\alpha_{\rm{B}}\alpha_{\rm{M}}]V_X'\nonumber\\
    &\quad\quad\quad\quad + \qty[6HH'+\frac{2H''}{a}+H^2\alpha_{\rm{B}}'+aH^3\alpha_{\rm{B}}\qty(4+\alpha_{\rm{M}})+H H'\qty(\alpha_{\rm{M}}+\alpha_{\rm{B}})]V_X\nonumber\\
    &\quad\quad\quad\quad+\frac{1}{M^2_*\mpls}\qty[(\rho+P)\qty(V_X'+aHV_X) +P'\,V_X],\label{eq:delta p de}\\
    &\frac{(\bar{\rho}_{\rm DE} + \bar{P}_{\rm DE})\theta_{\rm DE}}{\mpls}= \frac{(\bar{\rho}+\bar{P})\theta}{\mpls}\frac{(1-M_*^2)}{M_*^2}\nonumber\\
    &\quad\quad\qquad\qquad\qquad+k^2\qty[\frac{H}{a}\alphab V_X'+\qty(\frac{2H'}{a}+H^2\alpha_{\rm{B}} +\frac{\rho+P}{M^2_*\mpls })V_X]\label{eq:theta de}\\
    &\frac{(\bar{\rho}_{\rm{DE}} + \bar{P}_{\rm{DE}})\sigma_{\rm{DE}}}{\mpls} =\frac{(\rho+P)\sigma}{\mpls}\frac{(1-M^2_*)}{M^2_*} + \frac{H\alpha_{\rm{M}}}{3a}\qty(h'+6\eta'-2k^2V_X)\label{eq:shear de}.
\end{align} 

Hence, we have a set of equations that can be implemented in Einstein-Boltzmann solvers. One example of this is \texttt{EFCLASS} \cite{arjona_unraveling_2019, arjona_designing_2019}, where they implemented MG models with this approach, finding excellent agreement with \texttt{hi\_CLASS} \cite{zumalacarregui_hi_class_2017}. The latter is a patch of \class\ that includes Horndeski gravity. 

\section{Gravity models}
Since Horndeski encompasses a plethora of models, for the remainder of this thesis, we shift our focus to two models beyond the scope of GR. We will show how the Lagrangians of each theory of gravity look, and their DE counterparts.
\subsection{\texorpdfstring{$f(R)$}{f(R)} Hu-Sawicki}
The family of $f(R)$ theories is broad, as it generalises the Einstein–Hilbert action by allowing the gravitational Lagrangian to be an arbitrary (not necessarily linear) function of the Ricci scalar $R$. After considering the object $1+f_R$ as a scalar field $\phi$, with $f_R\equiv \dv*{f}{R}$, we see an action that looks like a scalar-tensor theory with no kinetic term and a potential. Even though $f(R)$ theories could naturally contain higher-order derivatives (and potentially be unstable), they are stable since they are a scalar-tensor theory in disguise. Thus, we can write these theories in terms of Horndeski functions: 
\begin{equation}
    K = \frac{f-Rf_R}{2}, \quad G_3 = 0, \quad  G_4=\frac{1+f_R}{2}=\frac{\phi}{2},
\end{equation}
or in terms of the $\alpha$ functions:
\begin{equation}
    M^2_* = 1+f_R,\quad \alpha_{\rm{M}}=\frac{1}{1+f_R}\dv{f_R}{\ln a},\quad \alpha_{\rm{B}}=-\alpha_{\rm{M}},\quad \alpha_{\rm{K}}=0.
\end{equation}
It is important to note that we recover GR by fixing $f(R)=0$, which is more clearly seen in the action:
\begin{align}
    S_\text{HS} &= \int\dd[4]{x}\sqrt{-g}\mpls\qty[\frac{f(R)-Rf_R}{2} + R\frac{1+f_R}{2}],\\
    &= \int\dd[4]{x}\sqrt{-g}\mpls\qty[\frac{f(R) + R}{2}].
\end{align}

At this point we can take two approaches, by either giving explicitly the expansion history and solving for $f(R)$, or vice versa. Here we take the first approach, and we specifically choose the famous Hu-Sawicki function \cite{hu_models_2007}. This function satisfies certain observationally desired properties, such as mimicking \lcdm\ at higher redshifts, self-acceleration at lower redshifts without a cosmological constant, etc. These requirements suggest that $f(R)$ should have the conditions:
\begin{equation}
    \lim_{R\to\infty}f(R)=\text{const,}\quad\lim_{R\to 0}f(R)=0,
\end{equation}
which are satisfied by broken power law models:
\begin{equation}
    f(R) = -m^2\frac{c_1(R/m^2)^n}{1+c_2(R/m^2)^n},
\end{equation}
with $n>0$, and for convenience the mass scale $m$ is set to:
\begin{equation}
    m^2 = \frac{1}{\mpls} \frac{\bar{\rho}_0}{3},
\end{equation}
where $\bar{\rho}_0$ is the average density today. In addition, the sign of $f(R)$ is taken such that its second derivative with respect to the Ricci scalar is positive, to ensure stability at high curvature \cite{song_large_2007}. If we demand convergence to a cosmological constant at early times, we fix one of the constants through:
\begin{equation}
    \frac{c_1}{c_2} = 6\frac{\Tilde{\Omega}_\Lambda}{\Omega_\text{m}},
\end{equation}
where $\Tilde{\Omega}_\Lambda =\lambda \Omega_\Lambda$ and $\lambda$ is a real positive number that informs of deviations from the value $\Omega_\Lambda = 1-\Omega_\text{m} - \Omega_\text{r}$, which holds at late times. Lastly, the combination of parameters $c_1/c_2^2$ can be rewritten in terms of the initial value $f_{R,\text{ini}}$:
\begin{equation}
    f(R) = -6m^2\frac{\Tilde{\Omega}_\Lambda}{\Omega_\text{m}} - \frac{f_{R,\text{ini}}}{n} R_\text{ini}\qty(\frac{R_\text{ini}}{R})^n.
    \label{eq:f(R) hu-sawicki}
\end{equation}
Now, we present what the effective fluid looks like for a theory of $f(R)$:
\begin{align}
    \bar{\rho}_{\rm DE} &= -3\mpls \qty[H^2\qty(f_R+\dv{f_R}{\ln a}) +\frac{f-Rf_R}{3}],\\
    \bar{P}_{\rm DE} &=\mpls\qty[\frac{f-Rf_R}{2}+H\qty(\dv{f_R}{\ln a}\frac{\ddot{f_R}}{\dot{f_R}}+2\dot{f_R}) + \qty(3H^2+2\dot{H})f_R],\\
    \frac{\delta\rho_{\rm DE}}{\mpls} &= \frac{\delta\rho}{\mpls}\frac{f_R}{1+f_R}- \frac{H}{2a}\frac{h'}{1+f_R}\dv{f_R}{\ln a} + \frac{2H^2}{1+f_R}\dv{f_R}{\ln a}V_X'\nonumber\\
    &\quad+\qty[\frac{1}{1+f_R}\dv{f_R}{\ln a}\qty(k^2 - 3aH'+2a^2H^2)-6aH'-\frac{3a^2(\rho+P)}{(1+f_R)\mpls}]\frac{H}{a}V_X,\\
    \frac{\delta P_{\rm DE}}{\mpls}&=\frac{\delta P}{\mpls}\frac{f_R}{1+f_R}+\frac{1}{1+f_R}\dv{f_R}{\ln a}\qty[\frac{H}{3a}h'-\frac{H}{a}V_X''-\frac{2H}{3a}k^2V_X]\nonumber\\
    &\quad+\qty[\frac{2H'}{a}-\frac{H}{a}\qty(\frac{1}{1+f_R}\dv{f_R}{\ln a})'-\frac{1}{1+f_R}\dv{f_R}{\ln a}\qty(4H^2 + \frac{H'}{a} + H^2\qty(\frac{1}{1+f_R}\dv{f_R}{\ln a}))]V_X'\nonumber\\
    &\quad+\qty[6HH'+\frac{2H'}{a}+H^2\frac{1}{1+f_R}\dv{f_R}{\ln a} - \frac{aH^3}{1+f_R}\dv{f_R}{\ln a}\qty(4+\frac{1}{1+f_R}\dv{f_R}{\ln a})]V_X\nonumber\\
    &\quad+\frac{1}{(1+f_R)\mpls}\qty[(\rho+P)(V_X'+aHV_X)+P'V_X],
 \end{align}
\begin{align}
    \frac{(\rho_{\rm DE} + P_{\rm DE})\theta_{\rm DE}}{\mpls}&= \frac{(\rho + P)\theta}{\mpls}\frac{f_R}{1+f_R} + \frac{k^2}{1+f_R}\dv{f_R}{\ln a}\qty(-\frac{H}{a}V_X'-H^2V_X)\nonumber\\
    &\quad+k^2\qty(\frac{2H'}{a}+\frac{\rho+P}{(1+f_R)\mpls})V_X,\\
    \frac{(\rho_{\rm DE} + P_{\rm DE})\sigma_{\rm DE}}{\mpls}&= \frac{(\rho + P)\sigma}{\mpls}\frac{f_R}{1+f_R}+\frac{H}{3a}\frac{1}{1+f_R}\dv{f_R}{\ln a}\qty(h'+6\eta'-2k^2V_X).\label{eq: shear f(R)}
\end{align}
It is important to note, that \cref{eq: shear f(R)} predicts an effective shear for this model of gravity. However, for late-time dynamics we have a universe mainly composed of matter and dark energy, such that the term proportional to $\sigma$\footnote{The term $(\bar{\rho} + \bar{P})\sigma$ corresponds to a sum over all species, except for dark energy.} is negligible. Additionally, the remaining term is expected to be zero, since in the high curvature regime we have $f_R\ll 1$, and it is accompanied by first-order perturbed quantities.
\subsection{Cubic Galileon}
Generalised Galileons correspond to a scalar-tensor theory of gravity, where the scalar field is invariant under Galilean transformations. Additionally, to avoid Ostrogradsky instabilities, the Lagrangian is written such that equations of motion are of second-order. It was first introduced in \cite{de_felice_cosmology_2010}, and as mentioned earlier, it is the exact same theory as Horndeski. One important subclass of this family is cubic Galileons, and its action is:
\begin{equation}
    S_\text{cGal} = \int\dd[4]x \sqrt{-g}\mpls\qty[\frac{R}{2}+c_2 X - 2c_3\frac{M_{\rm Pl}}{\Lambda_3^3}X\Box\phi],
\end{equation}
where $c_2$ and $c_3$ are dimensionless constants that parametrise deviations from General Relativity, and to obtain interesting dynamics for the field on cosmological scales. We take the common choice for $\Lambda_3^3=M_\text{Pl}H_0^2$. We can notice that the free functions of Horndeski are:
\begin{equation}
    G_2 = c_2 X, \quad G_3 = 2c_3\frac{M_{\rm Pl}}{\Lambda_3^3}X, \quad G_4 = \frac{1}{2}.
\end{equation}
Under a convenient redefinition of the scalar field, $c_2$ can be normalised to $\pm1$, where the sign plays an important role for the expansion history of the universe. For the positive sign, a cosmological constant is required to drive the accelerated expansion of the universe, whereas for the negative sign, the model self-accelerates. Because Galileon models are shift-symmetric by definition, the initial conditions chosen for the scalar field are irrelevant. Furthermore, the initial conditions for its derivative can be fixed by following the attractor solution $H\dot{\phi}=\text{const.}$

Lastly, the remaining parameter can be fixed through:
\begin{equation}
    6\Omega_{\rm DE,0} = \frac{1}{36c_3^2},
\end{equation}
such that the sum of all density parameters equals unity for a flat universe. Hence, the model has no additional parameters to be fixed.

Now we turn to the $\alpha$ functions, where the parameters are already fixed:
\begin{equation}
    M_*^2=1, \quad \alpham=0,\quad \alphak=3\alphab=\frac{6\Lambda_3^6\Omega_{\rm DE,0}}{\mpls H^4},
\end{equation}
which are useful to present the quantities related to the effective DE fluid for a cubic Galileon theory of gravity:
\begin{align}
    \bar{\rho}_{\rm DE} &= \frac{3\Lambda_3^6\Omega_{\rm DE,0}}{H^2},\\
    \bar{P}_{\rm DE} &= -\frac{\Lambda_3^6\Omega_{\rm DE,0}}{H^2}\qty(3 + \frac{2\ddot{\phi}}{\dot{\phi}}),\\
    \frac{\delta\rho_{\rm DE}}{\mpls} &= \frac{\Lambda_3^6\Omega_{\rm DE,0}h'}{aH^3\mpls}-\frac{12\Lambda_3^6\Omega_{\rm DE,0}}{H^2\mpls}V_X'\nonumber\\
    &\quad-\qty[6aH' + \frac{2\Lambda_3^6\Omega_{\rm DE,0}}{H^2\mpls}\qty(\frac{k^2}{H^2} - \frac{3aH'}{H^2}+6a^2)+\frac{3a^2(\rho+P)}{\mpls}]\frac{H}{a}V_X,
\end{align}
\begin{align}
    \frac{\delta P_{\rm DE}}{\mpls} &= \frac{2\Lambda_3^6\Omega_{\rm DE,0}}{aH^3\mpls}V_X''+\qty[\frac{2H'}{a}+\frac{\Lambda_3^6\Omega_{\rm DE,}}{H^4\mpls}\qty(8H^2-\frac{2H'}{a})]V_X'\nonumber\\
    &\quad+\qty[6HH'+\frac{2H''}{a}+\frac{2\Lambda_3^6\Omega_{\rm DE,0}}{H^4\mpls}\qty(4aH^3 - HH')]V_X\nonumber\\
    &\quad +\frac{1}{\mpls}\qty[(\rho+P)(V_X'+aHV_X)+P'V_X],\\
    \frac{(\bar{\rho}_{\rm DE} + \bar{P}_{\rm DE})\theta_{\rm DE}}{\mpls} &= k^2\qty[\frac{2\Lambda_3^6\Omega_{\rm DE,0}}{aH^3\mpls}V_X' + \qty(\frac{2H'}{a}+\frac{2\Lambda_3^6\Omega_{\rm DE,0}}{H^2\mpls}+\frac{\rho+P}{\mpls})V_X],\\
    \frac{(\bar{\rho}_{\rm DE} + \bar{P}_{\rm DE})\sigma_{\rm DE}}{\mpls} &= 0.
\end{align}
		\chapter{Numerical Tools}
\label{sec:numerics}
To obtain an accurate description of the universe, we saw in chapter \ref{sec:PT} that we must solve the Einstein field equations and characterise the energy sources using Boltzmann equations. In general, the system of equations is highly non-linear, but may be linearised on the largest scales, vastly simplifying the problem. Even then, high complexity renders a purely analytic treatment unfeasible. This is where numerical methods for solving the system come into play, allowing not only fast and accurate predictions of observables but also tinkering with cosmological parameters, different models, exotic species, etc. Previously, we have mentioned two of the most popular Einstein-Boltzmann solvers, \texttt{CAMB} and \class. The first was first introduced in 2000 \cite{lewis_efficient_2000}, while the second was released in 2011 \cite{lesgourgues_cosmic_2011, diego_blas_cosmic_2011, lesgourgues_classIII_2011}. Predictions from these codes might be considered virtually exact within the linear regime.

While the description of the linear regime has been extensively explored and tested, we need to go beyond it to describe the late universe. Gravitational collapse is a highly non-linear process that leads to the formation of stars, galaxies, clusters, and so on. This regime presents several analytical complications, including the mixing of distinct Fourier modes, non-Gaussian statistics, and the coupling among several species. For example, gas (baryon) pressure must be considered on very small scales due to its interactions with the electromagnetic field.

Certainly not everything is lost, and there have been remarkable analytic treatments for this regime, such as Effective Field Theories applied to LSS \cite{baumann_cosmological_2012, carrasco_effective_2012}, Dark Energy \cite{gubitosi_effective_2013, frusciante_effective_2020,} and their combination \cite{gergely_effective_2014, cusin_dark_2018, lewandowski_effective_2017}. They allow accurate prediction of the matter power spectrum in the mildly non-linear regime with arbitrary precision, at the cost of additional free coefficients that must be fitted from observations or numerical simulations. Additionally, they allow for an extensive exploration of Dark Energy models, exploiting the theory's symmetries and degrees of freedom. However, these analytic treatments are incredibly complex on their own, and in this work, we focus on numerical tools to study the highly non-linear regime. For that purpose, we focus on numerical cosmological N-body simulations and use the effective fluid approach described in the previous chapter.

\section{The Cosmic Linear Anisotropy Solving System}
The Cosmic Linear Anisotropy Solving System (\class) is an Einstein-Boltzmann solver code written in \texttt{C}, widely used by cosmologists. The main purpose of the code is to simulate linear perturbations and compute CMB and LSS observables. For a short tutorial on obtaining the code, running it, and considering modifications, refer to Annex \ref{annex: CLASS}.

The code was written with user-friendliness, flexibility, accuracy, and speed in mind. This is reflected by its ease of use, compilation and generalisation. For example, there is no hard-coding to maintain a clear structure, resulting in physical assumptions and equations being in a single well-identified place in the code. Also, dynamic index allocation allows developers to avoid memorising each index. This decision has led to several code patches that extend the capabilities of \class, enabling the implementation of non-standard reionisation histories, new species, or imperfect fluids. Additionally, the code is 2.5 times faster than \texttt{CAMB} in the \lcdm\ context \cite{lesgourgues_cosmic_2011}.

\subsection{Units}
The code uses units such that $c=\hbar=\kappa_{\rm{B}}=1$, and energy densities are rescaled such that the Friedmann equations read:
\begin{align}
    H^2&=\sum_{\rm{species}}\bar{\rho},\\
    \dot{H} &= -\frac{3}{2} \sum_{\rm{species}}(\bar{\rho} + \bar{P}),
\end{align}
such that if we compare a \class\ energy density with a physical one, they relate via:
\begin{equation}
    \rho^\class = \frac{8\pi G}{3}\rho^\text{physical} = \frac{1}{3\mpls}\rho^\text{physical}.
\end{equation}

The units for wavenumbers and the Hubble rate are inverse mega-parsecs, while the conformal time (time variable chosen by the code to solve differential equations) has units of mega-parsecs. The output of CMB multipoles is dimensionless, unlike that from \texttt{CAMB}, which has units of temperature squared. In general, the output tables include a comment indicating the unit of the quantity used or computed by the code.

\subsection{Structure of the Code}
The code is organised across multiple directories, but here we will focus on the most physically interesting one: the \texttt{source/} folder, which contains the most important modules. It is of special interest for implementing modifications, as we will discuss later. Other directories have numerical algorithms, the main module, tests, and include files.

There are 11 modules in the standard version, each associated with a structure (except for one) and at least two essential functions. The \texttt{module\_i\_init()} function will initialise the module, meaning that it will contain information about everything that needs to be computed, given that the previous modules were executed. The \texttt{module\_i\_free()} function frees the memory, erasing intermediate quantities that were necessary to arrive at the requested output. Arrays within a structure are sized and accessed through dynamic index allocation. There is no hardcoded index for the baryon density; it is assigned for each code run, easing the implementation of new quantities, such as an anisotropic stress term. Briefly, the role of each module is described as follows:

\begin{itemize}
    \item \texttt{input.c} is the first module, so it is unique. It is the only one without a structure, and it contains two \texttt{ini()} functions. Its main task is to convert the input parameter files into structures for the following modules and make them ready for interface with other codes. The other files this module reads are precision files that group accuracy parameters in one place. Additionally, since it lacks its own structure, it does not have a \texttt{free()} function.
    \item \texttt{background.c} is the one in charge of solving the background equations. This is the only place in the code where the Friedmann equations appear, in a single line. It also includes interpolators to make it easy for subsequent modules to access background quantities at specific values of conformal time or redshift.
    \item \texttt{thermodynamics.c} is quite similar to the background module; it solves for the thermodynamical evolution using RECFAST \cite{seager_new_1999,scott_matter_2009} and corrects it for reionisation at low redshifts.
    \item \texttt{perturbation.c}, as with the previous two modules, solves for all perturbations. It stores source functions in a table inside the structure \texttt{perturbs}.
    \item \texttt{bessel.c} is also a unique module since it is the only one that, in principle, could be run after input and before the transfer module. It computes spherical Bessel functions, so when running multiple instances of the code, it is recommended not to run this module every single time.
    \item \texttt{transfer.c} computes transfer functions by convolving the source functions and Bessel functions.
    \item \texttt{primordial.c} computes the primordial power spectra for each mode and initial condition.
    \item \texttt{spectra.c} computes observable power spectra as requested by the user. 
    \item \texttt{nonlinear.c} estimates the non-linear version of the power spectra according to a user-defined scheme. One example is HALOFIT.
    \item \texttt{lensing.c} is in charge of obtaining the lensed temperature and polarisation CMB power spectra, using the unlensed spectra and CMB lensing potential spectrum.
    \item \texttt{output.c} writes the output in files. When interfaced with other codes, it is not necessary to call this module. All the fields within the output structure can simply be freed at the very end of the \texttt{output\_init()} function itself, so it does not have a \texttt{free()} function.
\end{itemize}
After this short overview of \class, we now turn our attention to N-body simulations to study the non-linear regime. While \class\ excels at making predictions within the linear regime, N-body codes are able to reproduce structures that form in highly non-linear scales. However, simulations rely on linear transfer functions and the primordial power spectrum to generate initial conditions for the particles and fluids. Additionally, N-body codes use the Newtonian limit of GR, and to have observables consistent with GR on the largest scales, it is required to know and implement predictions from an Einstein-Boltzmann solver. To achieve this, output from \class\ must be transformed into the N-body gauge.

\section{Cosmological N-body Simulations}
Cosmological N-body simulations are the key to accessing non-linear scales where the density contrast is much larger than unity. It is of the utmost importance to study these scales, since they provide a large amount of information due to the large number of independent modes available. N-body simulations are often used for many purposes, such as testing models of dark matter, analysing the impact of initial conditions, and comparing with approximate solutions for the growth of density perturbations, etc. There have been numerous codes released through the years \cite{couchman_hydra_1995, xu_new_1995, springel_gadget_2001, springel_cosmological_2005, dubinski_gotpm_2004, teyssier_ramses_2010, angulo_scaling_2012, harnois-deraps_high-performance_2013}

Essentially, cosmological N-body simulations consist of evolving a large number of particles in a large box with periodic boundary conditions, with the background evolving dynamically. Particles from the simulation are considered as dark matter particles, and the reason for using dark matter particles is that they are dominant compared to baryonic matter, which makes simulations somewhat simpler since electromagnetic interactions are absent\footnote{It is possible to populate the dark matter halos that are simulated with galaxies, which is a topic on its own that goes beyond the scope of this manuscript.}.

While implementing a large number of particles that interact only gravitationally might seem simple, there are several aspects to consider. The first is to think through how to implement the interaction between two particles, which should mimic the interaction between two fluid elements, since each simulated particle is treated as a large number of dark matter particles. The evolution of the underlying fluid is dictated by the collisionless Boltzmann equation, coupled with the Poisson equation in the Newtonian limit. Consequently, given that fluid elements have a physical size, they should feel an external force much smaller compared to the case of two-point particles, at a scale comparable to the fluid size. Thus, in the simulation, each particle has a finite size and a density profile, leading to a softening of the force at small scales. The choice of this scale might lead to issues at smaller scales, and its form is important because it must reproduce the inverse-square law at the largest scales. Not implementing this softening scale would lead to two-body relaxation and scattering, which is completely unphysical for the case of cold dark matter.

Initial conditions for the simulation should be consistent with a primordial power spectrum predicted by inflation. Primordial density fluctuations form a Gaussian random field that is completely described by its power spectrum. Using linear transfer functions, this primordial power spectrum can be translated into a matter power spectrum, which must be evolved to the initial time of the simulation. However, we immediately notice that multiple different initial configurations of the system are consistent with the primordial power spectrum\footnote{In fact, in order to get rid of cosmic variance from simulations, older techniques relied on running multiple of these realisations and then averaging observables. The modern way to do it, is thorugh the fixed and paired initial conditions method \cite{angulo_cosmological_2016}.}, which are called realisations. To build the initial continuous density field, random numbers are drawn from a normal distribution for a given wavenumber, and the inverse Fourier transform is then performed. Then, to reach a discrete configuration, we use the Zel'dovich approximation, also known as first-order Lagrangian perturbation theory (1LPT), which involves displacing particles initialised in a perfectly homogeneous configuration. While this would give the initial positions, the velocities can be directly derived using the same scheme, now considering velocity transfer functions. Since in the linear regime, time evolution is dictated by the linear growth factor, velocities are described through the \textit{linear growth rate}, defined as:
\begin{equation}
    f\equiv\dv{\ln D}{\ln a},
\end{equation}
such that the relationship between the velocity and the density field is given by:
\begin{equation}
    \vb v = i\vb k \frac{\H f \delta}{k^2},
    \label{zeldovich app}
\end{equation}
which can be derived from \cref{delta cng2} for a pressurless matter component.

In general, to describe structure formation in this regime, we have to solve the Poisson equation in the Newtonian limit and make it consistent with GR on the largest scales by using linear transfer functions in the N-body gauge. Even though we are considering non-linear density perturbations that are much larger than unity, metric perturbations remain linear. The reason is that the Poisson equation in the Newtonian limit, for a matter-dominated universe, yields a relation:
\begin{equation}
    \Phi \sim \frac{\delta}{k^2},
\end{equation}
and even though $\delta\gg1$, the wavenumber is also a quantity with large values, resulting in $\Phi$ being a quantity well described in the linear regime. Thus, the relevant equations for the code are Hamilton equations, provided the gravitational potential through the Poisson equation:
\begin{align}
    \pdv{\vb x}{t} &= \frac{\vb q}{a^2(t)m},\label{hamilton1}\\
    \pdv{\vb q}{t} &= -m\grad \Phi(\vb x),\label{hamilton2}\\
    \laplacian \Phi(\vb x) &= 4\pi G_N a^2(t) \delta \rho(\vb x), \label{sim poisson}
\end{align}
where $\vb x$ and $\vb q$ are in comoving coordinates, and $\vb q = a^2(t)m\dot{\vb x}$ its the canonical momentum. This choice results in the background acting as a viscous term, which becomes clearer when combining equations \cref{hamilton1,hamilton2}:
\begin{equation}
    \ddot{\vb x} + 2H(t)\dot{\vb x} = -\frac{1}{a^2(t)}\grad\Phi(t,\vb x).
\end{equation}

\section{The COsmological N-body CodE in PyThon}
Several cosmological N-body codes have been developed over the years, but for this work, we focus on how the COsmological N-body CodE in PyThon (\concept) \cite{dakin_concept_2015} works. It is a free and open-source N-body simulation code, written in \texttt{Python}, yet achieves performance comparable to \texttt{C} by using a custom transpiler/optimiser and \texttt{Cython}\footnote{\url{https://cython.org/}}. The code is parallelised using the Message Passing Interface (MPI) and is tightly integrated with \class, from which it obtains initial conditions and transfer functions.

The essential setup consists of a cubic toroidal periodic box of constant comoving size $L_\text{box}$, containing $N$ matter particles of equal mass $m$, where each has a comoving position and canonical momentum. They are evolving in an expanding background in which they interact gravitationally.
\subsection{Gravitational Solver}
We previously stated that, to evolve the system, we must solve the Hamilton equations, given the Newtonian potential. Here we discuss how \concept\ solves the Poisson equation. There are 3 methods available, one of which is a novel approach compared to other codes.

\subsubsection{PP Method}
The Particle-Particle (PP) Method corresponds to a direct summation of the point particles, for which we could build the perturbed density field as:
\begin{equation}
    \delta\rho(\vb x) = \frac{m}{a^3}\sum_{\vb n \in \mathbb{Z}^3}\qty[-\frac{N}{L_\text{box}^3} + \sum_{j=1}^NS(\vb x -\vb x_{j\vb n})]\,,
    \label{reconstructed deltarho field}
\end{equation}
where the first term inside the first sum would correspond to the subtraction of the average density, and the second is a sum over the $N$ particles within the primary box, replicated over the infinite periodic boxes defined by integer triplets $\vb n$. In the latter, $S(\vb x)$ is a function that assigns a local shape to a given particle. For example, for point particles, one should use a 3-dimensional Dirac delta function, which is not sound since it would lead to numerical divergences, giving the particles high velocities. Additionally, the position $\vb x_{j\vb n} \equiv \vb x_j + L_\text{box}\vb n$ is the key for the periodicity of the box, when put together with the first sum over integer triplets $\vb n$. Essentially, for a given $\vb n$, the sum over $j$ goes over every one of the $N$ particles in a cubic box of length $L_\text{box}$. This would naturally lead to multiple boxes of the same length, each containing $N$ particles. We may think of one original box, where the real particles lie, and multiple other boxes where we have \textit{image} particles. Thus, the box's periodicity is effectively reproduced.

Introducing the generalised reciprocal distance $|\vb x|^{-1}_S $ for a shape $S$, the force is softened, effectively distributing the mass over a finite volume. This relates to the softening scale previously discussed, and common choices for the particle shape are the Plummer sphere \cite{plummer_problem_1911} and the B-spline profile \cite{monaghan_refined_1985}. Hence, the solution for the Newtonian potential may be found by using \cref{reconstructed deltarho field,sim poisson}:
\begin{equation}
    \Phi(\vb x) = -\frac{Gm}{a} \sum_{j=1}^N\sum_{\vb n \in \mathbb{Z}^3} |\vb x - \vb x_{j\vb n}|^{-1}_S,
\end{equation}
such that the comoving force acting on particle $i$ is given by:
\begin{equation}
    \vb f _i = -Gm^2 \sum_{\substack{j=1 \\ j \neq i}}^N\sum_{\vb n \in \mathbb{Z}^3} |\vb x_i - \vb x_{j\vb n}|^{-3}_S (\vb x_i - \vb x_{j\vb n}),
\end{equation}
where we have omitted the self-interacting particle that would lead to a divergence\footnote{Note that this does not happen for the images of the original $i$-th particle.}. Thus, in principle, it is possible to implement this in code, but it is mandatory to address how the computer performs the infinite sum over $\vb n$. The process involves splitting the reciprocal of the distance into a short-range and long-range part by using the complementary error function and the exponential function, respectively. Then the sum converges exponentially, but the long-range term must be Fourier-transformed.

This method would result in an exact computation of the potential, but it is an algorithm of order $\O(N^2)$. This makes it unfeasible to run for a large number of particles, as it is required for high-resolution cosmological simulations. 

\subsubsection{PM Method}
To overcome the scalability barrier, the Particle-Mesh (PM) method was developed. In \concept\ the potential is represented by a cubic grid of size $n_\Phi$ from which the particle forces are computed with numeric differentiation. To build the potential, it is necessary to have the density perturbation from \cref{reconstructed deltarho field}. To go from a discrete distribution to a continuous one, the general approach is to have a convolution of the particle distribution with a window function, where schemes such as the Cloud-in-Cell (CIC) and the Piecewise Cubic Spline (PCS) accomplish this. From this continuous field, we can transform it into a Fourier space. In \concept\ the $\vb k =0$ mode is set to zero, which removes the background density, allowing us to work with $\rho$ instead of $\delta\rho$. The potential in Fourier space reads:
\begin{equation}
    \Phi(\vb k) = -\frac{4\pi G_N a^2}{k^2}\delta\rho(\vb k).
\end{equation}

To correct for artificial smoothing introduced by interpolations, we must deconvolve the potential in Fourier space. This is performed by dividing the squared Fourier transformed window function: once for the mass assignment from particles to the grid, and once for the eventual force interpolation. This removes Fourier aliasing and improves the accuracy of the grid representation at small scales.

At this stage, it is possible to take the derivative of the potential in Fourier space and then inverse transform back to real space, but this process would involve 3 inverse Fourier transforms, which are computationally costly. Thus, for numerical efficiency, \concept\ first takes the inverse Fourier transform, and then takes the derivative in real space to arrive at the force. The final step is to interpolate the force from the grid back to the particle positions. To avoid self-interacting terms, it is required that this force interpolation uses the same window function as the initial mass assignment, and that the shapes for the particles should be multiple convolutions of top-hat functions.

The computational cost of this method scales as $\O(N\log N)$ inherited from the Fourier transforms. This is orders of magnitude faster than the PP method, but the drawback is that we are limited in the accuracy of the short-range force. Naturally, on scales comparable to the grid spacing, the force would be softened, leading to unphysical anisotropic terms and a loss of spatial resolution.
\subsubsection{\texorpdfstring{P$^{\vb 3}$M Method}{P3M Method}}
The main idea of the Particle-Particle Particle-Mesh (P$^{3}$M) method is to combine the two previously described approaches. For long-range forces, it relies on the PM method, which is considerably faster than the PP method, and for short-range forces, it uses the exact PP method to provide an accurate description at these scales. This hybrid approach is the default solver in \concept\ and is the main distinction from other codes, such as \texttt{GADGET}, which approximates short-range forces using a hierarchical tree structure. Here, we explain how the simulation code performs the separation and how the process is optimised. This method for solving the gravitational force includes a free parameter $x_r$, which balances the optimised computation of the PM method with the accuracy of the PP method. It effectively truncates the direct summation for particle separations greater than this scale.

As for the long-range force, the code follows the same steps previously described in the PM method, except that the Poisson equation kernel is replaced by $\G_{\rm lr} = 4\pi \exp(-x_s^2\vb k^2)/\vb k^2$, where $x_s$ is the split scale. This kernel allows for exponential suppression of Fourier modes with wavelengths much smaller than $x_s$, making this term negligible at small scales, as it should be. 

Then the short-range terms are computed via direct summation, with the potential $\G_{\rm sr}= |\vb x|^{-1}{\rm erfc} (|\vb x|/[2x_s])$, allowing for an exponential decay in real space for distances greater than $x_s$. Since this direct summation has the hard cut-off $x_r$, in \concept\ $x_r$ is set to $4.5x_s$ so that the contribution from these terms is negligible. A full expression for the computed force is presented in Annex \ref{annex:CONCEPT}, along with a detailed explanation.

To optimise the computation of short-range forces, a 3D data structure is needed for particle queries. To this end, the \concept\ partitions a domain\footnote{This corresponds to a region of the box handled by a single CPU core.} into smaller pieces called \textit{tiles}. The code chooses these cuboidal shapes with a minimum length of $x_r$, such that for a given particle, the direct summation can only take place in the neighbouring $3\times 3\times 3$ tiles. In general, neither the domains nor the tiles take a cubic shape, and the average number density in a tile is $N/n_{\rm tile}$. Hence, the time complexity for the tiled short-range computation is of order $\O(N)$ when taking the number of tiles $n_{\rm tile} \sim n^3_\Phi$, and the number of potential grids $n^3_\Phi \propto N$. This results in a computing time that beats the $\O(N\log N)$ time of tree methods. However, in practice, inhomogeneities might lead to different tiles requiring different computational effort, unbalancing the parallelisation scheme and decreasing performance. 

Even with tiling implemented, there are further optimisations that may be considered. One approach is to further divide tiles into \textit{subtiles}, yielding a better estimate of particle separation. As an example, a particle could lie in a tile, and another particle could be in the furthest possible corner of its surrounding tiles, such that their separation is larger than $x_r$. If we have subtiles that lead to earlier rejection of the pairing of these two particles, the algorithm takes less time to compute the short-range interaction in this particular domain. On average, the number of false-positive pairings depends on how two tiles are spaced, and it can reach 98 per cent when they share only a corner. Naturally, we could further divide subtiles, but there is a computational cost associated with this process. \concept\ by default automatically estimates the optimal subtile size, dynamically optimising with respect to space and time.

\concept\ chooses cosmic time as the time integration variable, and has a global time step that is divided into finer steps. Overall, the code follows the leapfrog scheme, as is typical for N-body simulations. Furthermore, \concept\ analyses particle accelerations, such that if their acceleration is higher than a particular fraction of its softening scale, then it is assigned to a particular \textit{rung} $\ell_i$, such that this parameter dictates the adaptive time stepping in the code. For a particle that was assigned a run $\ell_i$, its time is given by $\Delta t / 2^{\ell_i}$, so particles in rung 0 follow the global time of the simulation. 
		\chapter{Horndeski in \concept}
Implementing a Modified Gravity (MG) model in an N-body code is a non-trivial task, and has been done for particular models. The common choice is to include the $f(R)$ Hu-Sawicki model in these codes \cite{puchwein_modified-gravity-gadget_2013}, allowing for the exploration of deviations from General Relativity (GR) in the non-linear regime, and see their impact on observables. Given the plethora of models that deviate from GR, it becomes highly unfeasible to work out the equations for a particular model and try to implement it. There are several considerations to be made, such as the implementation of screening mechanisms, approximation schemes, numerical stability, etc.

In this chapter, we discuss how we implemented Horndeski gravity in a fully N-body code by using the effective fluid approach. The caveat is that we are only concerned with the linear effective Dark Energy perturbation and its impact on the matter power spectrum.

\section{Horndeski in \class}
Since our N-body code of choice is \concept, we must have a compatible version of the Einstein-Boltzmann solver \class. Fortunately, there are two versions of \class\ that include Horndeski gravity: \texttt{hi\_CLASS} \cite{bellini_hi_class_2020} and \mochi\ \cite{cataneo_mochi_class_2024}, where the latter was built on top of the first one. Although \texttt{hi\_CLASS} already considers Horndeski gravity in general, we chose to work with \mochi\ since it is a newer version of the code, with some advantages. As an example, the basis on which a particular model is written is chosen such that there are stability tests that can be performed immediately. Hence, in order to specify a gravity model, the user should give the four $\alpha$ functions that we discussed in \cref{sec:MG}, and the expansion history of the universe, through any function that describes the time evolution of Dark Energy (DE) or the Hubble parameter. Instead of working with those functions directly, the code considers a new basis with:
\begin{align}
    D_\text{kin} &\equiv \alphak + \frac{3}{2}\alphab^2,\\
    c_{\mathrm{s}}^{2} &\equiv \frac{1}{D_\text{kin}} \left[(2-\alpha_{\mathrm{B}}) \left(-{\frac{H^{\prime}}{a H^{2}}}+{\frac{1}{2}}\alpha_{\mathrm{B}}+\alpha_{\mathrm{M}}\right) -{\frac{3\left(\rho_{\mathrm{m}}+p_{\mathrm{m}}\right)}{H^{2}M_{\ast}^{2}}}+{\frac{\alpha_{\mathrm{B}}^{\prime}}{a H}}\right],
\end{align}
such that a gravity model is described by\footnote{Recall that we are fixing the $\alphat$ function to zero.} $\qty{D_\text{kin}, M^2_*, c^2_{\rm s}}$.
Thus, we arrive at an equivalent description of a Horndeski Lagrangian, but it has a clear computational advantage, since it guarantees no ghost and no gradient instabilities by imposing both $D_\text{kin}> 0$ and $c_{\rm{s}}^{2} > 0$.

The code also permits the user to input functions in several ways. The most popular models have already been implemented in \texttt{hi\_CLASS}, for which the user may configure an \texttt{.ini} file with the appropriate flags. Another option is to consider the already included \texttt{.dat} files, which contain tables with the new basis and a DE function of time that fixes the expansion history. Lastly, for further exploration, a user may construct their own models and store the necessary functions in \texttt{.dat} files, and provide the paths to those files to \mochi.

\section{Interfacing a new version of \class\ with \concept}
While these stand-alone Einstein-Boltzmann solvers exist and are publicly available, we have to modify them to interface them with \concept. The first consideration is that \concept\ uses version 2.7.2 of \class, so some aspects are outdated. There are two options as to how to interface the two codes: 1) would be to downgrade \mochi\ into vsion \texttt{2.7.2}; or 2) would correspond to identifying the necessary extensions of the version of \class\ that ships with \concept, and implement them in \mochi. In our case, we follow the second approach.

To integrate the N-body code with \class, three core modules of its original source code must be modified: \texttt{input}, \texttt{background}, and \texttt{perturbations}. Rather than rewriting the solver, these modifications focus primarily on extracting the necessary outputs in the right format to initialise an N-body simulation. One example is that in older versions of \class\ did not allow the output of perturbed variables to be in the N-body gauge.

\subsection{Initialisation and Background Dynamics}
The \texttt{input} and \texttt{background} modules are modified to track higher-order background variables and handle complex dark matter components. In the \texttt{input} module, CO$N$CEPT introduces parameters to read and define the default contributions of non-Cold Dark Matter (NCDM) species to the growth factor. Additionally, it initialises parameters specific to the parallelisation architecture used by the simulation wrapper.

Within the \texttt{background} module, the primary modification is the inclusion of second and third-order growth factors and their corresponding growth rates. Additionally, the solver must be updated to compute these higher-order functions along with the background expansion, with modified initial conditions. Some exotic models, such as decaying cold dark matter (DCDM) and NCDM, are considered, providing a clear way to include additional species. Finally, the output routines are adjusted to ensure that the new computed functions are written to background output arrays.

\subsection{Perturbations and the PPF Formalism}
The most extensive modifications occur within the \texttt{perturbations} module. Because \concept\ relies heavily on threaded environments to run large cosmological boxes, the entire module is parallelised using OpenMP.

A significant portion of the modifications in this module is dedicated to stabilising the effective dark energy fluid, particularly to prevent numerical instabilities when the equation of state crosses the phantom divide ($w=-1$). \concept\ uses the Parameterised Post-Friedmann (PPF) formalism, involving a fictitious potential and its derivative to control the divergences. In a way, this is similar to what is done with the Yukawa potential to avoid divergences with the Coulomb potential.

The perturbation output routines are heavily tailored to feed directly into the $N$-body gauge initialisation. Some extra output variables from the fluid perturbed variables are explicitly stored. The code also outputs the time derivatives of metric perturbations, $h'$ and $H_T'$, alongside the total velocity divergence $\theta_{\text{tot}}$, as can be seen from \cref{eq:enbody transf1,eq:enbody transf2}, they are needed to perform the gauge transformation into the N-body gauge. Then, when including a new non-standard species in \class, $\theta_{\rm tot}$ must be updated to consider the contribution from the new species. To optimise computational resources and memory allocation, the output is truncated to late times ($a \geq 3 \cdot 10^{-4}$), which are relevant for structure formation. Furthermore, the code explicitly prevents \class\ from converting the outputs into the Newtonian gauge.

Finally, there are further technical modifications related to the default values of \class, the \texttt{Python} wrapper, the Runge-Kutta evolver and adjustments to the \texttt{Makefile}.

\section{\mochi\ in \concept}
After reviewing the modifications needed to have \class\ interact with \concept, we turn to implement the effective fluid approach in \mochi. While  \cref{eq:delta rho de,eq:delta p de,eq:theta de,eq:shear de} are the key for considering an MG model within the Horndeski family as an effective DE fluid, their implementation is not a direct copy and paste of these functions. The source code of \class\ considers different approximation schemes that depend on the epoch, and there are different ways to obtain a specific metric perturbation. Additionally, \class\ has a particular choice of units, and in the literature, there is no convention as to how to write the Horndeski action, so it is expected to have different conventions for the perturbation equations. In consequence, we decided to rederive the equations from the \mochi\ source code and include them in the output. Implementing the \concept\ extensions to \class\ and our effective fluid approach to \mochi, results in\footnote{This version of \class\ is publicly available at \url{https://github.com/vPedreros/vp2_mochi}.}:
\begin{itemize}
    \item An updated version of \class\ used by \concept.
    \item New outputs in \mochi, related to the scalar modified gravity (smg) theory. While the original version outputs quantities like the average density of the effective DE fluid, we have included the perturbed DE density, its pressure, velocity divergence and shear, following the original naming scheme.
    \item A new non-standard species that contains linear perturbations from an MG model, ready to be used by \concept.
\end{itemize}

It must be highlighted that a choice of a model of gravity must be made through the flag \texttt{gravity\_model = stable\_params}. This tells \mochi\ that it will work with a new MG model for redshifts strictly less than \texttt{z\_gr\_smg}, which is set to 99 by default. Hence, the evolution with extensions to GR starts deep in the matter-dominated era. 

For redshifts smaller than \texttt{z\_gr\_smg}, the usual Friedmann equation is no longer valid, and the code does not dynamically solve for the Hubble parameter $H$. Instead, the user must supply a history expansion with the flag \texttt{expansion\_model}, for which they can choose to supply the equation of state parameter, or a DE density normalised such that $\rho_{\rm DE}(a=1) =1$.

For any of the possibilities for defining a gravity model and a history expansion, the user may provide files containing tables of $\ln a$ and the quantity of interest, or provide them as in-line arrays. It is recommended to provide time-spaced quantities with a step no larger than $\Delta\ln a = 7.5\cdot 10^{-4}$ for $\ln a \in [-2,0]$, for minimising interpolation errors.

\section{New species in \concept}
\concept\ is set up so that a user can easily include new species computed by \class. This support is at the three different levels relevant for cosmological analysis:
\begin{itemize}
    \item \textbf{Background}: This evolution is taken from \class, so there is no need to include it for a standard \concept\ simulation.
    \item \textbf{Linear perturbations}: linear perturbations of any species evolved by \class\ are taken by \concept, and realised in a grid during the simulation using the same random noise for particle initialisation. Then, the gravity contribution from this grid is applied to the particles, in addition to the non-linear gravitational interaction between them.
    \item \textbf{Non-linear perturbations}: Though highly unstable, \concept\ supports a form of non-linear fluid evolution for its density perturbation, keeping the evolution of its pressure and shear linear. 
\end{itemize}

For this work, we focus on implementing the new effective fluid at the linear level. For this perturbation to work in a simulation, several considerations apply. The label for the quantities arising from an MG theory, interpreted by \concept\ as a new fluid, will be dubbed smg, which stands for scalar modified gravity, and is the label used by \mochi. Then, the only modification to the source code of \concept\ is to register this new species, making it recognisable for the N-body code\footnote{The user then uses this name in a parameter file, so they can explicitly request it.}. Following the naming scheme of the code, where a fluid species labelled ``fld'' is called ``dark energy fluid'', we named our ``smg'' species ``scalar modified gravity''. This highlights the user-friendliness of the N-body code, inherited from \class, since the main workload is transferred to it.
		\chapter{Results \& Discussions}
In this chapter, we aim to present the results and discuss them for the three codes developed for this thesis. First, we take a close look at our custom solver for linear cosmological perturbations. Then, we focus on the results from our novel version of \class, interpreting the effects of Modified Gravity (MG) theories as an effective Dark Energy (DE) fluid. Finally, we examine how different MG theories impact the non-linear matter power spectrum using the same effective fluid approach. Our discussion is organised so that we first study the linear regime and then move on to the non-linear scales. For both linear and non-linear analysis, we will first focus on the $w_0w_a$CDM model, the $f(R)$ Hu-Sawicki model, and finish with the Cubic Galileon model. Our choices differ from \lcdm\ in that we include a dynamical DE fluid and consider massive neutrinos.

Given possible degeneracies between cosmological and MG parameters, we vary both. The only fixed parameter is the density parameter for baryons $\Omega_{\rm b}=0.049$. The rest of the cosmological parameters, with their respective values, are shown in \cref{tab:sim cosmo params}. These are present in all the models of gravity we consider, where $f(R)$ is the only one with extra parameters: $f_{R,0}$ and $n$. For this work we fix both $n=1$ and $f_{R,0}=10^{-4}$.

\begin{table}[ht]
\centering
\begin{tabular}{llll}
\hline
 Parameter& min & reference & max \\
\hline
$\Omega_{\rm c}$        & 0.2     & 0.27      & 0.34   \\
$A_{\rm s}\cdot10^{9}$  & $1.91$  & $2.1$     & $2.31$   \\
$n_{\rm s}$             & 0.87    & 0.96      & 1.056   \\
$\sum m_\nu$ (eV)       & 0.0     & 0.058     & 0.15   \\
$h$                     & 0.61    & 0.67      & 0.73   \\
$w_0$                   & $-1.3$  & $-1.0$    & $-0.7$   \\
$w_a$                   & $-0.6$  & 0.0       & 0.6      \\
\hline
\end{tabular}
\caption{Parameter space. The results presented will use the reference column, since it closely resembles Planck's best fit, considering massive neutrinos. For future work, it will be useful to make a fair comparison to EuclidEmulator 2 \cite{euclid_collaboration_euclid_2021}.}
\label{tab:sim cosmo params}
\end{table}

The reason for the CPL parameters to be in the \cref{tab:sim cosmo params} is that \mochi\ does not use the standard Friedmann equation to find the Hubble parameter when working with extensions to GR. Instead, the user must supply a history expansion, and we restrict ourselves to following this parametrisation. 

\section{Linear Cosmological Perturbations in \texttt{Python}}
As mentioned in \cref{sec:PT}, we built a solver for linear cosmological perturbations to study late-time dynamics. For that reason, we only consider a universe with matter and a dynamical Dark Energy fluid. It is written in \texttt{Python} and is publicly available at \url{https://github.com/vPedreros/perturbations}.

\subsection{Structure of the Solver}
Our custom solver chooses the scale factor as the time variable, such that the system of equations that needs to be solved is:
\begin{align}
\dv{\delta_{\rm m}}{a} &= -\frac{V_{\rm m}}{a^2 H} + 3\dv{\phi}{a},\\
\dv{V_{\rm m}}{a} &= -\frac{V_{\rm m}}{a} + \frac{k^2\phi}{a^2H},\\
\dv{\delta_\text{DE}}{a} &= -\frac{V_\text{DE}}{a^2H}+3(1+w_\text{DE})\dv{\phi}{a} - \frac{3}{a}\left(c_s^2-w_\text{DE}\right)\delta_\text{DE},\\
\dv{V_\text{DE}}{a} &= -\frac{1}{a}\left(1-3w_\text{DE}\right)V_\text{DE}+\frac{k^2}{a^2H}\qty[c_s^2\delta_\text{DE}+(1+w_{\rm DE}) \phi],\\
k^2\phi &= -\frac{3(aH)^2}{2}\left[\Omega_{\rm m}\left(\delta_{\rm m} + \frac{3aH}{k^2}V_{\rm m}\right) +\Omega_\text{DE} \left(\delta_\text{DE} + \frac{3aH}{k^2}V_\text{DE}\right) \right],\\
\dv{\phi}{a}&=\frac{3H}{2k^2}\left[\Omega_{\rm m}V_{\rm m} + \Omega_\text{DE} V_\text{DE}\right]-\frac{\phi}{a},\label{eq:dphida}
\end{align}
where the density parameters are functions of the scale factor. It is important to note that we are considering time-dependent equation of state parameters for DE. Furthermore, it is not necessary to consider \cref{eq:dphida}, since it is derivable from the other five equations.

To solve the initial value problem, we use the \texttt{scipy} library, with a Runge-Kutta solver of order 5. As for the initial conditions, we are starting at $a_{\rm ini} = 10^{-3}$ deep in the matter-dominated era, and we have adiabatic initial conditions, such that:
\begin{equation}
    \frac{\delta_{\rm r,ini}}{1+w_{\rm r}}=\frac{\delta_{\rm m,ini}}{1+w_{\rm m}},
\end{equation}
fixing $\delta_{\rm r,ini} = 10^{-5}$, according to CMB anisotropies. For the velocity perturbation $V_{\rm m}$, \cref{zeldovich app} allows for writing it in terms of $\delta_{\rm m,ini}$, while DE perturbations are set to zero, since they are negligible at the initial time. The values of cosmological parameters are found in \cref{tab:custom solver parameters}.
\begin{table}[ht]
\centering
\begin{tabular}{ll}
\hline
 Parameter& Value \\
\hline
$\Omega_{\rm b,0}$        & 0.049 \\
$\Omega_{\rm cdm, 0}$     & 0.27 \\
$\Omega_{\rm r, 0}$       & 0.0 \\
$H_0$                     & 67 (\rm{km/s/Mpc})\\
$c_s^2$                   & 1.0\\
\hline  
\end{tabular}
\caption{Parameters for custom linear solver.}
\label{tab:custom solver parameters}
\end{table}

A user can choose a DE model, defined by its equation of state parameter following the CPL parametrisation, and a sound speed. Additionally, they must set a scale for the perturbations to be computed. For this section, we have considered only three DE models, where we have chosen $w_0 \in\qty{-0.7, -1.0,-1.3}$, and fixed both $w_a=0$ and $c_{\rm s}^2=1$. As for the scale, we have chosen a sub-horizon scale to avoid concerns about relativistic effects on the largest scales, set to $k=200 H_0$ with $H_0$ in Mpc$^{-1}$ units. 

\subsection{Comparing results to \class}
\begin{figure}[ht]
    \centering
    \includegraphics[width=\linewidth]{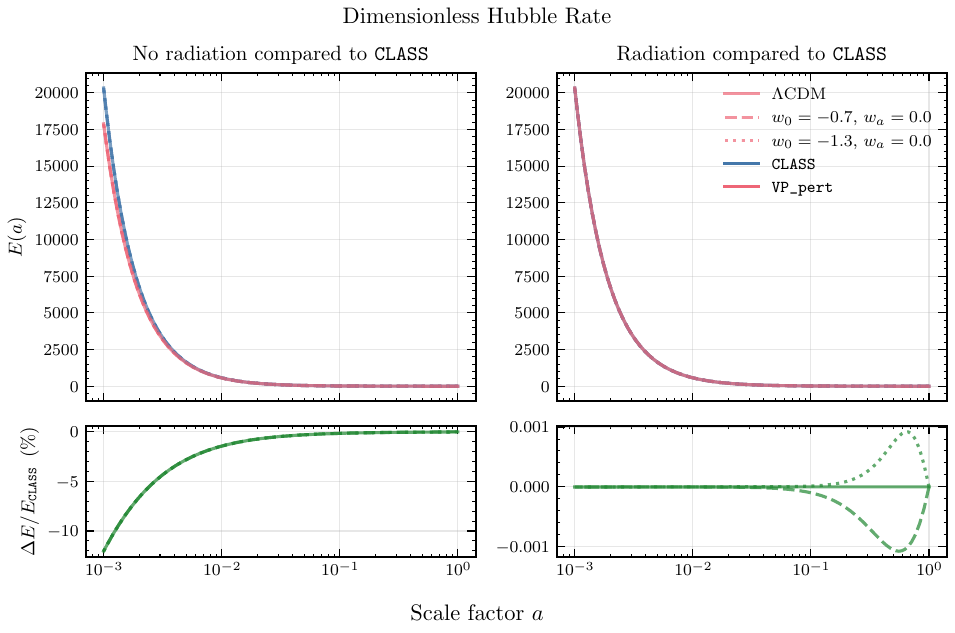}
    \caption{Comparison between dimensionless Hubble rates. In the left panel, we neglect radiation in our custom solver and compare it to \class, finding that they agree at the 1\% level for redshifts up to $z\simeq 68$ across models. On the right panel, we found excellent agreement when $\Omega_{\rm r,0}=9.319911 \cdot10^{-5}$.}
    \label{fig: E_z custom vs class}
\end{figure}

Since we are neglecting radiation, we take a look at the redshift when the dimensionless Hubble parameters $E(z)\equiv H(z)/H_0$ differ at the 1\% level. For $\Omega_{\rm r,0}$, we consider the null value and $9.319911 \cdot10^{-5}$. The results of comparing both Hubble rates to the one obtained from \class\ are in \cref{fig: E_z custom vs class}, where we can clearly see that across different DE models, we are underestimating the Hubble rate for early times when neglecting radiation. In contrast, in the right panel of the figure, we observe excellent agreement between the codes when considering radiation. 

Hence, we will restrict ourselves to presenting the perturbation results for redshifts up to $z_{\max} = 68$. The solver has 5 quantities as output, which are the overdensity $\delta_{\rm s}(a)$ and velocity perturbation $V_{\rm s}(a)\equiv(1+w_{\rm s})\theta_{\rm s}$ for both matter and DE, and the gravitational potential $\phi(a)$.

\begin{figure}[ht]
    \centering
    \includegraphics[width=\linewidth]{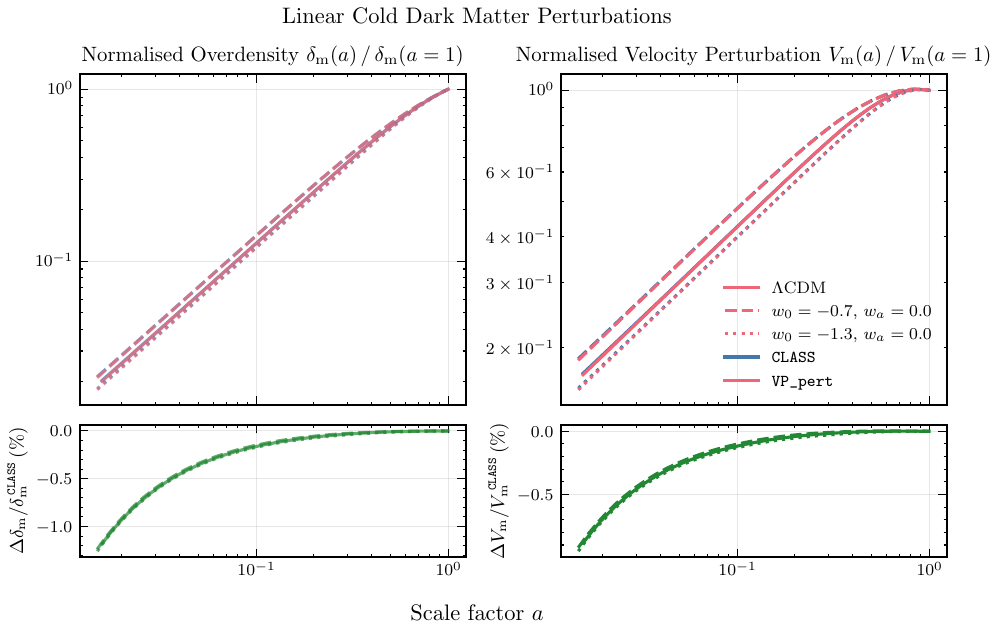}
    \caption{Normalised cold dark matter perturbations for different DE models, at the fixed scale of $k=200H_0$. The left panel shows the cold dark matter overdensity field normalised to its present-day value, while the right panel displays the correspondingly normalised velocity perturbation. The numerical integration exhibits a sub-1\% agreement with the standard \class\ output across the majority of the computed time steps.}
    \label{fig:vp_pert matter pert}
\end{figure}

In \cref{fig:vp_pert matter pert} we present a comparison between our results and \class\ predictions for the same cosmological models. We found excellent agreement between the codes, where as expected, discrepancies grow as we go back in time. It is important to highlight that, for this case, we have normalised the results to their present value. Technically, \class\ outputs transfer functions and not the physical overdensity field, since they are integrated with a normalisation where the initial primordial curvature is set to unity. Additionally, \class\ also might yield values of $\delta$ larger than unity, where linear perturbation theory breaks down, and the solver is blind to non-linear collapse, resulting in arbitrarily large values for the smallest scales.

\begin{figure}[ht]
    \centering
    \includegraphics[width=\linewidth]{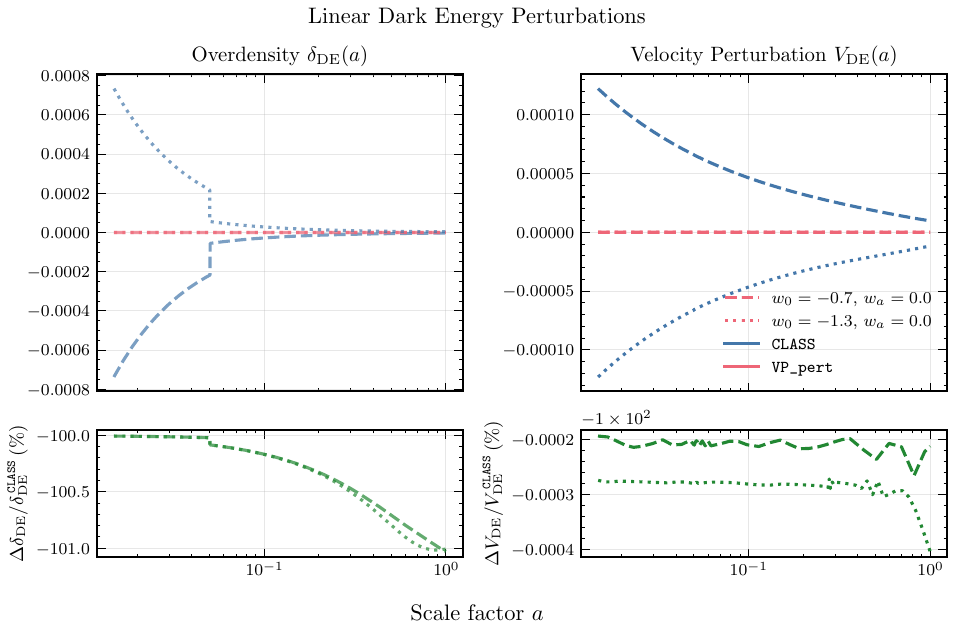}
    \caption{Normalised cold dark matter perturbations for different DE models, at the fixed scale of $k=200H_0$. The left panel shows the normalised overdensity field, while the right panel displays the normalised velocity perturbation. In the bottom-left panel, the large relative difference between the codes is due to the result of an exactly zero DE overdensity field, whereas \class\ returns small values for it. The \lcdm\ model is skipped since, by definition, it has no DE perturbations.}
    \label{fig:vp_pert de pert}
\end{figure}

As for DE perturbations, we present the corresponding comparison in \cref{fig:vp_pert de pert}. Given that we have fixed the sound speed of DE to unity at the chosen sub-horizon scale, we expect really small DE perturbations. This is the reason why we chose not to normalise this perturbation, since a small present-day value might lead to an incorrect analysis. Our solver found exactly null values for DE perturbations, yielding a high relative difference with \class, but by looking at the upper panels of both perturbations, we note that the values are small for \class.

\begin{figure}[ht]
    \centering
    \includegraphics[width=0.5\linewidth]{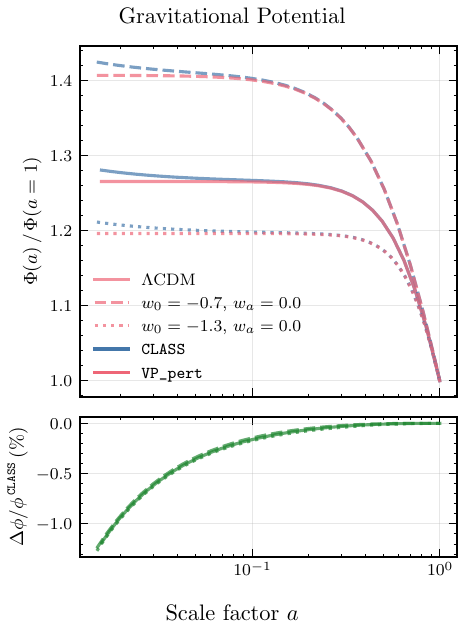}
    \caption{Gravitational potential for different DE models at the fixed scale of $k=200H_0$, normalised to its present-day value. The integration results in sub-1\% agreement with \class\ at most time points.}
    \label{fig:vp_pert phi pert}
\end{figure}

Lastly, the resulting gravitational potential also agrees at the sub-1\% level with \class, as seen in \cref{fig:vp_pert phi pert}. We only see that the codes disagree at earlier times, as expected, since our solver does not account for radiation effects. Interestingly, we observe that the potential is a constant of time for times corresponding to a matter-dominated universe, as expected.

Hence, we have successfully developed a simple and fast solver for linear cosmological perturbations. Now we present and discuss the main results of our novel version of \class.
\newpage
\section{\mochi\ as an effective dark energy fluid}
To reproduce the following results, we have set all cosmological parameters to default values selected by \class. We are considering 4 models, of which only one has all modifications to the standard version of \class\ disabled, by design of \mochi. The models are:
\begin{itemize}
    \item \class\ \lcdm: is the standard version of the \lcdm\ model, meaning that the only parameters passed to \class\ are to enable perturbations to be computed.
    \item \texttt{mochi} \lcdm: aims to mimic the same results as the previous model, by enabling \mochi\ extensions and choosing the $\alpha$ functions such that we end with General Relativity, i.e., all functions are set to $10 ^{-4}$, since smaller values lead to numerical instabilities. The initial condition for $\alpha_{\rm B,0}$ is fixed to $10^{-10}$. The expansion model is set to use the CPL parametrisation, with $ w_0 =- 1$ and $ w_a = 0$.
    \item \texttt{mochi} $f(R)$: is the $f(R)$ Hu-Sawicki model, where the chosen parameters are $f_{R,0}=10^{-4}$ and $n=1$. The files specifying the gravity model and expansion are included with the standard version of \mochi.
    \item \texttt{mochi} cGal: is the cubic Galileon mode, whose files for the gravity model and expansion are provided with the standard version of \mochi.
\end{itemize}

\begin{figure}[ht]
    \centering
    \includegraphics[width=0.5\linewidth]{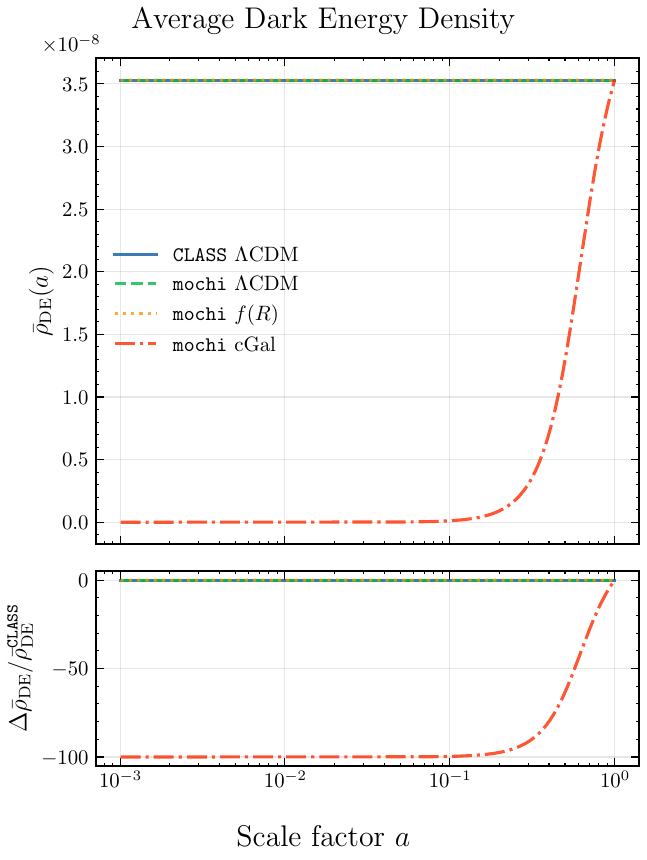}
    \caption{Background DE density evolution for \lcdm\ using the standard version of \class\ and \mochi, $f(R)$ Hu-Sawicki and cubic Galileon.}
    \label{fig:vp2mochi de_bk}
\end{figure}

Now, \cref{fig:vp2mochi de_bk} presents the background DE density for the different models. The two \lcdm\ models exhibit the exact same behaviour, while $f(R)$ only has a maximum 0.06\% relative difference to \lcdm, and cubic Galileon goes to exactly zero at early times. The small differences between the models are expected, since $f(R)$ Hu-Sawicki should mimic \lcdm\ at the background level, while the discrepancies for cubic Galileon are mainly related to the numerical stability of the model.

\begin{figure}[ht]
    \centering
    \includegraphics[width=\linewidth]{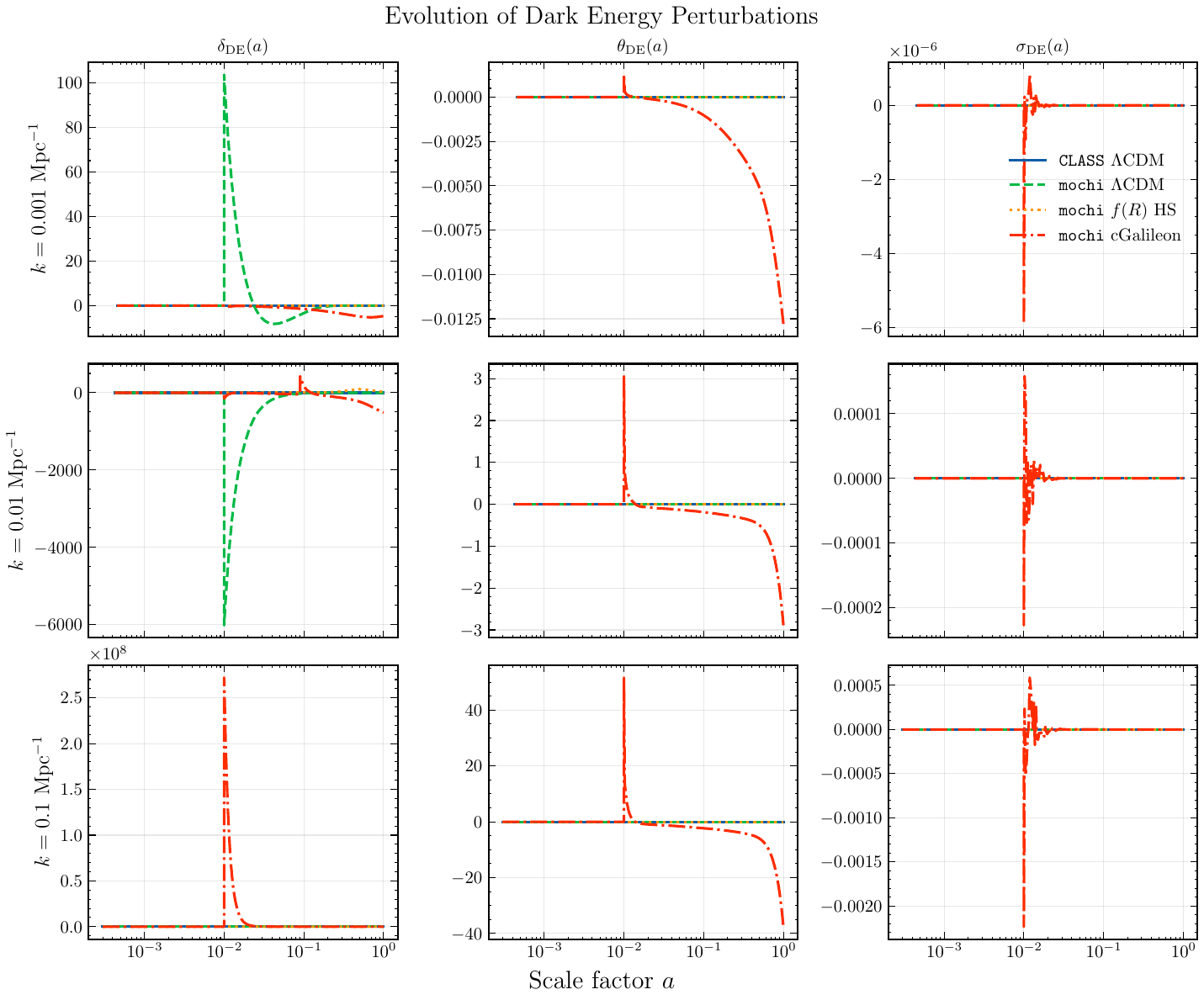}
    \caption{Perturbations of DE across time, for scales $k=10^{-3},10^{-2},10^{-1}\,\, \rm{Mpc}^{-1}$. The left column is for the overdensity field, the middle for velocity divergence and the right for anisotropic shear. The different models correspond to \lcdm\ using \class\ and \mochi, $f(R)$ Hu-Sawicki and cubic Galileon.}
    \label{fig:vp2mochi de_pt}
\end{figure}

As for the perturbed quantities, \cref{fig:vp2mochi de_pt} presents the different DE perturbations for the four models. The first thing to note is that for $a\simeq10^{-2}$ the perturbations of the models are abrupt, since at $z=99$ is where the extensions to GR are enabled. This is also the starting time of the simulations. As the scale decreases, we observe that the density contrast and velocity divergence increase in amplitude, as expected for a linear code. 

For the overdensity perturbation, we see that the limit that attempts to mimick GR with \lcdm, has non-zero perturbations for the effective fluid. This obviously is non-physical, and emerges from not choosing the $\alpha$ functions to be exactly null. Lastly, we see that for every model we have close to zero anisotropic term, which is expected for all of them except for $f(R)$. \cref{eq: shear f(R)} predicts a non-zero term for this model, but we observe an exactly zero result for this perturbation. The reason for this result is that in order to speed up the computation, the code decides to implement the quasi-static approximation (QSA) scheme, where time derivatives of metric perturbations are ignored, and the perturbation of the field is also vanishingly small.

\begin{figure}[ht]
    \centering
    \includegraphics[width=\linewidth]{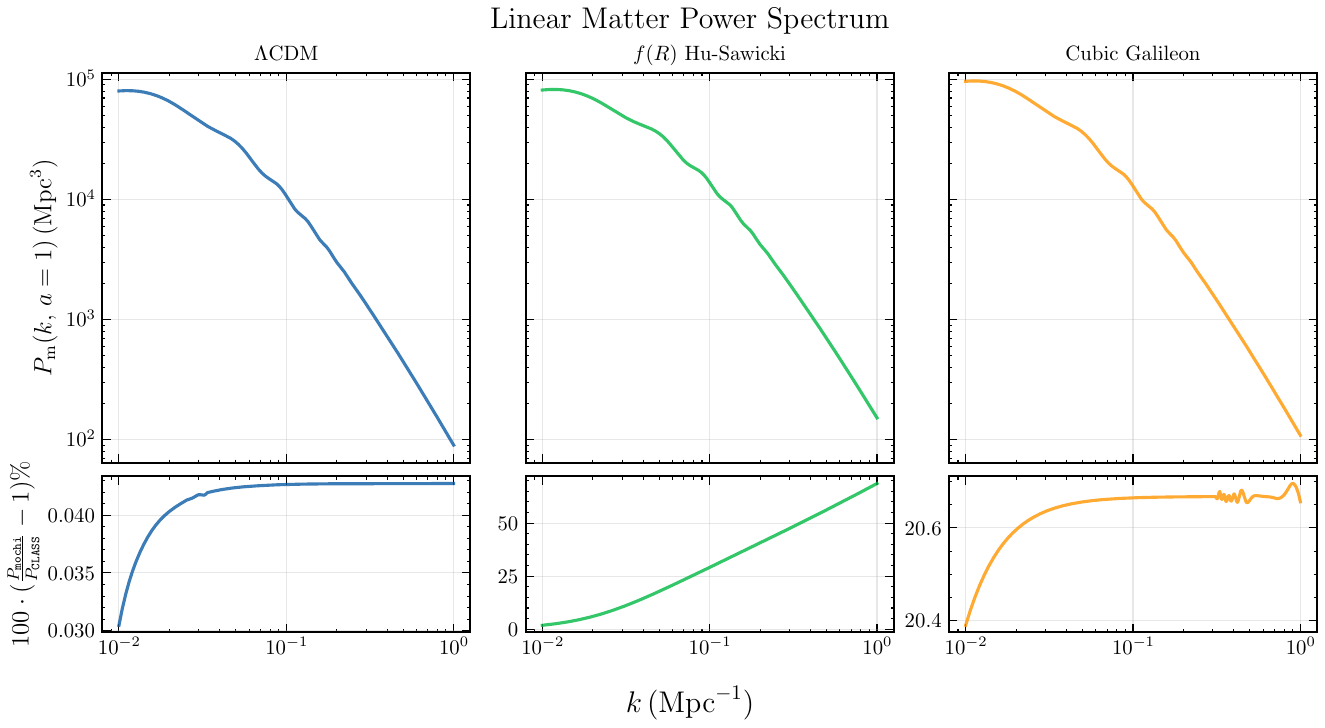}
    \caption{Linear matter power spectrum at the present time. The upper panels show the computed power spectra for the three different models run with \mochi, whereas the bottom presents the relative difference between each of them, with respect to the \lcdm\ model run with standard \class.}
    \label{fig:linear powerspec}
\end{figure}

Finally, we see how the linear matter power spectrum behaves across the different models. In \cref{fig:linear powerspec} we see the three different models run with \mochi. For each one of them, the bottom panels present the relative difference to the \lcdm\ model run with \class. On the left panel, we observe excellent agreement between the two ways in which the \lcdm\ model was computed. This is an expected behaviour for \mochi, since the chosen $\alpha$ functions are set to closely resemble the standard \lcdm\ model. As for the modified gravity models, we observe significant discrepancies with \lcdm\ for both. The chosen value for $f_{R,0}$ is two orders of magnitude larger than the baseline chosen for Euclid forecasts \cite{casas_euclid_2026}, thus yielding both large values for the relative difference, and a behaviour that appears to grow with the scale. The latter issue is more egregious, since we expect that, at smaller scales, gravity theory should agree with GR. However, given that \mochi\ is a strictly linear code, different Fourier modes are independent. Thus, the code lacks the capacity to implement a screening mechanism for a high-density region.

As for the cubic Galileon model, we observe a behaviour very similar to that of \lcdm, which could be explained by the idea that the change in the matter power spectrum is a simple amplitude enhancement.

\section{Horndeski in \concept}
The simulations that we present were run in a cubic box of length $L_{\rm box} = 512 \,(h^{-1}\rm{Mpc})$, with $N=352^3$ particles. This results in a simulation resolution of $\ell\equiv N/L \simeq0.68/(h^{-1}\rm{Mpc})$, which allows for fast enough simulations given our computational resources. 

\begin{figure}[ht]
    \centering
    \includegraphics[width=\linewidth]{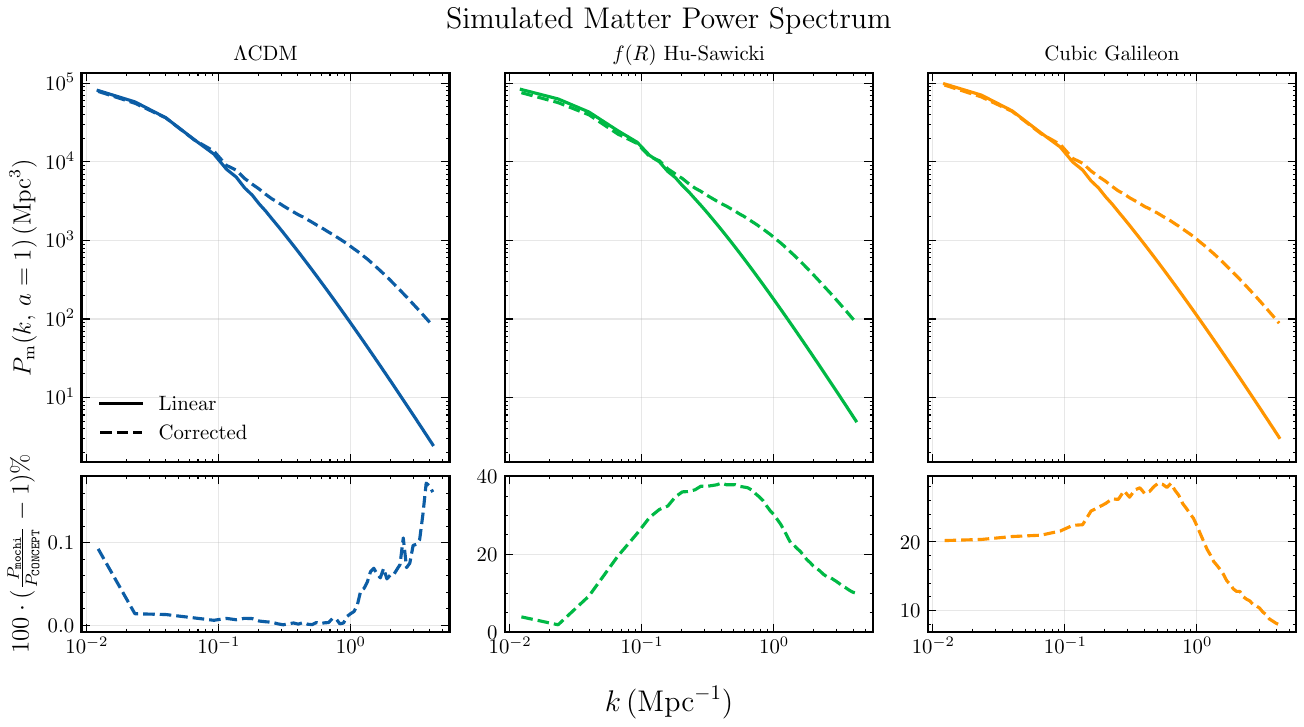}
    \caption{Simulated matter power spectra from our version of \concept, for three gravity models. The solid line corresponds to a linear computation, while the dashed line is a corrected non-linear one. The upper panels show the power spectra, whereas the bottom panels present the comparison to a \lcdm\ simulation using the standard version of \class.}
    \label{fig:sim powerspec}
\end{figure}

The simulation results are summarised in \cref{fig:sim powerspec}. The upper panels present the computed matter power spectrum at the present day, both linearly (solid line) and non-linearly (dashed line), for the three different gravity models. In the bottom panels, we show the relative difference between each non-linear matter power spectrum and a standard \concept\ run with the same cosmological parameters. We observe sub-0.2\% agreement between the two runs for the \lcdm\ model, whereas we see strong deviations in the modified gravity scenarios. This is in the same range as we previously saw for the linear computation using the standard version of \mochi, so this was expected. 

Interestingly, we found a decaying disagreement as the scale decreases, which someone might mistakenly interpret as the screening mechanism. In reality, \concept\ directly takes the $\delta(\vb k, \tau)$ transfer functions from \class, and realises them in the simulation grid over time, when applying the linear effects of the particular species to the simulation. Given that we have not explored the non-linear evolution of the effective fluid, due to high numerical instabilities. The reason for this decay is that the matter density contrast vastly outgrows that of the effective fluid, since it is non-linearly evolved. Thus, the modified gravity effects at these scales are diluted.

Lastly, we should highlight that the N-body gauge that allows for relativistic consistency on the largest scales has been explored for modified theories of gravity in this thesis. In particular, the metric perturbation associated to the local volume is crucial for this \cite{tram_fully_2019}. For GR, it reads:
\begin{equation}
    H_T'= 3\H \frac{1}{\bar{\rho} + \bar{P}}\qty[\delta P ^{\rm S/N} + \bar{P}'\frac{\theta^{\rm S/N}}{k^2} + (\bar{\rho}+\bar{P}) \sigma],
    \label{eq:H_T_prime}
\end{equation}
and it plays a role on applying relativistic corrections to the potential. Naturally, if we change the gravity theory \cref{eq:H_T_prime} should not hold anymore. For the computed simulations, we are not changing the form of $H_T'$, so it does not take into account the effects of the modified gravity model.

\begin{figure}[ht]
    \centering
    \includegraphics[width=\linewidth]{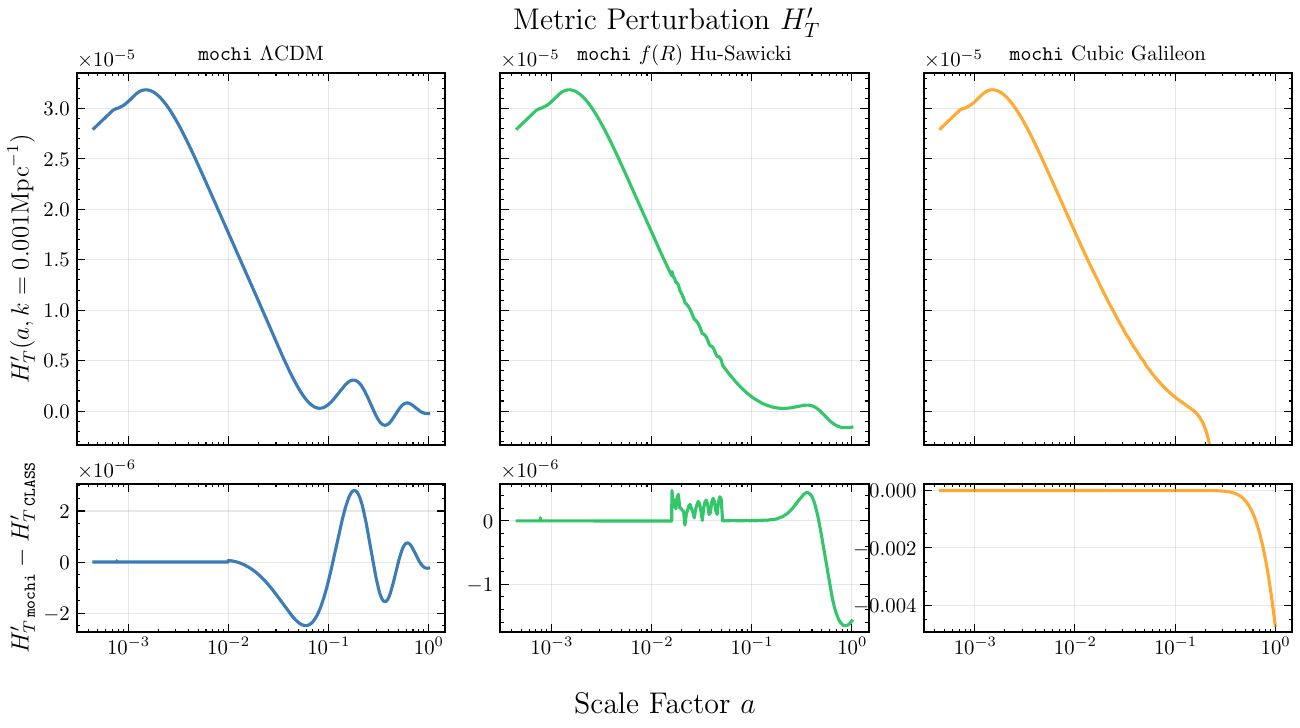}
    \caption{Metric perturbation of the N-body gauge. The upper panels show the actual value of the perturbation, whereas the lower panels present the difference relative to the first model, run with the standard version of \class. We found spurious oscillations in the $f(R)$ model and an unphysical decay in the cubic Galileon model. }
    \label{fig:H_T_prime}
\end{figure}

However, since we have access to information about the theory of gravity in the effective fluid description, it is possible to implement it in the Einstein-Boltzmann solver. In \cref{fig:H_T_prime} we present preliminary results for this correction, but recall that it did not affect the simulations or the results presented in this manuscript. 

We find that the behaviour of the \lcdm\ model is close to the standard version of \class, but we find strong deviations at late times. This is related to the same issue of the effective DE perturbation not being exactly zero for this model. Then, for $f(R)$, we see some spurious oscillations and a decaying perturbation for very late times. Finally, the cubic Galileon model shows a strongly negative perturbation as time gets closer to the present time. When running simulations with this implemented, the last model fails because the correction to the force must be applied with extremely small time steps, and the simulation abruptly ends.

		\chapter{Conclusions \& Future Prospects}
In this work, we explored how different theories of gravity impact one of the most important observables of the Large Scale Structure: the matter power spectrum. We specifically focus on the non-linear regime, where most of the information lies, since there are many independent Fourier modes. This is highly relevant for upcoming cosmological surveys, such as LSST and Euclid, which offer a window to test General Relativity (GR) on scales previously unexplored. 

\section{Summary}
In this manuscript, we have presented three codes that aim to study the impact of Dark Energy perturbations on one of the key observables of the Large Scale Structure of the universe: the matter power spectrum. 

We started with a very simple solver for linear cosmological perturbations, whose main purpose was to have a first approximation on how DE perturbations can be solved in the late-time universe. We found excellent agreement with \class\ for sub-horizon modes, despite the simplicity of our code.

Then, with the goal of implementing Horndeski gravity in an N-body code, we modified \mochi\ to include new outputs required by \concept, the N-body code of choice to run our simulations. This resulted in an organised task, since we first focused on the linear regime and making the new version of \class\ work with the simulation code. This had several considerations, since just updating the version of \class\ was not a trivial task, given the several modifications made from version \texttt{2.7.2} to \texttt{3.3.3}. Then, to include all modifications of \mochi\ could easily break the N-body code, as well as the new outputs for the effective fluid.

Finally, we are in the process of validating our N-body code, which has implemented Horndeski theories of gravity through the effective fluid approach. In essence, this allows for having modifications of GR as a fictitious fluid that, at the linear level, leaves Einstein's field equations identical to the Modified Gravity (MG) scenario. The main advantage of this is the ease of implementation in code, bypassing several geometric considerations that even GR struggles with when building a numerical code. Additionally, given the tight integration between \concept\ and \class, modifications to the first code are quite simple.

We found excellent agreement between the baseline \lcdm\ model, using the standard version of \class, and our version of \class\ in the limit where the model of gravity closely resembles GR. The sub-0.2\% difference could be lower, but we argue that this emerges from a difference in initial conditions or noise in the simulation.

\section{Future Work and Extensions}
Among the remaining tasks that need to be done prior to the proper code validation is to correctly implement the contributions of a modified gravity model into the metric perturbation $H_T'$ of \cref{eq:H_T_prime}. This has not been done through the effective fluid approach, so we cannot directly compare our preliminary results with the current literature. However, we must study how these corrections are applied in N-body codes for modified theories of gravity. Additionally, we should implement the screening mechanism to some level. In the fluid description, this would correspond to introducing a Jeans length, where the perturbations of the fluid start to oscillate instead of continuing to grow. We still need to figure out how to implement the Jeans suppression, since the fluid is not dynamically evolved, but instead inferred from metric perturbations. Thus, one approach would be to implement a hard freeze to the perturbations once the smaller modes pass the Jeans limit.

Given that our setup is the first to include Horndeski gravity in an N-body code, several validation tests remain to be performed. For instance, the only code remotely similar to ours in the sense of generality of a Modified Gravity theory is \texttt{Hi-COLA}\footnote{The code is publicly available at \url{https://github.com/Hi-COLACode/Hi-COLA}.} \cite{wright_hi-cola_2023,}, which stands for \textit{Horndeski in COmoving Lagrangian Acceleration}. COLA simulations are performed with a method that uses second-order Lagrangian Perturbation theory to reduce the number of time steps in a simulation \cite{tassev_solving_2013}. This is quite faster than a full N-body simulation, which comes with an expense on suppression on small scales when compared to an N-body power spectrum. This code was used to build emulators by the Dark Energy Science Collaboration, so it would be interesting to test the emulated power spectra with our results.

It would be of great value to focus on the non-linear regime of DE perturbations. Given that we restrict ourselves to studying the impact of linear perturbations, modifications to the N-body code were simple enough. However, going to the non-linear regime implies several aspects, such as implementing the screening mechanism in a robust way for the fluid, since with this approach, one would need to translate the screening into an effective Jeans length scale, where the perturbation is not allowed to evolve. This would open a window to study the equivalence principle that holds at the linear level, which is the foundation of our work, in the non-linear regime. It could very well be the case that this is more complicated at these scales, and that the relationship between the Modified Gravity and Dark Energy model is no longer simple.

Our novel code could be exploited in several ways, and one of them is to build emulators. Given that, in general, running a high-resolution N-body simulation is computationally costly, the community has focused on training emulators using machine learning techniques, such as neural networks and Gaussian processes. This would still require a large volume of simulation runs to perform simulation-based inference properly \cite{bairagi_how_2025}, but would greatly ease the generalisation. This could be done in two ways, since someone may be interested in setting cosmological parameters and producing accurate, fast non-linear matter power spectra for those parameters. On the other hand, it could also be quite interesting to extract cosmological parameters given a matter power spectrum, and an emulator could be constructed to estimate posterior distributions on the parameters. 

	\printbibliography[prenote=iadeclaration,heading=bibintoc]

@misc{lesgourgues_classIII_2011,
	title = {The Cosmic Linear Anisotropy Solving System ({CLASS}) {III}: Comparision with {CAMB} for {LambdaCDM}},
	url = {http://arxiv.org/abs/1104.2934},
	doi = {10.48550/arXiv.1104.2934},
	shorttitle = {The Cosmic Linear Anisotropy Solving System ({CLASS}) {III}},
	number = {{arXiv}:1104.2934},
	publisher = {{arXiv}},
	author = {Lesgourgues, Julien},
	urldate = {2026-03-26},
	date = {2011-05-16},
	eprinttype = {arxiv},
	eprint = {1104.2934 [astro-ph]},
}

@article{song_large_2007,
	title = {Large scale structure of f ( R ) gravity},
	volume = {75},
	rights = {http://link.aps.org/licenses/aps-default-license},
	issn = {1550-7998, 1550-2368},
	url = {https://link.aps.org/doi/10.1103/PhysRevD.75.044004},
	doi = {10.1103/PhysRevD.75.044004},
	pages = {044004},
	number = {4},
	journaltitle = {Physical Review D},
	shortjournal = {Phys. Rev. D},
	author = {Song, Yong-Seon and Hu, Wayne and Sawicki, Ignacy},
	urldate = {2026-07-15},
	date = {2007-02-02},
	langid = {english},
}

@misc{fanha_robustness_2026,
	title = {On the robustness of the angular homogeneity scale \$θ\_H\$: a comparative analysis of computational approaches},
	url = {http://arxiv.org/abs/2606.18975},
	doi = {10.48550/arXiv.2606.18975},
	shorttitle = {On the robustness of the angular homogeneity scale \$θ\_H\$},
	number = {{arXiv}:2606.18975},
	publisher = {{arXiv}},
	author = {Fanha, Pedro and Silva, António da and Fonseca, José and Mimoso, José Pedro and Sakr, Ziad},
	urldate = {2026-06-23},
	date = {2026-06-17},
	eprinttype = {arxiv},
	eprint = {2606.18975 [astro-ph.CO]},
}

@article{gaite_fractal_1999,
	title = {The Fractal Distribution of Galaxies and the Transition to Homogeneity},
	volume = {522},
	issn = {0004-637X},
	url = {https://iopscience.iop.org/article/10.1086/312204},
	doi = {10.1086/312204},
	pages = {L5},
	number = {1},
	journaltitle = {The Astrophysical Journal},
	shortjournal = {{ApJ}},
	publisher = {{IOP} Publishing},
	author = {Gaite, José and Domínguez, Alvaro and Pérez-Mercader, Juan},
	urldate = {2026-06-22},
	date = {1999-08-02},
	langid = {english},
}

@article{scott_matter_2009,
	title = {Matter temperature during cosmological recombination},
	volume = {397},
	issn = {0035-8711},
	url = {https://doi.org/10.1111/j.1365-2966.2009.14939.x},
	doi = {10.1111/j.1365-2966.2009.14939.x},
	pages = {445--446},
	number = {1},
	journaltitle = {Monthly Notices of the Royal Astronomical Society},
	shortjournal = {Mon Not R Astron Soc},
	author = {Scott, Douglas and Moss, Adam},
	urldate = {2026-06-22},
	date = {2009-07-21},
}

@article{tram_fully_2019,
	title = {Fully relativistic treatment of light neutrinos in N-body simulations},
	volume = {2019},
	issn = {1475-7516},
	url = {https://doi.org/10.1088/1475-7516/2019/03/022},
	doi = {10.1088/1475-7516/2019/03/022},
	pages = {022},
	number = {3},
	journaltitle = {Journal of Cosmology and Astroparticle Physics},
	shortjournal = {J. Cosmol. Astropart. Phys.},
	author = {Tram, Thomas and Brandbyge, Jacob and Dakin, Jeppe and Hannestad, Steen},
	urldate = {2026-05-22},
	date = {2019-03},
	langid = {english},
}

@article{valle_neutrino_2005,
	title = {Neutrino masses and oscillations},
	volume = {805},
	issn = {0094-243X},
	url = {https://doi.org/10.1063/1.2149688},
	doi = {10.1063/1.2149688},
	pages = {128--134},
	number = {1},
	journaltitle = {{AIP} Conference Proceedings},
	shortjournal = {{AIP} Conf. Proc.},
	author = {Valle, J. W. F.},
	urldate = {2026-05-19},
	date = {2005-12-02},
}

@article{schwetz_three-flavour_2008,
	title = {Three-flavour neutrino oscillation update},
	volume = {10},
	issn = {1367-2630},
	url = {https://doi.org/10.1088/1367-2630/10/11/113011},
	doi = {10.1088/1367-2630/10/11/113011},
	pages = {113011},
	number = {11},
	journaltitle = {New Journal of Physics},
	shortjournal = {New J. Phys.},
	author = {Schwetz, Thomas and Tórtola, Mariam and Valle, José W F},
	urldate = {2026-05-19},
	date = {2008-11},
	langid = {english},
}

@article{puchwein_modified-gravity-gadget_2013,
	title = {Modified-Gravity-gadget: a new code for cosmological hydrodynamical simulations of modified gravity models},
	volume = {436},
	issn = {0035-8711},
	url = {https://doi.org/10.1093/mnras/stt1575},
	doi = {10.1093/mnras/stt1575},
	shorttitle = {Modified-Gravity-gadget},
	pages = {348--360},
	number = {1},
	journaltitle = {Monthly Notices of the Royal Astronomical Society},
	shortjournal = {Mon Not R Astron Soc},
	author = {Puchwein, Ewald and Baldi, Marco and Springel, Volker},
	urldate = {2026-05-19},
	date = {2013-11-21},
}

@article{penzias_measurement_1965,
	title = {A Measurement of Excess Antenna Temperature at 4080 Mc/s.},
	volume = {142},
	issn = {0004-637X},
	url = {https://ui.adsabs.harvard.edu/abs/1965ApJ...142..419P},
	doi = {10.1086/148307},
	pages = {419--421},
	journaltitle = {The Astrophysical Journal},
	publisher = {{IOP}},
	author = {Penzias, A. A. and Wilson, R. W.},
	urldate = {2026-05-19},
	date = {1965-07-01},
	note = {{ADS} Bibcode: 1965ApJ...142..419P},
}

@article{teyssier_ramses_2010,
	title = {{RAMSES}: A new N-body and hydrodynamical code},
	url = {https://ui.adsabs.harvard.edu/abs/2010ascl.soft11007T},
	shorttitle = {{RAMSES}},
	pages = {ascl:1011.007},
	journaltitle = {Astrophysics Source Code Library},
	author = {Teyssier, Romain},
	urldate = {2026-05-19},
	date = {2010-11-01},
	note = {{ADS} Bibcode: 2010ascl.soft11007T},
}

@article{xu_new_1995,
	title = {A New Parallel N-Body Gravity Solver: {TPM}},
	volume = {98},
	issn = {0067-0049},
	url = {https://ui.adsabs.harvard.edu/abs/1995ApJS...98..355X},
	doi = {10.1086/192166},
	shorttitle = {A New Parallel N-Body Gravity Solver},
	pages = {355},
	journaltitle = {The Astrophysical Journal Supplement Series},
	publisher = {{IOP}},
	author = {Xu, Guohong},
	urldate = {2026-05-19},
	date = {1995-05-01},
	note = {{ADS} Bibcode: 1995ApJS...98..355X},
}

@article{couchman_hydra_1995,
	title = {Hydra: an Adaptive-Mesh Implementation of P 3M-{SPH}},
	volume = {452},
	issn = {0004-637X},
	url = {https://ui.adsabs.harvard.edu/abs/1995ApJ...452..797C},
	doi = {10.1086/176348},
	shorttitle = {Hydra},
	pages = {797},
	journaltitle = {The Astrophysical Journal},
	publisher = {{IOP}},
	author = {Couchman, H. M. P. and Thomas, P. A. and Pearce, F. R.},
	urldate = {2026-05-19},
	date = {1995-10-01},
	note = {{ADS} Bibcode: 1995ApJ...452..797C},
}

@article{dubinski_gotpm_2004,
	title = {{GOTPM}: a parallel hybrid particle-mesh treecode},
	volume = {9},
	issn = {1384-1076},
	url = {https://www.sciencedirect.com/science/article/pii/S1384107603001337},
	doi = {10.1016/j.newast.2003.08.002},
	shorttitle = {{GOTPM}},
	pages = {111--126},
	number = {2},
	journaltitle = {New Astronomy},
	shortjournal = {New Astronomy},
	author = {Dubinski, John and Kim, Juhan and Park, Changbom and Humble, Robin},
	urldate = {2026-05-19},
	date = {2004-02-01},
}

@article{angulo_scaling_2012,
	title = {Scaling relations for galaxy clusters in the Millennium-{XXL} simulation},
	volume = {426},
	issn = {0035-8711},
	url = {https://doi.org/10.1111/j.1365-2966.2012.21830.x},
	doi = {10.1111/j.1365-2966.2012.21830.x},
	pages = {2046--2062},
	number = {3},
	journaltitle = {Monthly Notices of the Royal Astronomical Society},
	shortjournal = {Mon Not R Astron Soc},
	author = {Angulo, R. E. and Springel, V. and White, S. D. M. and Jenkins, A. and Baugh, C. M. and Frenk, C. S.},
	urldate = {2026-05-19},
	date = {2012-11-01},
}

@article{springel_cosmological_2005,
	title = {The cosmological simulation code gadget-2},
	volume = {364},
	issn = {0035-8711},
	url = {https://doi.org/10.1111/j.1365-2966.2005.09655.x},
	doi = {10.1111/j.1365-2966.2005.09655.x},
	pages = {1105--1134},
	number = {4},
	journaltitle = {Monthly Notices of the Royal Astronomical Society},
	shortjournal = {Mon Not R Astron Soc},
	author = {Springel, Volker},
	urldate = {2026-05-19},
	date = {2005-12-21},
}

@article{springel_gadget_2001,
	title = {{GADGET}: a code for collisionless and gasdynamical cosmological simulations},
	volume = {6},
	issn = {1384-1076},
	url = {https://www.sciencedirect.com/science/article/pii/S1384107601000422},
	doi = {10.1016/S1384-1076(01)00042-2},
	shorttitle = {{GADGET}},
	pages = {79--117},
	number = {2},
	journaltitle = {New Astronomy},
	shortjournal = {New Astronomy},
	author = {Springel, Volker and Yoshida, Naoki and White, Simon D. M.},
	urldate = {2026-05-19},
	date = {2001-04-01},
}

@article{harnois-deraps_high-performance_2013,
	title = {High-performance P3M N-body code: cubep3m},
	volume = {436},
	issn = {0035-8711},
	url = {https://doi.org/10.1093/mnras/stt1591},
	doi = {10.1093/mnras/stt1591},
	shorttitle = {High-performance P3M N-body code},
	pages = {540--559},
	number = {1},
	journaltitle = {Monthly Notices of the Royal Astronomical Society},
	shortjournal = {Mon Not R Astron Soc},
	author = {Harnois-Déraps, Joachim and Pen, Ue-Li and Iliev, Ilian T. and Merz, Hugh and Emberson, J. D. and Desjacques, Vincent},
	urldate = {2026-05-19},
	date = {2013-11-21},
}

@article{angulo_cosmological_2016,
	title = {Cosmological N-body simulations with suppressed variance},
	volume = {462},
	issn = {1745-3925},
	url = {https://doi.org/10.1093/mnrasl/slw098},
	doi = {10.1093/mnrasl/slw098},
	pages = {L1--L5},
	number = {1},
	journaltitle = {Monthly Notices of the Royal Astronomical Society: Letters},
	shortjournal = {Mon Not R Astron Soc Lett},
	author = {Angulo, Raul E. and Pontzen, Andrew},
	urldate = {2026-05-18},
	date = {2016-07-01},
}

@article{seager_new_1999,
	title = {A New Calculation of the Recombination Epoch},
	volume = {523},
	issn = {0004-637X},
	url = {https://iopscience.iop.org/article/10.1086/312250},
	doi = {10.1086/312250},
	pages = {L1},
	number = {1},
	journaltitle = {The Astrophysical Journal},
	shortjournal = {{ApJ}},
	publisher = {{IOP} Publishing},
	author = {Seager, S. and Sasselov, D. D. and Scott, D.},
	urldate = {2026-05-18},
	date = {1999-08-23},
	langid = {english},
}

@article{baumann_probing_2009,
	title = {Probing Inflation with {CMB} Polarization},
	volume = {1141},
	issn = {0094-243X},
	url = {https://doi.org/10.1063/1.3160885},
	doi = {10.1063/1.3160885},
	pages = {10--120},
	number = {1},
	journaltitle = {{AIP} Conference Proceedings},
	shortjournal = {{AIP} Conf. Proc.},
	author = {Baumann, Daniel and Jackson, Mark G. and Adshead, Peter and Amblard, Alexandre and Ashoorioon, Amjad and Bartolo, Nicola and Bean, Rachel and Beltrán, Maria and de Bernardis, Francesco and Bird, Simeon and Chen, Xingang and Chung, Daniel J. H. and Colombo, Loris and Cooray, Asantha and Creminelli, Paolo and Dodelson, Scott and Dunkley, Joanna and Dvorkin, Cora and Easther, Richard and Finelli, Fabio and Flauger, Raphael and Hertzberg, Mark P. and Jones‐Smith, Katherine and Kachru, Shamit and Kadota, Kenji and Khoury, Justin and Kinney, William H. and Komatsu, Eiichiro and Krauss, Lawrence M. and Lesgourgues, Julien and Liddle, Andrew and Liguori, Michele and Lim, Eugene and Linde, Andrei and Matarrese, Sabino and Mathur, Harsh and {McAllister}, Liam and Melchiorri, Alessandro and Nicolis, Alberto and Pagano, Luca and Peiris, Hiranya V. and Peloso, Marco and Pogosian, Levon and Pierpaoli, Elena and Riotto, Antonio and Seljak, Uroš and Senatore, Leonardo and Shandera, Sarah and Silverstein, Eva and Smith, Tristan and Vaudrevange, Pascal and Verde, Licia and Wandelt, Ben and Wands, David and Watson, Scott and Wyman, Mark and Yadav, Amit and Valkenburg, Wessel and Zaldarriaga, Matias},
	urldate = {2026-05-18},
	date = {2009-06-09},
}

@article{casas_euclid_2026,
	title = {Euclid: Constraints on f(R) cosmologies from the spectroscopic and photometric primary probes},
	volume = {707},
	rights = {© The Authors 2026},
	issn = {0004-6361, 1432-0746},
	url = {https://www.aanda.org/articles/aa/abs/2026/03/aa48713-23/aa48713-23.html},
	doi = {10.1051/0004-6361/202348713},
	shorttitle = {Euclid},
	pages = {A176},
	journaltitle = {Astronomy \& Astrophysics},
	shortjournal = {A\&A},
	publisher = {{EDP} Sciences},
	author = {Casas, S. and Cardone, V. F. and Sapone, D. and Frusciante, N. and Pace, F. and Parimbelli, G. and Archidiacono, M. and Koyama, K. and Tutusaus, I. and Camera, S. and Martinelli, M. and Pettorino, V. and Sakr, Z. and Lombriser, L. and Silvestri, A. and Pietroni, M. and Vernizzi, F. and Kunz, M. and Ntelis, P. and Kitching, T. and Pourtsidou, A. and Lacasa, F. and Carbone, C. and Garcia-Bellido, J. and Aghanim, N. and Altieri, B. and Amara, A. and Auricchio, N. and Baldi, M. and Bodendorf, C. and Branchini, E. and Brescia, M. and Brinchmann, J. and Capobianco, V. and Carretero, J. and Castellano, M. and Cavuoti, S. and Cimatti, A. and Cledassou, R. and Congedo, G. and Conselice, C. J. and Conversi, L. and Copin, Y. and Corcione, L. and Courbin, F. and Courtois, H. M. and Silva, A. Da and Degaudenzi, H. and Dubath, F. and Duncan, C. a. J. and Dupac, X. and Dusini, S. and Farrens, S. and Ferriol, S. and Fosalba, P. and Frailis, M. and Franceschi, E. and Fumana, M. and Galeotta, S. and Garilli, B. and Gillard, W. and Gillis, B. and Giocoli, C. and Grazian, A. and Grupp, F. and Guzzo, L. and Haugan, S. V. H. and Hormuth, F. and Hornstrup, A. and Hudelot, P. and Jahnke, K. and Kermiche, S. and Kiessling, A. and Kilbinger, M. and Kurki-Suonio, H. and Ligori, S. and Lilje, P. B. and Lloro, I. and Maiorano, E. and Mansutti, O. and Marggraf, O. and Marulli, F. and Massey, R. and Medinaceli, E. and Meneghetti, M. and Merlin, E. and Meylan, G. and Moresco, M. and Moscardini, L. and Munari, E. and Niemi, S.-M. and Padilla, C. and Paltani, S. and Pasian, F. and Pedersen, K. and Percival, W. J. and Pires, S. and Polenta, G. and Poncet, M. and Popa, L. A. and Raison, F. and Renzi, A. and Rhodes, J. and Riccio, G. and Romelli, E. and Roncarelli, M. and Rossetti, E. and Saglia, R. and Sartoris, B. and Secroun, A. and Seidel, G. and Serrano, S. and Sirignano, C. and Sirri, G. and Stanco, L. and Starck, J.-L. and Surace, C. and Tallada-Crespí, P. and Taylor, A. N. and Tereno, I. and Toledo-Moreo, R. and Torradeflot, F. and Valentijn, E. A. and Valenziano, L. and Vassallo, T. and Wang, Y. and Weller, J. and Zoubian, J. and Mellier, Y. and Scottez, V.},
	urldate = {2026-05-11},
	date = {2026-03-01},
	langid = {english},
}

@article{wright_hi-cola_2023,
	title = {Hi-{COLA}: fast, approximate simulations of structure formation in Horndeski gravity},
	volume = {2023},
	issn = {1475-7516},
	url = {https://doi.org/10.1088/1475-7516/2023/03/040},
	doi = {10.1088/1475-7516/2023/03/040},
	shorttitle = {Hi-{COLA}},
	pages = {040},
	number = {3},
	journaltitle = {Journal of Cosmology and Astroparticle Physics},
	shortjournal = {J. Cosmol. Astropart. Phys.},
	publisher = {{IOP} Publishing},
	author = {Wright, Bill S. and Sen Gupta, Ashim and Baker, Tessa and Valogiannis, Georgios and Fiorini, Bartolomeo and collaboration, The {LSST} Dark Energy Science},
	urldate = {2026-05-11},
	date = {2023-03},
	langid = {english},
}

@article{tassev_solving_2013,
	title = {Solving large scale structure in ten easy steps with {COLA}},
	volume = {2013},
	issn = {1475-7516},
	url = {https://doi.org/10.1088/1475-7516/2013/06/036},
	doi = {10.1088/1475-7516/2013/06/036},
	pages = {036},
	number = {6},
	journaltitle = {Journal of Cosmology and Astroparticle Physics},
	shortjournal = {J. Cosmol. Astropart. Phys.},
	author = {Tassev, Svetlin and Zaldarriaga, Matias and Eisenstein, Daniel J.},
	urldate = {2026-05-11},
	date = {2013-06},
	langid = {english},
}

@article{de_felice_cosmology_2010,
	title = {Cosmology of a Covariant Galileon Field},
	volume = {105},
	url = {https://link.aps.org/doi/10.1103/PhysRevLett.105.111301},
	doi = {10.1103/PhysRevLett.105.111301},
	pages = {111301},
	number = {11},
	journaltitle = {Physical Review Letters},
	shortjournal = {Phys. Rev. Lett.},
	publisher = {American Physical Society},
	author = {De Felice, Antonio and Tsujikawa, Shinji},
	urldate = {2026-05-08},
	date = {2010-09-07},
}

@article{lewis_efficient_2000,
	title = {Efficient Computation of Cosmic Microwave Background Anisotropies {inClosed} Friedmann-Robertson-Walker Models},
	volume = {538},
	issn = {0004-637X},
	url = {https://iopscience.iop.org/article/10.1086/309179},
	doi = {10.1086/309179},
	pages = {473},
	number = {2},
	journaltitle = {The Astrophysical Journal},
	shortjournal = {{ApJ}},
	publisher = {{IOP} Publishing},
	author = {Lewis, Antony and Challinor, Anthony and Lasenby, Anthony},
	urldate = {2026-05-04},
	date = {2000-08-01},
	langid = {english},
}

@article{desi_collaboration_desi_2025,
	title = {{DESI} {DR}2 results. {II}. Measurements of baryon acoustic oscillations and cosmological constraints},
	volume = {112},
	url = {https://link.aps.org/doi/10.1103/tr6y-kpc6},
	doi = {10.1103/tr6y-kpc6},
	pages = {083515},
	number = {8},
	journaltitle = {Physical Review D},
	shortjournal = {Phys. Rev. D},
	publisher = {American Physical Society},
	author = {{DESI Collaboration} and Abdul Karim, M. and Aguilar, J. and Ahlen, S. and Alam, S. and Allen, L. and Prieto, C. Allende and Alves, O. and Anand, A. and Andrade, U. and Armengaud, E. and Aviles, A. and Bailey, S. and Baltay, C. and Bansal, P. and Bault, A. and Behera, J. and BenZvi, S. and Bianchi, D. and Blake, C. and Brieden, S. and Brodzeller, A. and Brooks, D. and Buckley-Geer, E. and Burtin, E. and Calderon, R. and Canning, R. and Rosell, A. Carnero and Carrilho, P. and Casas, L. and Castander, F. J. and Charles, M. and Chaussidon, E. and Chaves-Montero, J. and Chebat, D. and Chen, X. and Claybaugh, T. and Cole, S. and Cooper, A. P. and Cuceu, A. and Dawson, K. S. and de la Macorra, A. and de Mattia, A. and Deiosso, N. and Della Costa, J. and Demina, R. and Dey, A. and Dey, B. and Ding, Z. and Doel, P. and Edelstein, J. and Eisenstein, D. J. and Elbers, W. and Fagrelius, P. and Fanning, K. and Fernández-García, E. and Ferraro, S. and Font-Ribera, A. and Forero-Romero, J. E. and Frenk, C. S. and Garcia-Quintero, C. and Garrison, L. H. and Gaztañaga, E. and Gil-Marín, H. and Gontcho, S. Gontcho A. and Gonzalez, D. and Gonzalez-Morales, A. X. and Gordon, C. and Green, D. and Gutierrez, G. and Guy, J. and Hadzhiyska, B. and Hahn, C. and He, S. and Herbold, M. and Herrera-Alcantar, H. K. and Ho, M.-F. and Honscheid, K. and Howlett, C. and Huterer, D. and Ishak, M. and Juneau, S. and Kamble, N. V. and Karaçayl𝚤, N. G. and Kehoe, R. and Kent, S. and Kim, A. G. and Kirkby, D. and Kisner, T. and Koposov, S. E. and Kremin, A. and Krolewski, A. and Lahav, O. and Lamman, C. and Landriau, M. and Lang, D. and Lasker, J. and Le Goff, J. M. and Le Guillou, L. and Leauthaud, A. and Levi, M. E. and Li, Q. and Li, T. S. and Lodha, K. and Lokken, M. and Lozano-Rodríguez, F. and Magneville, C. and Manera, M. and Martini, P. and Matthewson, W. L. and Meisner, A. and Mena-Fernández, J. and Menegas, A. and Mergulhão, T. and Miquel, R. and Moustakas, J. and Muñoz-Gutiérrez, A. and Muñoz-Santos, D. and Myers, A. D. and Nadathur, S. and Naidoo, K. and Napolitano, L. and Newman, J. A. and Niz, G. and Noriega, H. E. and Paillas, E. and Palanque-Delabrouille, N. and Pan, J. and Peacock, J. A. and Ibanez, M. P. and Percival, W. J. and Pérez-Fernández, A. and Pérez-Ràfols, I. and Pieri, M. M. and Poppett, C. and Prada, F. and Rabinowitz, D. and Raichoor, A. and Ramírez-Pérez, C. and Rashkovetskyi, M. and Ravoux, C. and Rich, J. and Rocher, A. and Rockosi, C. and Rohlf, J. and Román-Herrera, J. O. and Ross, A. J. and Rossi, G. and Ruggeri, R. and Ruhlmann-Kleider, V. and Samushia, L. and Sanchez, E. and Sanders, N. and Schlegel, D. and Schubnell, M. and Seo, H. and Shafieloo, A. and Sharples, R. and Silber, J. and Sinigaglia, F. and Sprayberry, D. and Tan, T. and Tarlé, G. and Taylor, P. and Turner, W. and Ureña-López, L. A. and Vaisakh, R. and Valdes, F. and Valogiannis, G. and Vargas-Magaña, M. and Verde, L. and Walther, M. and Weaver, B. A. and Weinberg, D. H. and White, M. and Wolfson, M. and Yèche, C. and Yu, J. and Zaborowski, E. A. and Zarrouk, P. and Zhai, Z. and Zhang, H. and Zhao, C. and Zhao, G. B. and Zhou, R. and Zou, H.},
	urldate = {2026-05-04},
	date = {2025-10-06},
}

@article{einstein_feldgleichungen_1915,
	title = {Die Feldgleichungen der Gravitation},
	url = {https://ui.adsabs.harvard.edu/abs/1915SPAW.......844E},
	pages = {844--847},
	journaltitle = {Sitzungsberichte der Königlich Preussischen Akademie der Wissenschaften},
	author = {Einstein, Albert},
	urldate = {2026-05-04},
	date = {1915-01-01},
	note = {{ADS} Bibcode: 1915SPAW.......844E},
}

@book{weinberg_cosmology_2008,
	location = {Oxford, New York},
	title = {Cosmology},
	isbn = {978-0-19-852682-7},
	pagetotal = {624},
	publisher = {Oxford University Press},
	author = {Weinberg, Steven},
	date = {2008-04-28},
}

@article{monaghan_refined_1985,
	title = {A refined particle method for astrophysical problems},
	volume = {149},
	issn = {0004-6361},
	url = {https://ui.adsabs.harvard.edu/abs/1985A&A...149..135M},
	pages = {135--143},
	journaltitle = {Astronomy and Astrophysics},
	publisher = {{EDP}},
	author = {Monaghan, J. J. and Lattanzio, J. C.},
	urldate = {2026-04-02},
	date = {1985-08-01},
	note = {{ADS} Bibcode: 1985A\&A...149..135M},
}

@article{plummer_problem_1911,
	title = {On the Problem of Distribution in Globular Star Clusters: (Plate 8.)},
	volume = {71},
	issn = {0035-8711},
	url = {https://doi.org/10.1093/mnras/71.5.460},
	doi = {10.1093/mnras/71.5.460},
	shorttitle = {On the Problem of Distribution in Globular Star Clusters},
	pages = {460--470},
	number = {5},
	journaltitle = {Monthly Notices of the Royal Astronomical Society},
	shortjournal = {Mon Not R Astron Soc},
	author = {Plummer, H. C.},
	urldate = {2026-04-02},
	date = {1911-03-10},
}

@article{bozorgnia_dark_2025,
	title = {Dark matter candidates and searches},
	volume = {103},
	issn = {0008-4204},
	url = {https://cdnsciencepub.com/doi/10.1139/cjp-2024-0128},
	doi = {10.1139/cjp-2024-0128},
	pages = {671--703},
	number = {8},
	journaltitle = {Canadian Journal of Physics},
	shortjournal = {Can. J. Phys.},
	publisher = {{NRC} Research Press},
	author = {Bozorgnia, Nassim and Bramante, Joseph and Cline, James M. and Curtin, David and {McKeen}, David and Morrissey, David E. and Ritz, Adam and Viel, Simon and Vincent, Aaron C. and Zhang, Yue},
	urldate = {2026-03-19},
	date = {2025-08},
}

@article{starobinsky_new_1980,
	title = {A new type of isotropic cosmological models without singularity},
	volume = {91},
	issn = {0370-2693},
	url = {https://www.sciencedirect.com/science/article/pii/037026938090670X},
	doi = {10.1016/0370-2693(80)90670-X},
	pages = {99--102},
	number = {1},
	journaltitle = {Physics Letters B},
	shortjournal = {Physics Letters B},
	author = {Starobinsky, A. A.},
	urldate = {2026-03-19},
	date = {1980-03-24},
}

@article{charmousis_general_2012,
	title = {General Second-Order Scalar-Tensor Theory and Self-Tuning},
	volume = {108},
	url = {https://link.aps.org/doi/10.1103/PhysRevLett.108.051101},
	doi = {10.1103/PhysRevLett.108.051101},
	pages = {051101},
	number = {5},
	journaltitle = {Physical Review Letters},
	shortjournal = {Phys. Rev. Lett.},
	publisher = {American Physical Society},
	author = {Charmousis, Christos and Copeland, Edmund J. and Padilla, Antonio and Saffin, Paul M.},
	urldate = {2026-03-19},
	date = {2012-01-30},
}

@article{kobayashi_generalized_2011,
	title = {Generalized G-Inflation: —Inflation with the Most General Second-Order Field Equations—},
	volume = {126},
	issn = {0033-068X},
	url = {https://dx.doi.org/10.1143/PTP.126.511},
	doi = {10.1143/PTP.126.511},
	shorttitle = {Generalized G-Inflation},
	pages = {511--529},
	number = {3},
	journaltitle = {Progress of Theoretical Physics},
	shortjournal = {Prog Theor Phys},
	publisher = {Oxford Academic},
	author = {Kobayashi, Tsutomu and Yamaguchi, Masahide and Yokoyama, Jun'ichi},
	urldate = {2026-03-19},
	date = {2011-09-01},
	langid = {english},
}

@article{deffayet_k-essence_2011,
	title = {From \$k\$-essence to generalized Galileons},
	volume = {84},
	url = {https://link.aps.org/doi/10.1103/PhysRevD.84.064039},
	doi = {10.1103/PhysRevD.84.064039},
	pages = {064039},
	number = {6},
	journaltitle = {Physical Review D},
	shortjournal = {Phys. Rev. D},
	publisher = {American Physical Society},
	author = {Deffayet, C. and Gao, Xian and Steer, D. A. and Zahariade, G.},
	urldate = {2026-03-19},
	date = {2011-09-27},
}

@article{aihara_hyper_2018,
	title = {The Hyper Suprime-Cam {SSP} Survey: Overview and survey design},
	volume = {70},
	issn = {0004-6264},
	url = {https://doi.org/10.1093/pasj/psx066},
	doi = {10.1093/pasj/psx066},
	shorttitle = {The Hyper Suprime-Cam {SSP} Survey},
	pages = {S4},
	issue = {{SP}1},
	journaltitle = {Publications of the Astronomical Society of Japan},
	shortjournal = {Publ Astron Soc Jpn Nihon Tenmon Gakkai},
	author = {Aihara, Hiroaki and Arimoto, Nobuo and Armstrong, Robert and Arnouts, Stéphane and Bahcall, Neta A and Bickerton, Steven and Bosch, James and Bundy, Kevin and Capak, Peter L and Chan, James H H and Chiba, Masashi and Coupon, Jean and Egami, Eiichi and Enoki, Motohiro and Finet, Francois and Fujimori, Hiroki and Fujimoto, Seiji and Furusawa, Hisanori and Furusawa, Junko and Goto, Tomotsugu and Goulding, Andy and Greco, Johnny P and Greene, Jenny E and Gunn, James E and Hamana, Takashi and Harikane, Yuichi and Hashimoto, Yasuhiro and Hattori, Takashi and Hayashi, Masao and Hayashi, Yusuke and Hełminiak, Krzysztof G and Higuchi, Ryo and Hikage, Chiaki and Ho, Paul T P and Hsieh, Bau-Ching and Huang, Kuiyun and Huang, Song and Ikeda, Hiroyuki and Imanishi, Masatoshi and Inoue, Akio K and Iwasawa, Kazushi and Iwata, Ikuru and Jaelani, Anton T and Jian, Hung-Yu and Kamata, Yukiko and Karoji, Hiroshi and Kashikawa, Nobunari and Katayama, Nobuhiko and Kawanomoto, Satoshi and Kayo, Issha and Koda, Jin and Koike, Michitaro and Kojima, Takashi and Komiyama, Yutaka and Konno, Akira and Koshida, Shintaro and Koyama, Yusei and Kusakabe, Haruka and Leauthaud, Alexie and Lee, Chien-Hsiu and Lin, Lihwai and Lin, Yen-Ting and Lupton, Robert H and Mandelbaum, Rachel and Matsuoka, Yoshiki and Medezinski, Elinor and Mineo, Sogo and Miyama, Shoken and Miyatake, Hironao and Miyazaki, Satoshi and Momose, Rieko and More, Anupreeta and More, Surhud and Moritani, Yuki and Moriya, Takashi J and Morokuma, Tomoki and Mukae, Shiro and Murata, Ryoma and Murayama, Hitoshi and Nagao, Tohru and Nakata, Fumiaki and Niida, Mana and Niikura, Hiroko and Nishizawa, Atsushi J and Obuchi, Yoshiyuki and Oguri, Masamune and Oishi, Yukie and Okabe, Nobuhiro and Okamoto, Sakurako and Okura, Yuki and Ono, Yoshiaki and Onodera, Masato and Onoue, Masafusa and Osato, Ken and Ouchi, Masami and Price, Paul A and Pyo, Tae-Soo and Sako, Masao and Sawicki, Marcin and Shibuya, Takatoshi and Shimasaku, Kazuhiro and Shimono, Atsushi and Shirasaki, Masato and Silverman, John D and Simet, Melanie and Speagle, Joshua and Spergel, David N and Strauss, Michael A and Sugahara, Yuma and Sugiyama, Naoshi and Suto, Yasushi and Suyu, Sherry H and Suzuki, Nao and Tait, Philip J and Takada, Masahiro and Takata, Tadafumi and Tamura, Naoyuki and Tanaka, Manobu M and Tanaka, Masaomi and Tanaka, Masayuki and Tanaka, Yoko and Terai, Tsuyoshi and Terashima, Yuichi and Toba, Yoshiki and Tominaga, Nozomu and Toshikawa, Jun and Turner, Edwin L and Uchida, Tomohisa and Uchiyama, Hisakazu and Umetsu, Keiichi and Uraguchi, Fumihiro and Urata, Yuji and Usuda, Tomonori and Utsumi, Yousuke and Wang, Shiang-Yu and Wang, Wei-Hao and Wong, Kenneth C and Yabe, Kiyoto and Yamada, Yoshihiko and Yamanoi, Hitomi and Yasuda, Naoki and Yeh, Sherry and Yonehara, Atsunori and Yuma, Suraphong},
	urldate = {2026-03-19},
	date = {2018-01-01},
}

@article{amon_non-linear_2022,
	title = {A non-linear solution to the S8 tension?},
	volume = {516},
	issn = {0035-8711},
	url = {https://doi.org/10.1093/mnras/stac2429},
	doi = {10.1093/mnras/stac2429},
	pages = {5355--5366},
	number = {4},
	journaltitle = {Monthly Notices of the Royal Astronomical Society},
	shortjournal = {Mon Not R Astron Soc},
	author = {Amon, Alexandra and Efstathiou, George},
	urldate = {2026-03-19},
	date = {2022-11-11},
}

@article{di_valentino_snowmass2021_2021,
	title = {Snowmass2021 - Letter of interest cosmology intertwined I: Perspectives for the next decade},
	volume = {131},
	issn = {09276505},
	url = {https://linkinghub.elsevier.com/retrieve/pii/S0927650521000505},
	doi = {10.1016/j.astropartphys.2021.102606},
	shorttitle = {Snowmass2021 - Letter of interest cosmology intertwined I},
	pages = {102606},
	journaltitle = {Astroparticle Physics},
	shortjournal = {Astroparticle Physics},
	author = {Di Valentino, Eleonora and Anchordoqui, Luis A. and Akarsu, Özgür and Ali-Haimoud, Yacine and Amendola, Luca and Arendse, Nikki and Asgari, Marika and Ballardini, Mario and Basilakos, Spyros and Battistelli, Elia and Benetti, Micol and Birrer, Simon and Bouchet, François R. and Bruni, Marco and Calabrese, Erminia and Camarena, David and Capozziello, Salvatore and Chen, Angela and Chluba, Jens and Chudaykin, Anton and Colgáin, Eoin Ó and Cyr-Racine, Francis-Yan and De Bernardis, Paolo and De Cruz Pérez, Javier and Delabrouille, Jacques and Dunkley, Jo and Escamilla-Rivera, Celia and Ferté, Agnès and Finelli, Fabio and Freedman, Wendy and Frusciante, Noemi and Giusarma, Elena and Gómez-Valent, Adrià and Handley, Will and Harrison, Ian and Hart, Luke and Heavens, Alan and Hildebrandt, Hendrik and Holz, Daniel and Huterer, Dragan and Ivanov, Mikhail M. and Joudaki, Shahab and Kamionkowski, Marc and Karwal, Tanvi and Knox, Lloyd and Kumar, Suresh and Lamagna, Luca and Lesgourgues, Julien and Lucca, Matteo and Marra, Valerio and Masi, Silvia and Matarrese, Sabino and Mazumdar, Arindam and Melchiorri, Alessandro and Mena, Olga and Mersini-Houghton, Laura and Miranda, Vivian and Moreno-Pulido, Cristian and Mota, David F. and Muir, Jessica and Mukherjee, Ankan and Niedermann, Florian and Notari, Alessio and Nunes, Rafael C. and Pace, Francesco and Paliathanasis, Andronikos and Palmese, Antonella and Pan, Supriya and Paoletti, Daniela and Pettorino, Valeria and Piacentini, Francesco and Poulin, Vivian and Raveri, Marco and Riess, Adam G. and Salzano, Vincenzo and Saridakis, Emmanuel N. and Sen, Anjan A. and Shafieloo, Arman and Shajib, Anowar J. and Silk, Joseph and Silvestri, Alessandra and Sloth, Martin S. and Smith, Tristan L. and Solà Peracaula, Joan and Van De Bruck, Carsten and Verde, Licia and Visinelli, Luca and Wandelt, Benjamin D. and Wang, Deng and Wang, Jian-Min and Yadav, Anil K. and Yang, Weiqiang},
	urldate = {2026-03-19},
	date = {2021-09},
	langid = {english},
}

@article{dalal_hyper_2023,
	title = {Hyper Suprime-Cam Year 3 results: Cosmology from cosmic shear power spectra},
	volume = {108},
	url = {https://link.aps.org/doi/10.1103/PhysRevD.108.123519},
	doi = {10.1103/PhysRevD.108.123519},
	shorttitle = {Hyper Suprime-Cam Year 3 results},
	pages = {123519},
	number = {12},
	journaltitle = {Physical Review D},
	shortjournal = {Phys. Rev. D},
	publisher = {American Physical Society},
	author = {Dalal, Roohi and Li, Xiangchong and Nicola, Andrina and Zuntz, Joe and Strauss, Michael A. and Sugiyama, Sunao and Zhang, Tianqing and Rau, Markus M. and Mandelbaum, Rachel and Takada, Masahiro and More, Surhud and Miyatake, Hironao and Kannawadi, Arun and Shirasaki, Masato and Taniguchi, Takanori and Takahashi, Ryuichi and Osato, Ken and Hamana, Takashi and Oguri, Masamune and Nishizawa, Atsushi J. and Malagón, Andrés A. Plazas and Sunayama, Tomomi and Alonso, David and Slosar, Anže and Luo, Wentao and Armstrong, Robert and Bosch, James and Hsieh, Bau-Ching and Komiyama, Yutaka and Lupton, Robert H. and Lust, Nate B. and {MacArthur}, Lauren A. and Miyazaki, Satoshi and Murayama, Hitoshi and Nishimichi, Takahiro and Okura, Yuki and Price, Paul A. and Tait, Philip J. and Tanaka, Masayuki and Wang, Shiang-Yu},
	urldate = {2026-03-19},
	date = {2023-12-11},
}

@misc{preston_non-linear_2023,
	title = {A non-linear solution to the \$S\_8\$ tension {II}: Analysis of {DES} Year 3 cosmic shear},
	url = {http://arxiv.org/abs/2305.09827},
	doi = {10.48550/arXiv.2305.09827},
	shorttitle = {A non-linear solution to the \$S\_8\$ tension {II}},
	number = {{arXiv}:2305.09827},
	publisher = {{arXiv}},
	author = {Preston, Calvin and Amon, Alexandra and Efstathiou, George},
	urldate = {2026-03-19},
	date = {2023-05-16},
	eprinttype = {arxiv},
	eprint = {2305.09827 [astro-ph]},
}

@article{di_valentino_exploring_2018,
	title = {Exploring the Tension between Current Cosmic Microwave Background and Cosmic Shear Data},
	volume = {10},
	rights = {http://creativecommons.org/licenses/by/3.0/},
	issn = {2073-8994},
	url = {https://www.mdpi.com/2073-8994/10/11/585},
	doi = {10.3390/sym10110585},
	pages = {585},
	number = {11},
	journaltitle = {Symmetry},
	publisher = {Multidisciplinary Digital Publishing Institute},
	author = {Di Valentino, Eleonora and Bridle, Sarah},
	urldate = {2026-03-19},
	date = {2018-11},
	langid = {english},
}

@article{des_collaboration_dark_2022,
	title = {Dark Energy Survey Year 3 results: Cosmology from cosmic shear and robustness to data calibration},
	volume = {105},
	url = {https://link.aps.org/doi/10.1103/PhysRevD.105.023514},
	doi = {10.1103/PhysRevD.105.023514},
	shorttitle = {Dark Energy Survey Year 3 results},
	pages = {023514},
	number = {2},
	journaltitle = {Physical Review D},
	shortjournal = {Phys. Rev. D},
	publisher = {American Physical Society},
	author = {{DES Collaboration} and Amon, A. and Gruen, D. and Troxel, M. A. and MacCrann, N. and Dodelson, S. and Choi, A. and Doux, C. and Secco, L. F. and Samuroff, S. and Krause, E. and Cordero, J. and Myles, J. and DeRose, J. and Wechsler, R. H. and Gatti, M. and Navarro-Alsina, A. and Bernstein, G. M. and Jain, B. and Blazek, J. and Alarcon, A. and Ferté, A. and Lemos, P. and Raveri, M. and Campos, A. and Prat, J. and Sánchez, C. and Jarvis, M. and Alves, O. and Andrade-Oliveira, F. and Baxter, E. and Bechtol, K. and Becker, M. R. and Bridle, S. L. and Camacho, H. and Carnero Rosell, A. and Carrasco Kind, M. and Cawthon, R. and Chang, C. and Chen, R. and Chintalapati, P. and Crocce, M. and Davis, C. and Diehl, H. T. and Drlica-Wagner, A. and Eckert, K. and Eifler, T. F. and Elvin-Poole, J. and Everett, S. and Fang, X. and Fosalba, P. and Friedrich, O. and Gaztanaga, E. and Giannini, G. and Gruendl, R. A. and Harrison, I. and Hartley, W. G. and Herner, K. and Huang, H. and Huff, E. M. and Huterer, D. and Kuropatkin, N. and Leget, P. and Liddle, A. R. and McCullough, J. and Muir, J. and Pandey, S. and Park, Y. and Porredon, A. and Refregier, A. and Rollins, R. P. and Roodman, A. and Rosenfeld, R. and Ross, A. J. and Rykoff, E. S. and Sanchez, J. and Sevilla-Noarbe, I. and Sheldon, E. and Shin, T. and Troja, A. and Tutusaus, I. and Tutusaus, I. and Varga, T. N. and Weaverdyck, N. and Yanny, B. and Yin, B. and Zhang, Y. and Zuntz, J. and Aguena, M. and Allam, S. and Annis, J. and Bacon, D. and Bertin, E. and Bhargava, S. and Brooks, D. and Buckley-Geer, E. and Burke, D. L. and Carretero, J. and Costanzi, M. and da Costa, L. N. and Pereira, M. E. S. and De Vicente, J. and Desai, S. and Dietrich, J. P. and Doel, P. and Ferrero, I. and Flaugher, B. and Frieman, J. and García-Bellido, J. and Gaztanaga, E. and Gerdes, D. W. and Giannantonio, T. and Gschwend, J. and Gutierrez, G. and Hinton, S. R. and Hollowood, D. L. and Honscheid, K. and Hoyle, B. and James, D. J. and Kron, R. and Kuehn, K. and Lahav, O. and Lima, M. and Lin, H. and Maia, M. A. G. and Marshall, J. L. and Martini, P. and Melchior, P. and Menanteau, F. and Miquel, R. and Mohr, J. J. and Morgan, R. and Ogando, R. L. C. and Palmese, A. and Paz-Chinchón, F. and Petravick, D. and Pieres, A. and Romer, A. K. and Sanchez, E. and Scarpine, V. and Schubnell, M. and Serrano, S. and Smith, M. and Soares-Santos, M. and Tarle, G. and Thomas, D. and To, C. and Weller, J.},
	urldate = {2026-03-19},
	date = {2022-01-13},
}

@article{di_valentino_reducing_2018,
	title = {Reducing the \$\{H\}\_\{0\}\$ and \$\{{\textbackslash}ensuremath\{{\textbackslash}sigma\}\}\_\{8\}\$ tensions with dark matter-neutrino interactions},
	volume = {97},
	url = {https://link.aps.org/doi/10.1103/PhysRevD.97.043513},
	doi = {10.1103/PhysRevD.97.043513},
	pages = {043513},
	number = {4},
	journaltitle = {Physical Review D},
	shortjournal = {Phys. Rev. D},
	publisher = {American Physical Society},
	author = {Di Valentino, Eleonora and Bœhm, Céline and Hivon, Eric and Bouchet, François R.},
	urldate = {2026-03-19},
	date = {2018-02-13},
}

@article{dark_energy_survey_collaboration_dark_2016,
	title = {The Dark Energy Survey: more than dark energy – an overview},
	volume = {460},
	issn = {0035-8711},
	url = {https://doi.org/10.1093/mnras/stw641},
	doi = {10.1093/mnras/stw641},
	shorttitle = {The Dark Energy Survey},
	pages = {1270--1299},
	number = {2},
	journaltitle = {Monthly Notices of the Royal Astronomical Society},
	shortjournal = {Mon Not R Astron Soc},
	author = {{Dark Energy Survey Collaboration:} and Abbott, T. and Abdalla, F. B. and Aleksić, J. and Allam, S. and Amara, A. and Bacon, D. and Balbinot, E. and Banerji, M. and Bechtol, K. and Benoit-Lévy, A. and Bernstein, G. M. and Bertin, E. and Blazek, J. and Bonnett, C. and Bridle, S. and Brooks, D. and Brunner, R. J. and Buckley-Geer, E. and Burke, D. L. and Caminha, G. B. and Capozzi, D. and Carlsen, J. and Carnero-Rosell, A. and Carollo, M. and Carrasco-Kind, M. and Carretero, J. and Castander, F. J. and Clerkin, L. and Collett, T. and Conselice, C. and Crocce, M. and Cunha, C. E. and D'Andrea, C. B. and da Costa, L. N. and Davis, T. M. and Desai, S. and Diehl, H. T. and Dietrich, J. P. and Dodelson, S. and Doel, P. and Drlica-Wagner, A. and Estrada, J. and Etherington, J. and Evrard, A. E. and Fabbri, J. and Finley, D. A. and Flaugher, B. and Foley, R. J. and Fosalba, P. and Frieman, J. and García-Bellido, J. and Gaztanaga, E. and Gerdes, D. W. and Giannantonio, T. and Goldstein, D. A. and Gruen, D. and Gruendl, R. A. and Guarnieri, P. and Gutierrez, G. and Hartley, W. and Honscheid, K. and Jain, B. and James, D. J. and Jeltema, T. and Jouvel, S. and Kessler, R. and King, A. and Kirk, D. and Kron, R. and Kuehn, K. and Kuropatkin, N. and Lahav, O. and Li, T. S. and Lima, M. and Lin, H. and Maia, M. A. G. and Makler, M. and Manera, M. and Maraston, C. and Marshall, J. L. and Martini, P. and McMahon, R. G. and Melchior, P. and Merson, A. and Miller, C. J. and Miquel, R. and Mohr, J. J. and Morice-Atkinson, X. and Naidoo, K. and Neilsen, E. and Nichol, R. C. and Nord, B. and Ogando, R. and Ostrovski, F. and Palmese, A. and Papadopoulos, A. and Peiris, H. V. and Peoples, J. and Percival, W. J. and Plazas, A. A. and Reed, S. L. and Refregier, A. and Romer, A. K. and Roodman, A. and Ross, A. and Rozo, E. and Rykoff, E. S. and Sadeh, I. and Sako, M. and Sánchez, C. and Sanchez, E. and Santiago, B. and Scarpine, V. and Schubnell, M. and Sevilla-Noarbe, I. and Sheldon, E. and Smith, M. and Smith, R. C. and Soares-Santos, M. and Sobreira, F. and Soumagnac, M. and Suchyta, E. and Sullivan, M. and Swanson, M. and Tarle, G. and Thaler, J. and Thomas, D. and Thomas, R. C. and Tucker, D. and Vieira, J. D. and Vikram, V. and Walker, A. R. and Wechsler, R. H. and Weller, J. and Wester, W. and Whiteway, L. and Wilcox, H. and Yanny, B. and Zhang, Y. and Zuntz, J.},
	urldate = {2026-03-19},
	date = {2016-08-01},
}

@misc{yao_can_2024,
	title = {Can interacting dark energy with dynamical coupling resolve the Hubble tension},
	url = {http://arxiv.org/abs/2207.05955},
	doi = {10.48550/arXiv.2207.05955},
	number = {{arXiv}:2207.05955},
	publisher = {{arXiv}},
	author = {Yao, Yan-Hong and Meng, Xin-He},
	urldate = {2026-03-19},
	date = {2024-05-30},
	eprinttype = {arxiv},
	eprint = {2207.05955 [astro-ph]},
}

@misc{kamionkowski_hubble_2022,
	title = {The Hubble Tension and Early Dark Energy},
	url = {http://arxiv.org/abs/2211.04492},
	doi = {10.48550/arXiv.2211.04492},
	number = {{arXiv}:2211.04492},
	publisher = {{arXiv}},
	author = {Kamionkowski, Marc and Riess, Adam G.},
	urldate = {2026-03-19},
	date = {2022-11-08},
	eprinttype = {arxiv},
	eprint = {2211.04492 [astro-ph]},
}

@article{liu_cosmological-model-independent_2023,
	title = {Cosmological-model-independent Determination of Hubble Constant from Fast Radio Bursts and Hubble Parameter Measurements},
	volume = {946},
	issn = {2041-8205},
	url = {https://doi.org/10.3847/2041-8213/acc650},
	doi = {10.3847/2041-8213/acc650},
	pages = {L49},
	number = {2},
	journaltitle = {The Astrophysical Journal Letters},
	shortjournal = {{ApJL}},
	publisher = {The American Astronomical Society},
	author = {Liu, Yang and Yu, Hongwei and Wu, Puxun},
	urldate = {2026-03-19},
	date = {2023-04},
	langid = {english},
}

@article{perivolaropoulos_challenges_2022,
	title = {Challenges for Λ {CDM}: An update},
	volume = {95},
	issn = {13876473},
	url = {https://linkinghub.elsevier.com/retrieve/pii/S1387647322000185},
	doi = {10.1016/j.newar.2022.101659},
	shorttitle = {Challenges for Λ {CDM}},
	pages = {101659},
	journaltitle = {New Astronomy Reviews},
	shortjournal = {New Astronomy Reviews},
	author = {Perivolaropoulos, L. and Skara, F.},
	urldate = {2026-03-19},
	date = {2022-12},
	langid = {english},
}

@article{di_valentino_snowmass2021_2021-1,
	title = {Snowmass2021 - Letter of interest cosmology intertwined {II}: The hubble constant tension},
	volume = {131},
	issn = {09276505},
	url = {https://linkinghub.elsevier.com/retrieve/pii/S0927650521000499},
	doi = {10.1016/j.astropartphys.2021.102605},
	shorttitle = {Snowmass2021 - Letter of interest cosmology intertwined {II}},
	pages = {102605},
	journaltitle = {Astroparticle Physics},
	shortjournal = {Astroparticle Physics},
	author = {Di Valentino, Eleonora and Anchordoqui, Luis A. and Akarsu, Özgür and Ali-Haimoud, Yacine and Amendola, Luca and Arendse, Nikki and Asgari, Marika and Ballardini, Mario and Basilakos, Spyros and Battistelli, Elia and Benetti, Micol and Birrer, Simon and Bouchet, François R. and Bruni, Marco and Calabrese, Erminia and Camarena, David and Capozziello, Salvatore and Chen, Angela and Chluba, Jens and Chudaykin, Anton and Colgáin, Eoin Ó and Cyr-Racine, Francis-Yan and De Bernardis, Paolo and De Cruz Pérez, Javier and Delabrouille, Jacques and Dunkley, Jo and Escamilla-Rivera, Celia and Ferté, Agnès and Finelli, Fabio and Freedman, Wendy and Frusciante, Noemi and Giusarma, Elena and Gómez-Valent, Adrià and Guy, Julien and Handley, Will and Harrison, Ian and Hart, Luke and Heavens, Alan and Hildebrandt, Hendrik and Holz, Daniel and Huterer, Dragan and Ivanov, Mikhail M. and Joudaki, Shahab and Kamionkowski, Marc and Karwal, Tanvi and Knox, Lloyd and Kumar, Suresh and Lamagna, Luca and Lesgourgues, Julien and Lucca, Matteo and Marra, Valerio and Masi, Silvia and Matarrese, Sabino and Mazumdar, Arindam and Melchiorri, Alessandro and Mena, Olga and Mersini-Houghton, Laura and Miranda, Vivian and Moreno-Pulido, Cristian and Mota, David F. and Muir, Jessica and Mukherjee, Ankan and Niedermann, Florian and Notari, Alessio and Nunes, Rafael C. and Pace, Francesco and Paliathanasis, Andronikos and Palmese, Antonella and Pan, Supriya and Paoletti, Daniela and Pettorino, Valeria and Piacentini, Francesco and Poulin, Vivian and Raveri, Marco and Riess, Adam G. and Salzano, Vincenzo and Saridakis, Emmanuel N. and Sen, Anjan A. and Shafieloo, Arman and Shajib, Anowar J. and Silk, Joseph and Silvestri, Alessandra and Sloth, Martin S. and Smith, Tristan L. and Solà Peracaula, Joan and Van De Bruck, Carsten and Verde, Licia and Visinelli, Luca and Wandelt, Benjamin D. and Wang, Deng and Wang, Jian-Min and Yadav, Anil K. and Yang, Weiqiang},
	urldate = {2026-03-19},
	date = {2021-09},
	langid = {english},
}

@article{brout_pantheon_2022,
	title = {The Pantheon+ Analysis: Cosmological Constraints},
	volume = {938},
	issn = {0004-637X},
	url = {https://doi.org/10.3847/1538-4357/ac8e04},
	doi = {10.3847/1538-4357/ac8e04},
	shorttitle = {The Pantheon+ Analysis},
	pages = {110},
	number = {2},
	journaltitle = {The Astrophysical Journal},
	shortjournal = {{ApJ}},
	publisher = {The American Astronomical Society},
	author = {Brout, Dillon and Scolnic, Dan and Popovic, Brodie and Riess, Adam G. and Carr, Anthony and Zuntz, Joe and Kessler, Rick and Davis, Tamara M. and Hinton, Samuel and Jones, David and Kenworthy, W. D’Arcy and Peterson, Erik R. and Said, Khaled and Taylor, Georgie and Ali, Noor and Armstrong, Patrick and Charvu, Pranav and Dwomoh, Arianna and Meldorf, Cole and Palmese, Antonella and Qu, Helen and Rose, Benjamin M. and Sanchez, Bruno and Stubbs, Christopher W. and Vincenzi, Maria and Wood, Charlotte M. and Brown, Peter J. and Chen, Rebecca and Chambers, Ken and Coulter, David A. and Dai, Mi and Dimitriadis, Georgios and Filippenko, Alexei V. and Foley, Ryan J. and Jha, Saurabh W. and Kelsey, Lisa and Kirshner, Robert P. and Möller, Anais and Muir, Jessie and Nadathur, Seshadri and Pan, Yen-Chen and Rest, Armin and Rojas-Bravo, Cesar and Sako, Masao and Siebert, Matthew R. and Smith, Mat and Stahl, Benjamin E. and Wiseman, Phil},
	urldate = {2026-03-19},
	date = {2022-10},
	langid = {english},
}

@article{di_valentino_challenges_2022,
	title = {Challenges of the Standard Cosmological Model},
	volume = {8},
	issn = {2218-1997},
	url = {https://www.mdpi.com/2218-1997/8/8/399},
	doi = {10.3390/universe8080399},
	pages = {399},
	number = {8},
	journaltitle = {Universe},
	shortjournal = {Universe},
	author = {Di Valentino, Eleonora},
	urldate = {2026-03-19},
	date = {2022-07-29},
	langid = {english},
}

@article{carroll_cosmological_2001,
	title = {The Cosmological Constant},
	volume = {4},
	issn = {1433-8351},
	url = {https://doi.org/10.12942/lrr-2001-1},
	doi = {10.12942/lrr-2001-1},
	pages = {1},
	number = {1},
	journaltitle = {Living Reviews in Relativity},
	shortjournal = {Living Rev. Relativ.},
	author = {Carroll, Sean M.},
	urldate = {2026-03-19},
	date = {2001-02-07},
	langid = {english},
}

@article{hu_models_2007,
	title = {Models of \$f(R)\$ cosmic acceleration that evade solar system tests},
	volume = {76},
	rights = {http://link.aps.org/licenses/aps-default-license},
	issn = {1550-7998, 1550-2368},
	url = {https://link.aps.org/doi/10.1103/PhysRevD.76.064004},
	doi = {10.1103/PhysRevD.76.064004},
	pages = {064004},
	number = {6},
	journaltitle = {Physical Review D},
	shortjournal = {Phys. Rev. D},
	author = {Hu, Wayne and Sawicki, Ignacy},
	urldate = {2025-05-19},
	date = {2007-09-10},
	langid = {english},
}

@article{bardeen_statistics_1986,
	title = {The Statistics of Peaks of Gaussian Random Fields},
	volume = {304},
	issn = {0004-637X},
	url = {https://ui.adsabs.harvard.edu/abs/1986ApJ...304...15B},
	doi = {10.1086/164143},
	pages = {15},
	journaltitle = {The Astrophysical Journal},
	publisher = {{IOP}},
	author = {Bardeen, J. M. and Bond, J. R. and Kaiser, N. and Szalay, A. S.},
	urldate = {2026-03-18},
	date = {1986-05-01},
	note = {{ADS} Bibcode: 1986ApJ...304...15B},
}

@article{york_sloan_2000,
	title = {The Sloan Digital Sky Survey: Technical Summary},
	volume = {120},
	issn = {1538-3881},
	url = {https://iopscience.iop.org/article/10.1086/301513},
	doi = {10.1086/301513},
	shorttitle = {The Sloan Digital Sky Survey},
	pages = {1579},
	number = {3},
	journaltitle = {The Astronomical Journal},
	shortjournal = {{AJ}},
	publisher = {{IOP} Publishing},
	author = {York, Donald G. and Adelman, J. and John E. Anderson, Jr and Anderson, Scott F. and Annis, James and Bahcall, Neta A. and Bakken, J. A. and Barkhouser, Robert and Bastian, Steven and Berman, Eileen and Boroski, William N. and Bracker, Steve and Briegel, Charlie and Briggs, John W. and Brinkmann, J. and Brunner, Robert and Burles, Scott and Carey, Larry and Carr, Michael A. and Castander, Francisco J. and Chen, Bing and Colestock, Patrick L. and Connolly, A. J. and Crocker, J. H. and Csabai, István and Czarapata, Paul C. and Davis, John Eric and Doi, Mamoru and Dombeck, Tom and Eisenstein, Daniel and Ellman, Nancy and Elms, Brian R. and Evans, Michael L. and Fan, Xiaohui and Federwitz, Glenn R. and Fiscelli, Larry and Friedman, Scott and Frieman, Joshua A. and Fukugita, Masataka and Gillespie, Bruce and Gunn, James E. and Gurbani, Vijay K. and Haas, Ernst de and Haldeman, Merle and Harris, Frederick H. and Hayes, J. and Heckman, Timothy M. and Hennessy, G. S. and Hindsley, Robert B. and Holm, Scott and Holmgren, Donald J. and Huang, Chi-hao and Hull, Charles and Husby, Don and Ichikawa, Shin-Ichi and Ichikawa, Takashi and Ivezić, Željko and Kent, Stephen and Kim, Rita S. J. and Kinney, E. and Klaene, Mark and Kleinman, A. N. and Kleinman, S. and Knapp, G. R. and Korienek, John and Kron, Richard G. and Kunszt, Peter Z. and Lamb, D. Q. and Lee, B. and Leger, R. French and Limmongkol, Siriluk and Lindenmeyer, Carl and Long, Daniel C. and Loomis, Craig and Loveday, Jon and Lucinio, Rich and Lupton, Robert H. and {MacKinnon}, Bryan and Mannery, Edward J. and Mantsch, P. M. and Margon, Bruce and {McGehee}, Peregrine and {McKay}, Timothy A. and Meiksin, Avery and Merelli, Aronne and Monet, David G. and Munn, Jeffrey A. and Narayanan, Vijay K. and Nash, Thomas and Neilsen, Eric and Neswold, Rich and Newberg, Heidi Jo and Nichol, R. C. and Nicinski, Tom and Nonino, Mario and Okada, Norio and Okamura, Sadanori and Ostriker, Jeremiah P. and Owen, Russell and Pauls, A. George and Peoples, John and Peterson, R. L. and Petravick, Donald and Pier, Jeffrey R. and Pope, Adrian and Pordes, Ruth and Prosapio, Angela and Rechenmacher, Ron and Quinn, Thomas R. and Richards, Gordon T. and Richmond, Michael W. and Rivetta, Claudio H. and Rockosi, Constance M. and Ruthmansdorfer, Kurt and Sandford, Dale and Schlegel, David J. and Schneider, Donald P. and Sekiguchi, Maki and Sergey, Gary and Shimasaku, Kazuhiro and Siegmund, Walter A. and Smee, Stephen and Smith, J. Allyn and Snedden, S. and Stone, R. and Stoughton, Chris and Strauss, Michael A. and Stubbs, Christopher and {SubbaRao}, Mark and Szalay, Alexander S. and Szapudi, Istvan and Szokoly, Gyula P. and Thakar, Anirudda R. and Tremonti, Christy and Tucker, Douglas L. and Uomoto, Alan and Berk, Dan Vanden and Vogeley, Michael S. and Waddell, Patrick and Wang, Shu-i and Watanabe, Masaru and Weinberg, David H. and Yanny, Brian and Yasuda, Naoki},
	urldate = {2026-03-18},
	date = {2000-09-01},
	langid = {english},
}

@article{colless_2df_2001,
	title = {The 2dF Galaxy Redshift Survey: spectra and redshifts},
	volume = {328},
	issn = {0035-8711},
	url = {https://doi.org/10.1046/j.1365-8711.2001.04902.x},
	doi = {10.1046/j.1365-8711.2001.04902.x},
	shorttitle = {The 2dF Galaxy Redshift Survey},
	pages = {1039--1063},
	number = {4},
	journaltitle = {Monthly Notices of the Royal Astronomical Society},
	shortjournal = {Mon Not R Astron Soc},
	author = {Colless, Matthew and Dalton, Gavin and Maddox, Steve and Sutherland, Will and Norberg, Peder and Cole, Shaun and Bland-Hawthorn, Joss and Bridges, Terry and Cannon, Russell and Collins, Chris and Couch, Warrick and Cross, Nicholas and Deeley, Kathryn and de Propris, Roberto and Driver, Simon P. and Efstathiou, George and Ellis, Richard S. and Frenk, Carlos S. and Glazebrook, Karl and Jackson, Carole and Lahav, Ofer and Lewis, Ian and Lumsden, Stuart and Madgwick, Darren and Peacock, John A. and Peterson, Bruce A. and Price, Ian and Seaborne, Mark and Taylor, Keith},
	urldate = {2026-03-18},
	date = {2001-12-21},
}

@article{hu_parametrized_2008,
	title = {Parametrized post-Friedmann signatures of acceleration in the {CMB}},
	volume = {77},
	url = {https://link.aps.org/doi/10.1103/PhysRevD.77.103524},
	doi = {10.1103/PhysRevD.77.103524},
	pages = {103524},
	number = {10},
	journaltitle = {Physical Review D},
	shortjournal = {Phys. Rev. D},
	publisher = {American Physical Society},
	author = {Hu, Wayne},
	urldate = {2026-03-18},
	date = {2008-05-27},
}

@article{guth_inflationary_1981,
	title = {Inflationary universe: A possible solution to the horizon and flatness problems},
	volume = {23},
	url = {https://link.aps.org/doi/10.1103/PhysRevD.23.347},
	doi = {10.1103/PhysRevD.23.347},
	shorttitle = {Inflationary universe},
	pages = {347--356},
	number = {2},
	journaltitle = {Physical Review D},
	shortjournal = {Phys. Rev. D},
	publisher = {American Physical Society},
	author = {Guth, Alan H.},
	urldate = {2026-03-18},
	date = {1981-01-15},
}

@misc{achucarro_inflation_2022,
	title = {Inflation: Theory and Observations},
	url = {http://arxiv.org/abs/2203.08128},
	doi = {10.48550/arXiv.2203.08128},
	shorttitle = {Inflation},
	number = {{arXiv}:2203.08128},
	publisher = {{arXiv}},
	author = {Achúcarro, Ana and Biagetti, Matteo and Braglia, Matteo and Cabass, Giovanni and Caldwell, Robert and Castorina, Emanuele and Chen, Xingang and Coulton, William and Flauger, Raphael and Fumagalli, Jacopo and Ivanov, Mikhail M. and Lee, Hayden and Maleknejad, Azadeh and Meerburg, P. Daniel and Dizgah, Azadeh Moradinezhad and Palma, Gonzalo A. and Pimentel, Guilherme L. and Renaux-Petel, Sébastien and Wallisch, Benjamin and Wandelt, Benjamin D. and Witkowski, Lukas T. and Wu, W. L. Kimmy},
	urldate = {2026-03-18},
	date = {2022-09-29},
	eprinttype = {arxiv},
	eprint = {2203.08128 [astro-ph]},
}

@article{linder_exploring_2003,
	title = {Exploring the Expansion History of the Universe},
	volume = {90},
	url = {https://link.aps.org/doi/10.1103/PhysRevLett.90.091301},
	doi = {10.1103/PhysRevLett.90.091301},
	pages = {091301},
	number = {9},
	journaltitle = {Physical Review Letters},
	shortjournal = {Phys. Rev. Lett.},
	publisher = {American Physical Society},
	author = {Linder, Eric V.},
	urldate = {2026-03-18},
	date = {2003-03-03},
}

@misc{laureijs_euclid_2011,
	title = {Euclid Definition Study Report},
	url = {http://arxiv.org/abs/1110.3193},
	doi = {10.48550/arXiv.1110.3193},
	number = {{arXiv}:1110.3193},
	publisher = {{arXiv}},
	author = {Laureijs, R. and Amiaux, J. and Arduini, S. and Auguères, J.-L. and Brinchmann, J. and Cole, R. and Cropper, M. and Dabin, C. and Duvet, L. and Ealet, A. and Garilli, B. and Gondoin, P. and Guzzo, L. and Hoar, J. and Hoekstra, H. and Holmes, R. and Kitching, T. and Maciaszek, T. and Mellier, Y. and Pasian, F. and Percival, W. and Rhodes, J. and Criado, G. Saavedra and Sauvage, M. and Scaramella, R. and Valenziano, L. and Warren, S. and Bender, R. and Castander, F. and Cimatti, A. and Fèvre, O. Le and Kurki-Suonio, H. and Levi, M. and Lilje, P. and Meylan, G. and Nichol, R. and Pedersen, K. and Popa, V. and Lopez, R. Rebolo and Rix, H.-W. and Rottgering, H. and Zeilinger, W. and Grupp, F. and Hudelot, P. and Massey, R. and Meneghetti, M. and Miller, L. and Paltani, S. and Paulin-Henriksson, S. and Pires, S. and Saxton, C. and Schrabback, T. and Seidel, G. and Walsh, J. and Aghanim, N. and Amendola, L. and Bartlett, J. and Baccigalupi, C. and Beaulieu, J.-P. and Benabed, K. and Cuby, J.-G. and Elbaz, D. and Fosalba, P. and Gavazzi, G. and Helmi, A. and Hook, I. and Irwin, M. and Kneib, J.-P. and Kunz, M. and Mannucci, F. and Moscardini, L. and Tao, C. and Teyssier, R. and Weller, J. and Zamorani, G. and Osorio, M. R. Zapatero and Boulade, O. and Foumond, J. J. and Giorgio, A. Di and Guttridge, P. and James, A. and Kemp, M. and Martignac, J. and Spencer, A. and Walton, D. and Blümchen, T. and Bonoli, C. and Bortoletto, F. and Cerna, C. and Corcione, L. and Fabron, C. and Jahnke, K. and Ligori, S. and Madrid, F. and Martin, L. and Morgante, G. and Pamplona, T. and Prieto, E. and Riva, M. and Toledo, R. and Trifoglio, M. and Zerbi, F. and Abdalla, F. and Douspis, M. and Grenet, C. and Borgani, S. and Bouwens, R. and Courbin, F. and Delouis, J.-M. and Dubath, P. and Fontana, A. and Frailis, M. and Grazian, A. and Koppenhöfer, J. and Mansutti, O. and Melchior, M. and Mignoli, M. and Mohr, J. and Neissner, C. and Noddle, K. and Poncet, M. and Scodeggio, M. and Serrano, S. and Shane, N. and Starck, J.-L. and Surace, C. and Taylor, A. and Verdoes-Kleijn, G. and Vuerli, C. and Williams, O. R. and Zacchei, A. and Altieri, B. and Sanz, I. Escudero and Kohley, R. and Oosterbroek, T. and Astier, P. and Bacon, D. and Bardelli, S. and Baugh, C. and Bellagamba, F. and Benoist, C. and Bianchi, D. and Biviano, A. and Branchini, E. and Carbone, C. and Cardone, V. and Clements, D. and Colombi, S. and Conselice, C. and Cresci, G. and Deacon, N. and Dunlop, J. and Fedeli, C. and Fontanot, F. and Franzetti, P. and Giocoli, C. and Garcia-Bellido, J. and Gow, J. and Heavens, A. and Hewett, P. and Heymans, C. and Holland, A. and Huang, Z. and Ilbert, O. and Joachimi, B. and Jennins, E. and Kerins, E. and Kiessling, A. and Kirk, D. and Kotak, R. and Krause, O. and Lahav, O. and Leeuwen, F. van and Lesgourgues, J. and Lombardi, M. and Magliocchetti, M. and Maguire, K. and Majerotto, E. and Maoli, R. and Marulli, F. and Maurogordato, S. and {McCracken}, H. and {McLure}, R. and Melchiorri, A. and Merson, A. and Moresco, M. and Nonino, M. and Norberg, P. and Peacock, J. and Pello, R. and Penny, M. and Pettorino, V. and Porto, C. Di and Pozzetti, L. and Quercellini, C. and Radovich, M. and Rassat, A. and Roche, N. and Ronayette, S. and Rossetti, E. and Sartoris, B. and Schneider, P. and Semboloni, E. and Serjeant, S. and Simpson, F. and Skordis, C. and Smadja, G. and Smartt, S. and Spano, P. and Spiro, S. and Sullivan, M. and Tilquin, A. and Trotta, R. and Verde, L. and Wang, Y. and Williger, G. and Zhao, G. and Zoubian, J. and Zucca, E.},
	urldate = {2026-03-18},
	date = {2011-10-14},
	eprinttype = {arxiv},
	eprint = {1110.3193 [astro-ph]},
}

@article{fang_crossing_2008,
	title = {Crossing the phantom divide with parametrized post-Friedmann dark energy},
	volume = {78},
	url = {https://link.aps.org/doi/10.1103/PhysRevD.78.087303},
	doi = {10.1103/PhysRevD.78.087303},
	pages = {087303},
	number = {8},
	journaltitle = {Physical Review D},
	shortjournal = {Phys. Rev. D},
	publisher = {American Physical Society},
	author = {Fang, Wenjuan and Hu, Wayne and Lewis, Antony},
	urldate = {2026-03-18},
	date = {2008-10-24},
}

@article{chevallier_accelerating_2001,
	title = {Accelerating Universes with Scaling Dark Matter},
	volume = {10},
	issn = {0218-2718, 1793-6594},
	url = {http://arxiv.org/abs/gr-qc/0009008},
	doi = {10.1142/S0218271801000822},
	pages = {213--223},
	number = {2},
	journaltitle = {International Journal of Modern Physics D},
	shortjournal = {Int. J. Mod. Phys. D},
	author = {Chevallier, M. and Polarski, D.},
	urldate = {2026-03-18},
	date = {2001-04},
	eprinttype = {arxiv},
	eprint = {gr-qc/0009008},
}

@book{dodelson_modern_2003,
	title = {Modern cosmology},
	isbn = {978-0-12-219141-1},
	url = {http://archive.org/details/moderncosmology0000dode},
	pagetotal = {474},
	publisher = {San Diego, Calif. : Academic Press},
	author = {Dodelson, Scott},
	editora = {{Internet Archive}},
	editoratype = {collaborator},
	urldate = {2026-03-17},
	date = {2003},
}

@article{weinberg_observational_2013,
	title = {Observational probes of cosmic acceleration},
	volume = {530},
	issn = {0370-1573},
	url = {https://www.sciencedirect.com/science/article/pii/S0370157313001592},
	doi = {10.1016/j.physrep.2013.05.001},
	series = {Observational Probes of Cosmic Acceleration},
	pages = {87--255},
	number = {2},
	journaltitle = {Physics Reports},
	shortjournal = {Physics Reports},
	author = {Weinberg, David H. and Mortonson, Michael J. and Eisenstein, Daniel J. and Hirata, Christopher and Riess, Adam G. and Rozo, Eduardo},
	urldate = {2026-03-17},
	date = {2013-09-10},
}

@article{lovelock_four-dimensionality_1972,
	title = {The Four-Dimensionality of Space and the Einstein Tensor},
	volume = {13},
	issn = {0022-2488, 1089-7658},
	url = {https://pubs.aip.org/jmp/article/13/6/874/223527/The-Four-Dimensionality-of-Space-and-the-Einstein},
	doi = {10.1063/1.1666069},
	pages = {874--876},
	number = {6},
	journaltitle = {Journal of Mathematical Physics},
	author = {Lovelock, David},
	urldate = {2026-03-17},
	date = {1972-06-01},
	langid = {english},
}

@article{kuijken_gravitational_2015,
	title = {Gravitational lensing analysis of the Kilo-Degree Survey},
	volume = {454},
	issn = {0035-8711},
	url = {https://doi.org/10.1093/mnras/stv2140},
	doi = {10.1093/mnras/stv2140},
	pages = {3500--3532},
	number = {4},
	journaltitle = {Monthly Notices of the Royal Astronomical Society},
	shortjournal = {Mon Not R Astron Soc},
	author = {Kuijken, Konrad and Heymans, Catherine and Hildebrandt, Hendrik and Nakajima, Reiko and Erben, Thomas and de Jong, Jelte T. A. and Viola, Massimo and Choi, Ami and Hoekstra, Henk and Miller, Lance and van Uitert, Edo and Amon, Alexandra and Blake, Chris and Brouwer, Margot and Buddendiek, Axel and Conti, Ian Fenech and Eriksen, Martin and Grado, Aniello and Harnois-Déraps, Joachim and Helmich, Ewout and Herbonnet, Ricardo and Irisarri, Nancy and Kitching, Thomas and Klaes, Dominik and La Barbera, Francesco and Napolitano, Nicola and Radovich, Mario and Schneider, Peter and Sifón, Cristóbal and Sikkema, Gert and Simon, Patrick and Tudorica, Alexandru and Valentijn, Edwin and Verdoes Kleijn, Gijs and van Waerbeke, Ludovic},
	urldate = {2026-03-17},
	date = {2015-12-21},
}

@article{weinberg_cosmological_1989,
	title = {The cosmological constant problem},
	volume = {61},
	url = {https://link.aps.org/doi/10.1103/RevModPhys.61.1},
	doi = {10.1103/RevModPhys.61.1},
	pages = {1--23},
	number = {1},
	journaltitle = {Reviews of Modern Physics},
	shortjournal = {Rev. Mod. Phys.},
	publisher = {American Physical Society},
	author = {Weinberg, Steven},
	urldate = {2026-03-17},
	date = {1989-01-01},
}

@article{aghanim_planck_2020,
	title = {Planck 2018 results - I. Overview and the cosmological legacy of Planck},
	volume = {641},
	rights = {© {ESO} 2020},
	issn = {0004-6361, 1432-0746},
	url = {https://www.aanda.org/articles/aa/abs/2020/09/aa33880-18/aa33880-18.html},
	doi = {10.1051/0004-6361/201833880},
	pages = {A1},
	journaltitle = {Astronomy \& Astrophysics},
	shortjournal = {A\&A},
	publisher = {{EDP} Sciences},
	author = {Aghanim, N. and Akrami, Y. and Arroja, F. and Ashdown, M. and Aumont, J. and Baccigalupi, C. and Ballardini, M. and Banday, A. J. and Barreiro, R. B. and Bartolo, N. and Basak, S. and Battye, R. and Benabed, K. and Bernard, J.-P. and Bersanelli, M. and Bielewicz, P. and Bock, J. J. and Bond, J. R. and Borrill, J. and Bouchet, F. R. and Boulanger, F. and Bucher, M. and Burigana, C. and Butler, R. C. and Calabrese, E. and Cardoso, J.-F. and Carron, J. and Casaponsa, B. and Challinor, A. and Chiang, H. C. and Colombo, L. P. L. and Combet, C. and Contreras, D. and Crill, B. P. and Cuttaia, F. and Bernardis, P. de and Zotti, G. de and Delabrouille, J. and Delouis, J.-M. and Désert, F.-X. and Valentino, E. Di and Dickinson, C. and Diego, J. M. and Donzelli, S. and Doré, O. and Douspis, M. and Ducout, A. and Dupac, X. and Efstathiou, G. and Elsner, F. and Enßlin, T. A. and Eriksen, H. K. and Falgarone, E. and Fantaye, Y. and Fergusson, J. and Fernandez-Cobos, R. and Finelli, F. and Forastieri, F. and Frailis, M. and Franceschi, E. and Frolov, A. and Galeotta, S. and Galli, S. and Ganga, K. and Génova-Santos, R. T. and Gerbino, M. and Ghosh, T. and González-Nuevo, J. and Górski, K. M. and Gratton, S. and Gruppuso, A. and Gudmundsson, J. E. and Hamann, J. and Handley, W. and Hansen, F. K. and Helou, G. and Herranz, D. and Hildebrandt, S. R. and Hivon, E. and Huang, Z. and Jaffe, A. H. and Jones, W. C. and Karakci, A. and Keihänen, E. and Keskitalo, R. and Kiiveri, K. and Kim, J. and Kisner, T. S. and Knox, L. and Krachmalnicoff, N. and Kunz, M. and Kurki-Suonio, H. and Lagache, G. and Lamarre, J.-M. and Langer, M. and Lasenby, A. and Lattanzi, M. and Lawrence, C. R. and Jeune, M. Le and Leahy, J. P. and Lesgourgues, J. and Levrier, F. and Lewis, A. and Liguori, M. and Lilje, P. B. and Lilley, M. and Lindholm, V. and López-Caniego, M. and Lubin, P. M. and Ma, Y.-Z. and Macías-Pérez, J. F. and Maggio, G. and Maino, D. and Mandolesi, N. and Mangilli, A. and Marcos-Caballero, A. and Maris, M. and Martin, P. G. and Martinelli, M. and Martínez-González, E. and Matarrese, S. and Mauri, N. and {McEwen}, J. D. and Meerburg, P. D. and Meinhold, P. R. and Melchiorri, A. and Mennella, A. and Migliaccio, M. and Millea, M. and Mitra, S. and Miville-Deschênes, M.-A. and Molinari, D. and Moneti, A. and Montier, L. and Morgante, G. and Moss, A. and Mottet, S. and Münchmeyer, M. and Natoli, P. and Nørgaard-Nielsen, H. U. and Oxborrow, C. A. and Pagano, L. and Paoletti, D. and Partridge, B. and Patanchon, G. and Pearson, T. J. and Peel, M. and Peiris, H. V. and Perrotta, F. and Pettorino, V. and Piacentini, F. and Polastri, L. and Polenta, G. and Puget, J.-L. and Rachen, J. P. and Reinecke, M. and Remazeilles, M. and Renault, C. and Renzi, A. and Rocha, G. and Rosset, C. and Roudier, G. and Rubiño-Martín, J. A. and Ruiz-Granados, B. and Salvati, L. and Sandri, M. and Savelainen, M. and Scott, D. and Shellard, E. P. S. and Shiraishi, M. and Sirignano, C. and Sirri, G. and Spencer, L. D. and Sunyaev, R. and Suur-Uski, A.-S. and Tauber, J. A. and Tavagnacco, D. and Tenti, M. and Terenzi, L. and Toffolatti, L. and Tomasi, M. and Trombetti, T. and Valiviita, J. and Tent, B. Van and Vibert, L. and Vielva, P. and Villa, F. and Vittorio, N. and Wandelt, B. D. and Wehus, I. K. and White, M. and White, S. D. M. and Zacchei, A. and Zonca, A.},
	urldate = {2026-03-08},
	date = {2020-09-01},
	langid = {english},
}

@article{planck_collaboration_planck_2020,
	title = {\textit{Planck} 2018 results: {VI}. Cosmological parameters},
	volume = {641},
	rights = {https://www.edpsciences.org/en/authors/copyright-and-licensing},
	issn = {0004-6361, 1432-0746},
	url = {https://www.aanda.org/10.1051/0004-6361/201833910},
	doi = {10.1051/0004-6361/201833910},
	shorttitle = {\textit{Planck} 2018 results},
	pages = {A6},
	journaltitle = {Astronomy \& Astrophysics},
	shortjournal = {A\&A},
	author = {{Planck Collaboration} and Aghanim, N. and Akrami, Y. and Ashdown, M. and Aumont, J. and Baccigalupi, C. and Ballardini, M. and Banday, A. J. and Barreiro, R. B. and Bartolo, N. and Basak, S. and Battye, R. and Benabed, K. and Bernard, J.-P. and Bersanelli, M. and Bielewicz, P. and Bock, J. J. and Bond, J. R. and Borrill, J. and Bouchet, F. R. and Boulanger, F. and Bucher, M. and Burigana, C. and Butler, R. C. and Calabrese, E. and Cardoso, J.-F. and Carron, J. and Challinor, A. and Chiang, H. C. and Chluba, J. and Colombo, L. P. L. and Combet, C. and Contreras, D. and Crill, B. P. and Cuttaia, F. and De Bernardis, P. and De Zotti, G. and Delabrouille, J. and Delouis, J.-M. and Di Valentino, E. and Diego, J. M. and Doré, O. and Douspis, M. and Ducout, A. and Dupac, X. and Dusini, S. and Efstathiou, G. and Elsner, F. and Enßlin, T. A. and Eriksen, H. K. and Fantaye, Y. and Farhang, M. and Fergusson, J. and Fernandez-Cobos, R. and Finelli, F. and Forastieri, F. and Frailis, M. and Fraisse, A. A. and Franceschi, E. and Frolov, A. and Galeotta, S. and Galli, S. and Ganga, K. and Génova-Santos, R. T. and Gerbino, M. and Ghosh, T. and González-Nuevo, J. and Górski, K. M. and Gratton, S. and Gruppuso, A. and Gudmundsson, J. E. and Hamann, J. and Handley, W. and Hansen, F. K. and Herranz, D. and Hildebrandt, S. R. and Hivon, E. and Huang, Z. and Jaffe, A. H. and Jones, W. C. and Karakci, A. and Keihänen, E. and Keskitalo, R. and Kiiveri, K. and Kim, J. and Kisner, T. S. and Knox, L. and Krachmalnicoff, N. and Kunz, M. and Kurki-Suonio, H. and Lagache, G. and Lamarre, J.-M. and Lasenby, A. and Lattanzi, M. and Lawrence, C. R. and Le Jeune, M. and Lemos, P. and Lesgourgues, J. and Levrier, F. and Lewis, A. and Liguori, M. and Lilje, P. B. and Lilley, M. and Lindholm, V. and López-Caniego, M. and Lubin, P. M. and Ma, Y.-Z. and Macías-Pérez, J. F. and Maggio, G. and Maino, D. and Mandolesi, N. and Mangilli, A. and Marcos-Caballero, A. and Maris, M. and Martin, P. G. and Martinelli, M. and Martínez-González, E. and Matarrese, S. and Mauri, N. and McEwen, J. D. and Meinhold, P. R. and Melchiorri, A. and Mennella, A. and Migliaccio, M. and Millea, M. and Mitra, S. and Miville-Deschênes, M.-A. and Molinari, D. and Montier, L. and Morgante, G. and Moss, A. and Natoli, P. and Nørgaard-Nielsen, H. U. and Pagano, L. and Paoletti, D. and Partridge, B. and Patanchon, G. and Peiris, H. V. and Perrotta, F. and Pettorino, V. and Piacentini, F. and Polastri, L. and Polenta, G. and Puget, J.-L. and Rachen, J. P. and Reinecke, M. and Remazeilles, M. and Renzi, A. and Rocha, G. and Rosset, C. and Roudier, G. and Rubiño-Martín, J. A. and Ruiz-Granados, B. and Salvati, L. and Sandri, M. and Savelainen, M. and Scott, D. and Shellard, E. P. S. and Sirignano, C. and Sirri, G. and Spencer, L. D. and Sunyaev, R. and Suur-Uski, A.-S. and Tauber, J. A. and Tavagnacco, D. and Tenti, M. and Toffolatti, L. and Tomasi, M. and Trombetti, T. and Valenziano, L. and Valiviita, J. and Van Tent, B. and Vibert, L. and Vielva, P. and Villa, F. and Vittorio, N. and Wandelt, B. D. and Wehus, I. K. and White, M. and White, S. D. M. and Zacchei, A. and Zonca, A.},
	urldate = {2026-03-08},
	date = {2020-09},
	langid = {english},
}

@article{akrami_planck_2020,
	title = {Planck 2018 results - X. Constraints on inflation},
	volume = {641},
	rights = {© Planck Collaboration 2020},
	issn = {0004-6361, 1432-0746},
	url = {https://www.aanda.org/articles/aa/abs/2020/09/aa33887-18/aa33887-18.html},
	doi = {10.1051/0004-6361/201833887},
	pages = {A10},
	journaltitle = {Astronomy \& Astrophysics},
	shortjournal = {A\&A},
	publisher = {{EDP} Sciences},
	author = {Akrami, Y. and Arroja, F. and Ashdown, M. and Aumont, J. and Baccigalupi, C. and Ballardini, M. and Banday, A. J. and Barreiro, R. B. and Bartolo, N. and Basak, S. and Benabed, K. and Bernard, J.-P. and Bersanelli, M. and Bielewicz, P. and Bock, J. J. and Bond, J. R. and Borrill, J. and Bouchet, F. R. and Boulanger, F. and Bucher, M. and Burigana, C. and Butler, R. C. and Calabrese, E. and Cardoso, J.-F. and Carron, J. and Challinor, A. and Chiang, H. C. and Colombo, L. P. L. and Combet, C. and Contreras, D. and Crill, B. P. and Cuttaia, F. and Bernardis, P. de and Zotti, G. de and Delabrouille, J. and Delouis, J.-M. and Valentino, E. Di and Diego, J. M. and Donzelli, S. and Doré, O. and Douspis, M. and Ducout, A. and Dupac, X. and Dusini, S. and Efstathiou, G. and Elsner, F. and Enßlin, T. A. and Eriksen, H. K. and Fantaye, Y. and Fergusson, J. and Fernandez-Cobos, R. and Finelli, F. and Forastieri, F. and Frailis, M. and Franceschi, E. and Frolov, A. and Galeotta, S. and Galli, S. and Ganga, K. and Gauthier, C. and Génova-Santos, R. T. and Gerbino, M. and Ghosh, T. and González-Nuevo, J. and Górski, K. M. and Gratton, S. and Gruppuso, A. and Gudmundsson, J. E. and Hamann, J. and Handley, W. and Hansen, F. K. and Herranz, D. and Hivon, E. and Hooper, D. C. and Huang, Z. and Jaffe, A. H. and Jones, W. C. and Keihänen, E. and Keskitalo, R. and Kiiveri, K. and Kim, J. and Kisner, T. S. and Krachmalnicoff, N. and Kunz, M. and Kurki-Suonio, H. and Lagache, G. and Lamarre, J.-M. and Lasenby, A. and Lattanzi, M. and Lawrence, C. R. and Jeune, M. Le and Lesgourgues, J. and Levrier, F. and Lewis, A. and Liguori, M. and Lilje, P. B. and Lindholm, V. and López-Caniego, M. and Lubin, P. M. and Ma, Y.-Z. and Macías-Pérez, J. F. and Maggio, G. and Maino, D. and Mandolesi, N. and Mangilli, A. and Marcos-Caballero, A. and Maris, M. and Martin, P. G. and Martínez-González, E. and Matarrese, S. and Mauri, N. and {McEwen}, J. D. and Meerburg, P. D. and Meinhold, P. R. and Melchiorri, A. and Mennella, A. and Migliaccio, M. and Mitra, S. and Miville-Deschênes, M.-A. and Molinari, D. and Moneti, A. and Montier, L. and Morgante, G. and Moss, A. and Münchmeyer, M. and Natoli, P. and Nørgaard-Nielsen, H. U. and Pagano, L. and Paoletti, D. and Partridge, B. and Patanchon, G. and Peiris, H. V. and Perrotta, F. and Pettorino, V. and Piacentini, F. and Polastri, L. and Polenta, G. and Puget, J.-L. and Rachen, J. P. and Reinecke, M. and Remazeilles, M. and Renzi, A. and Rocha, G. and Rosset, C. and Roudier, G. and Rubiño-Martín, J. A. and Ruiz-Granados, B. and Salvati, L. and Sandri, M. and Savelainen, M. and Scott, D. and Shellard, E. P. S. and Shiraishi, M. and Sirignano, C. and Sirri, G. and Spencer, L. D. and Sunyaev, R. and Suur-Uski, A.-S. and Tauber, J. A. and Tavagnacco, D. and Tenti, M. and Toffolatti, L. and Tomasi, M. and Trombetti, T. and Valiviita, J. and Tent, B. Van and Vielva, P. and Villa, F. and Vittorio, N. and Wandelt, B. D. and Wehus, I. K. and White, S. D. M. and Zacchei, A. and Zibin, J. P. and Zonca, A.},
	urldate = {2026-03-08},
	date = {2020-09-01},
	langid = {english},
}

@article{fidler_general_2015,
	title = {General relativistic corrections to N -body simulations and the Zel’dovich approximation},
	volume = {92},
	rights = {http://link.aps.org/licenses/aps-default-license},
	issn = {1550-7998, 1550-2368},
	url = {https://link.aps.org/doi/10.1103/PhysRevD.92.123517},
	doi = {10.1103/PhysRevD.92.123517},
	pages = {123517},
	number = {12},
	journaltitle = {Physical Review D},
	shortjournal = {Phys. Rev. D},
	author = {Fidler, Christian and Rampf, Cornelius and Tram, Thomas and Crittenden, Robert and Koyama, Kazuya and Wands, David},
	urldate = {2026-02-19},
	date = {2015-12-15},
	langid = {english},
}

@article{bardeen_gauge-invariant_1980,
	title = {Gauge-invariant cosmological perturbations},
	volume = {22},
	rights = {http://link.aps.org/licenses/aps-default-license},
	issn = {0556-2821},
	url = {https://link.aps.org/doi/10.1103/PhysRevD.22.1882},
	doi = {10.1103/PhysRevD.22.1882},
	pages = {1882--1905},
	number = {8},
	journaltitle = {Physical Review D},
	shortjournal = {Phys. Rev. D},
	author = {Bardeen, James M.},
	urldate = {2026-02-19},
	date = {1980-10-15},
	langid = {english},
}

@article{lifshitz_republication_2017,
	title = {Republication of: On the gravitational stability of the expanding universe},
	volume = {49},
	issn = {1572-9532},
	url = {https://doi.org/10.1007/s10714-016-2165-8},
	doi = {10.1007/s10714-016-2165-8},
	shorttitle = {Republication of},
	pages = {18},
	number = {2},
	journaltitle = {General Relativity and Gravitation},
	shortjournal = {Gen Relativ Gravit},
	author = {Lifshitz, E.},
	urldate = {2026-02-19},
	date = {2017-01-18},
	langid = {english},
}

@article{lemaitre_homogeneous_1931,
	title = {A Homogeneous Universe of Constant Mass and Increasing Radius accounting for the Radial Velocity of Extra-galactic Nebulæ},
	volume = {91},
	issn = {0035-8711},
	url = {https://doi.org/10.1093/mnras/91.5.483},
	doi = {10.1093/mnras/91.5.483},
	pages = {483--490},
	number = {5},
	journaltitle = {Monthly Notices of the Royal Astronomical Society},
	shortjournal = {Mon Not R Astron Soc},
	author = {Lemaître, Abbé G.},
	urldate = {2026-02-19},
	date = {1931-03-13},
}

@article{hubble_relation_1929,
	title = {A relation between distance and radial velocity among extra-galactic nebulae},
	volume = {15},
	url = {https://www.pnas.org/doi/10.1073/pnas.15.3.168},
	doi = {10.1073/pnas.15.3.168},
	pages = {168--173},
	number = {3},
	journaltitle = {Proceedings of the National Academy of Sciences},
	publisher = {Proceedings of the National Academy of Sciences},
	author = {Hubble, Edwin},
	urldate = {2026-02-08},
	date = {1929-03-15},
}

@article{barrera-hinojosa_looking_2022,
	title = {Looking for a twist: probing the cosmological gravitomagnetic effect via weak lensing-{kSZ} cross correlations},
	volume = {510},
	issn = {0035-8711, 1365-2966},
	url = {http://arxiv.org/abs/2109.02632},
	doi = {10.1093/mnras/stab3657},
	shorttitle = {Looking for a twist},
	pages = {3589--3604},
	number = {3},
	journaltitle = {Monthly Notices of the Royal Astronomical Society},
	author = {Barrera-Hinojosa, Cristian and Li, Baojiu and Cai, Yan-Chuan},
	urldate = {2025-09-05},
	date = {2022-01-11},
	langid = {english},
	eprinttype = {arxiv},
	eprint = {2109.02632 [astro-ph]},
}

@misc{bairagi_how_2025,
	title = {How many simulations do we need for simulation-based inference in cosmology?},
	url = {http://arxiv.org/abs/2503.13755},
	doi = {10.48550/arXiv.2503.13755},
	number = {{arXiv}:2503.13755},
	publisher = {{arXiv}},
	author = {Bairagi, Anirban and Wandelt, Benjamin and Villaescusa-Navarro, Francisco},
	urldate = {2025-07-08},
	date = {2025-03-17},
	langid = {english},
	eprinttype = {arxiv},
	eprint = {2503.13755 [astro-ph]},
}

@article{cataneo_mochi_class_2024,
	title = {mochi\_class: Modelling Optimisation to Compute Horndeski In class},
	volume = {7},
	issn = {2565-6120},
	url = {http://arxiv.org/abs/2407.11968},
	doi = {10.33232/001c.123470},
	shorttitle = {mochi\_class},
	journaltitle = {The Open Journal of Astrophysics},
	author = {Cataneo, Matteo and Bellini, Emilio},
	urldate = {2025-03-05},
	date = {2024-09-09},
	langid = {english},
	eprinttype = {arxiv},
	eprint = {2407.11968 [astro-ph]},
}

@article{perlmutter_measurements_1999,
	title = {Measurements of \${\textbackslash}Omega\$ and \${\textbackslash}Lambda\$ from 42 High-Redshift Supernovae},
	volume = {517},
	issn = {0004-637X},
	url = {https://iopscience.iop.org/article/10.1086/307221/meta},
	doi = {10.1086/307221},
	pages = {565},
	number = {2},
	journaltitle = {The Astrophysical Journal},
	shortjournal = {{ApJ}},
	publisher = {{IOP} Publishing},
	author = {Perlmutter, S. and Aldering, G. and Goldhaber, G. and Knop, R. A. and Nugent, P. and Castro, P. G. and Deustua, S. and Fabbro, S. and Goobar, A. and Groom, D. E. and Hook, I. M. and Kim, A. G. and Kim, M. Y. and Lee, J. C. and Nunes, N. J. and Pain, R. and Pennypacker, C. R. and Quimby, R. and Lidman, C. and Ellis, R. S. and Irwin, M. and {McMahon}, R. G. and Ruiz-Lapuente, P. and Walton, N. and Schaefer, B. and Boyle, B. J. and Filippenko, A. V. and Matheson, T. and Fruchter, A. S. and Panagia, N. and Newberg, H. J. M. and Couch, W. J. and Project, The Supernova Cosmology},
	urldate = {2025-03-04},
	date = {1999-06-01},
	langid = {english},
}

@article{friedman_uber_1922,
	title = {Über die Krümmung des Raumes},
	volume = {10},
	rights = {http://www.springer.com/tdm},
	issn = {1434-6001, 1434-601X},
	url = {http://link.springer.com/10.1007/BF01332580},
	doi = {10.1007/BF01332580},
	pages = {377--386},
	number = {1},
	journaltitle = {Zeitschrift für Physik},
	shortjournal = {Z. Physik},
	author = {Friedman, A.},
	urldate = {2025-02-19},
	date = {1922-12},
	langid = {german},
}

@article{robertson_kinematics_1935,
	title = {Kinematics and World-Structure},
	volume = {82},
	issn = {0004-637X},
	url = {https://ui.adsabs.harvard.edu/abs/1935ApJ....82..284R},
	doi = {10.1086/143681},
	pages = {284},
	journaltitle = {The Astrophysical Journal},
	publisher = {{IOP}},
	author = {Robertson, H. P.},
	urldate = {2025-02-19},
	date = {1935-11-01},
	note = {{ADS} Bibcode: 1935ApJ....82..284R},
}

@article{walker_milnes_1937,
	title = {On Milne's Theory of World-Structure},
	volume = {s2-42},
	issn = {1460-244X},
	url = {https://onlinelibrary.wiley.com/doi/abs/10.1112/plms/s2-42.1.90},
	doi = {10.1112/plms/s2-42.1.90},
	pages = {90--127},
	number = {1},
	journaltitle = {Proceedings of the London Mathematical Society},
	author = {Walker, A. G.},
	urldate = {2025-02-19},
	date = {1937},
	langid = {english},
	note = {\_eprint: https://onlinelibrary.wiley.com/doi/pdf/10.1112/plms/s2-42.1.90},
}

@article{joyce_beyond_2015,
	title = {Beyond the cosmological standard model},
	volume = {568},
	issn = {0370-1573},
	url = {https://www.sciencedirect.com/science/article/pii/S0370157314004487},
	doi = {10.1016/j.physrep.2014.12.002},
	series = {Beyond the cosmological standard model},
	pages = {1--98},
	journaltitle = {Physics Reports},
	shortjournal = {Physics Reports},
	author = {Joyce, Austin and Jain, Bhuvnesh and Khoury, Justin and Trodden, Mark},
	urldate = {2025-02-13},
	date = {2015-03-22},
}

@article{riess_observational_1998,
	title = {Observational Evidence from Supernovae for an Accelerating Universe and a Cosmological Constant},
	volume = {116},
	issn = {00046256},
	url = {https://iopscience.iop.org/article/10.1086/300499},
	doi = {10.1086/300499},
	pages = {1009--1038},
	number = {3},
	journaltitle = {The Astronomical Journal},
	author = {Riess, Adam G. and Filippenko, Alexei V. and Challis, Peter and Clocchiatti, Alejandro and Diercks, Alan and Garnavich, Peter M. and Gilliland, Ron L. and Hogan, Craig J. and Jha, Saurabh and Kirshner, Robert P. and Leibundgut, B. and Phillips, M. M. and Reiss, David and Schmidt, Brian P. and Schommer, Robert A. and Smith, R. Chris and Spyromilio, J. and Stubbs, Christopher and Suntzeff, Nicholas B. and Tonry, John},
	urldate = {2025-02-13},
	date = {1998-09},
	langid = {english},
}

@article{gergely_effective_2014,
	title = {Effective field theory of modified gravity with two scalar fields: Dark energy and dark matter},
	volume = {89},
	rights = {http://link.aps.org/licenses/aps-default-license},
	issn = {1550-7998, 1550-2368},
	url = {https://link.aps.org/doi/10.1103/PhysRevD.89.064059},
	doi = {10.1103/PhysRevD.89.064059},
	shorttitle = {Effective field theory of modified gravity with two scalar fields},
	pages = {064059},
	number = {6},
	journaltitle = {Physical Review D},
	shortjournal = {Phys. Rev. D},
	author = {Gergely, László Á. and Tsujikawa, Shinji},
	urldate = {2025-01-21},
	date = {2014-03-25},
	langid = {english},
}

@article{barrera-hinojosa_vector_2021,
	title = {Vector modes in \$Λ\${CDM}: the gravitomagnetic potential in dark matter haloes from relativistic \$N\$-body simulations},
	issn = {0035-8711, 1365-2966},
	url = {http://arxiv.org/abs/2010.08257},
	doi = {10.1093/mnras/staa4025},
	shorttitle = {Vector modes in \$Λ\${CDM}},
	pages = {staa4025},
	journaltitle = {Monthly Notices of the Royal Astronomical Society},
	author = {Barrera-Hinojosa, Cristian and Li, Baojiu and Bruni, Marco and He, Jian-hua},
	urldate = {2025-01-17},
	date = {2021-01-05},
	langid = {english},
	eprinttype = {arxiv},
	eprint = {2010.08257 [astro-ph]},
}

@article{bellini_maximal_2014,
	title = {Maximal freedom at minimum cost: linear large-scale structure in general modifications of gravity},
	volume = {2014},
	issn = {1475-7516},
	url = {https://iopscience.iop.org/article/10.1088/1475-7516/2014/07/050},
	doi = {10.1088/1475-7516/2014/07/050},
	shorttitle = {Maximal freedom at minimum cost},
	pages = {050--050},
	number = {7},
	journaltitle = {Journal of Cosmology and Astroparticle Physics},
	shortjournal = {J. Cosmol. Astropart. Phys.},
	author = {Bellini, Emilio and Sawicki, Ignacy},
	urldate = {2024-12-26},
	date = {2014-07-01},
	langid = {english},
}

@article{bellini_hi_class_2020,
	title = {hi\_class background evolution, initial conditions and approximation schemes},
	volume = {2020},
	rights = {http://iopscience.iop.org/info/page/text-and-data-mining},
	issn = {1475-7516},
	url = {https://iopscience.iop.org/article/10.1088/1475-7516/2020/02/008},
	doi = {10.1088/1475-7516/2020/02/008},
	pages = {008--008},
	number = {2},
	journaltitle = {Journal of Cosmology and Astroparticle Physics},
	shortjournal = {J. Cosmol. Astropart. Phys.},
	author = {Bellini, Emilio and Sawicki, Ignacy and Zumalacárregui, Miguel},
	urldate = {2024-11-27},
	date = {2020-02-10},
	langid = {english},
}

@article{zumalacarregui_hi_class_2017,
	title = {hi\_class: Horndeski in the Cosmic Linear Anisotropy Solving System},
	volume = {2017},
	rights = {http://iopscience.iop.org/info/page/text-and-data-mining},
	issn = {1475-7516},
	url = {https://iopscience.iop.org/article/10.1088/1475-7516/2017/08/019},
	doi = {10.1088/1475-7516/2017/08/019},
	shorttitle = {hi\_class},
	pages = {019--019},
	number = {8},
	journaltitle = {Journal of Cosmology and Astroparticle Physics},
	shortjournal = {J. Cosmol. Astropart. Phys.},
	author = {Zumalacárregui, Miguel and Bellini, Emilio and Sawicki, Ignacy and Lesgourgues, Julien and Ferreira, Pedro G.},
	urldate = {2024-11-27},
	date = {2017-08-16},
	langid = {english},
}

@article{ezquiaga_dark_2017,
	title = {Dark Energy after {GW}170817: dead ends and the road ahead},
	volume = {119},
	issn = {0031-9007, 1079-7114},
	url = {http://arxiv.org/abs/1710.05901},
	doi = {10.1103/PhysRevLett.119.251304},
	shorttitle = {Dark Energy after {GW}170817},
	pages = {251304},
	number = {25},
	journaltitle = {Physical Review Letters},
	shortjournal = {Phys. Rev. Lett.},
	author = {Ezquiaga, Jose María and Zumalacárregui, Miguel},
	urldate = {2024-11-13},
	date = {2017-12-18},
	langid = {english},
	eprinttype = {arxiv},
	eprint = {1710.05901 [astro-ph]},
}

@article{arjona_unraveling_2019,
	title = {Unraveling the effective fluid approach for \$f(R)\$ models in the subhorizon approximation},
	volume = {99},
	issn = {2470-0010, 2470-0029},
	url = {http://arxiv.org/abs/1811.02469},
	doi = {10.1103/PhysRevD.99.043516},
	pages = {043516},
	number = {4},
	journaltitle = {Physical Review D},
	shortjournal = {Phys. Rev. D},
	author = {Arjona, Rubén and Cardona, Wilmar and Nesseris, Savvas},
	urldate = {2024-11-06},
	date = {2019-02-12},
	langid = {english},
	eprinttype = {arxiv},
	eprint = {1811.02469 [astro-ph]},
}

@article{horndeski_second-order_1974,
	title = {Second-order scalar-tensor field equations in a four-dimensional space},
	volume = {10},
	rights = {http://www.springer.com/tdm},
	issn = {0020-7748, 1572-9575},
	url = {http://link.springer.com/10.1007/BF01807638},
	doi = {10.1007/BF01807638},
	pages = {363--384},
	number = {6},
	journaltitle = {International Journal of Theoretical Physics},
	shortjournal = {Int J Theor Phys},
	author = {Horndeski, Gregory Walter},
	urldate = {2024-10-14},
	date = {1974-09},
	langid = {english},
}

@misc{dakin_concept_2015,
	title = {{CONCEPT} - The {COsmological} N-body {CodE} in {PyThon}},
	url = {http://arxiv.org/abs/1510.07621},
	number = {{arXiv}:1510.07621},
	publisher = {{arXiv}},
	author = {Dakin, Jeppe},
	urldate = {2024-09-02},
	date = {2015-10-26},
	langid = {english},
	eprinttype = {arxiv},
	eprint = {1510.07621 [astro-ph]},
}

@article{baumann_cosmological_2012,
	title = {Cosmological Non-Linearities as an Effective Fluid},
	volume = {2012},
	issn = {1475-7516},
	url = {http://arxiv.org/abs/1004.2488},
	doi = {10.1088/1475-7516/2012/07/051},
	pages = {051--051},
	number = {7},
	journaltitle = {Journal of Cosmology and Astroparticle Physics},
	shortjournal = {J. Cosmol. Astropart. Phys.},
	author = {Baumann, Daniel and Nicolis, Alberto and Senatore, Leonardo and Zaldarriaga, Matias},
	urldate = {2024-05-27},
	date = {2012-07-30},
	langid = {english},
	eprinttype = {arxiv},
	eprint = {1004.2488 [astro-ph, physics:gr-qc, physics:hep-th]},
}

@article{creminelli_dark_2017,
	title = {Dark Energy after {GW}170817 and {GRB}170817A},
	volume = {119},
	url = {https://link.aps.org/doi/10.1103/PhysRevLett.119.251302},
	doi = {10.1103/PhysRevLett.119.251302},
	pages = {251302},
	number = {25},
	journaltitle = {Physical Review Letters},
	shortjournal = {Phys. Rev. Lett.},
	publisher = {American Physical Society},
	author = {Creminelli, Paolo and Vernizzi, Filippo},
	urldate = {2024-10-01},
	date = {2017-12-18},
}

@article{carrasco_effective_2012,
	title = {The effective field theory of cosmological large scale structures},
	volume = {2012},
	issn = {1029-8479},
	url = {https://doi.org/10.1007/JHEP09(2012)082},
	doi = {10.1007/JHEP09(2012)082},
	pages = {82},
	number = {9},
	journaltitle = {Journal of High Energy Physics},
	shortjournal = {J. High Energ. Phys.},
	author = {Carrasco, John Joseph M. and Hertzberg, Mark P. and Senatore, Leonardo},
	urldate = {2024-09-30},
	date = {2012-09-20},
	langid = {english},
}

@article{arjona_designing_2019,
	title = {Designing Horndeski and the effective fluid approach},
	volume = {100},
	issn = {2470-0010, 2470-0029},
	url = {http://arxiv.org/abs/1904.06294},
	doi = {10.1103/PhysRevD.100.063526},
	pages = {063526},
	number = {6},
	journaltitle = {Physical Review D},
	shortjournal = {Phys. Rev. D},
	author = {Arjona, Rubén and Cardona, Wilmar and Nesseris, Savvas},
	urldate = {2024-09-23},
	date = {2019-09-20},
	langid = {english},
	eprinttype = {arxiv},
	eprint = {1904.06294 [astro-ph, physics:gr-qc]},
}

@article{kunz_dark_2007,
	title = {Dark Energy versus Modified Gravity},
	volume = {98},
	issn = {0031-9007, 1079-7114},
	url = {http://arxiv.org/abs/astro-ph/0612452},
	doi = {10.1103/PhysRevLett.98.121301},
	pages = {121301},
	number = {12},
	journaltitle = {Physical Review Letters},
	shortjournal = {Phys. Rev. Lett.},
	author = {Kunz, Martin and Sapone, Domenico},
	urldate = {2024-09-04},
	date = {2007-03-21},
	langid = {english},
	eprinttype = {arxiv},
	eprint = {astro-ph/0612452},
}

@article{kunz_crossing_2006,
	title = {Crossing the phantom divide},
	volume = {74},
	rights = {http://link.aps.org/licenses/aps-default-license},
	issn = {1550-7998, 1550-2368},
	url = {https://link.aps.org/doi/10.1103/PhysRevD.74.123503},
	doi = {10.1103/PhysRevD.74.123503},
	pages = {123503},
	number = {12},
	journaltitle = {Physical Review D},
	shortjournal = {Phys. Rev. D},
	author = {Kunz, Martin and Sapone, Domenico},
	urldate = {2024-08-27},
	date = {2006-12-08},
	langid = {english},
}

@article{diego_blas_cosmic_2011,
	title = {The Cosmic Linear Anisotropy Solving System ({CLASS}). Part {II}: Approximation schemes},
	volume = {2011},
	issn = {1475-7516},
	url = {https://iopscience.iop.org/article/10.1088/1475-7516/2011/07/034},
	doi = {10.1088/1475-7516/2011/07/034},
	shorttitle = {The Cosmic Linear Anisotropy Solving System ({CLASS}). Part {II}},
	pages = {034--034},
	number = {7},
	journaltitle = {Journal of Cosmology and Astroparticle Physics},
	shortjournal = {J. Cosmol. Astropart. Phys.},
	author = {{Diego Blas} and {Julien Lesgourgues} and {Thomas Tram}},
	urldate = {2024-08-27},
	date = {2011-07-01},
	langid = {english},
}

@article{gubitosi_effective_2013,
	title = {The Effective Field Theory of Dark Energy},
	volume = {2013},
	issn = {1475-7516},
	url = {http://arxiv.org/abs/1210.0201},
	doi = {10.1088/1475-7516/2013/02/032},
	pages = {032--032},
	number = {2},
	journaltitle = {Journal of Cosmology and Astroparticle Physics},
	shortjournal = {J. Cosmol. Astropart. Phys.},
	author = {Gubitosi, Giulia and Piazza, Federico and Vernizzi, Filippo},
	urldate = {2024-08-26},
	date = {2013-02-19},
	langid = {english},
	eprinttype = {arxiv},
	eprint = {1210.0201 [astro-ph, physics:gr-qc, physics:hep-th]},
}

@misc{lewandowski_effective_2017,
	title = {An effective description of dark matter and dark energy in the mildly non-linear regime},
	url = {http://arxiv.org/abs/1611.07966},
	doi = {10.1088/1475-7516/2017/05/038},
	author = {Lewandowski, Matthew and Maleknejad, Azadeh and Senatore, Leonardo},
	urldate = {2024-08-21},
	date = {2017-09-12},
	langid = {english},
	eprinttype = {arxiv},
	eprint = {1611.07966 [astro-ph, physics:gr-qc, physics:hep-ph, physics:hep-th]},
}

@article{frusciante_effective_2020,
	title = {Effective Field Theory of Dark Energy: a Review},
	volume = {857},
	issn = {03701573},
	url = {http://arxiv.org/abs/1907.03150},
	doi = {10.1016/j.physrep.2020.02.004},
	shorttitle = {Effective Field Theory of Dark Energy},
	pages = {1--63},
	journaltitle = {Physics Reports},
	shortjournal = {Physics Reports},
	author = {Frusciante, Noemi and Perenon, Louis},
	urldate = {2024-08-16},
	date = {2020-05},
	langid = {english},
	eprinttype = {arxiv},
	eprint = {1907.03150 [astro-ph, physics:gr-qc]},
}

@article{cusin_dark_2018,
	title = {Dark Energy and Modified Gravity in the Effective Field Theory of Large-Scale Structure},
	volume = {2018},
	issn = {1475-7516},
	url = {http://arxiv.org/abs/1712.02783},
	doi = {10.1088/1475-7516/2018/04/005},
	pages = {005--005},
	number = {4},
	journaltitle = {Journal of Cosmology and Astroparticle Physics},
	shortjournal = {J. Cosmol. Astropart. Phys.},
	author = {Cusin, Giulia and Lewandowski, Matthew and Vernizzi, Filippo},
	urldate = {2024-08-16},
	date = {2018-04-03},
	langid = {english},
	eprinttype = {arxiv},
	eprint = {1712.02783 [astro-ph, physics:gr-qc, physics:hep-ph]},
}

@article{euclid_collaboration_euclid_2021,
	title = {Euclid preparation: {IX}. {EuclidEmulator}2 -- Power spectrum emulation with massive neutrinos and self-consistent dark energy perturbations},
	volume = {505},
	issn = {0035-8711, 1365-2966},
	url = {http://arxiv.org/abs/2010.11288},
	doi = {10.1093/mnras/stab1366},
	shorttitle = {Euclid preparation},
	pages = {2840--2869},
	number = {2},
	journaltitle = {Monthly Notices of the Royal Astronomical Society},
	author = {Euclid Collaboration and Knabenhans, M. and Stadel, J. and Potter, D. and Dakin, J. and Hannestad, S. and Tram, T. and Marelli, S. and Schneider, A. and Teyssier, R. and Andreon, S. and Auricchio, N. and Baccigalupi, C. and Balaguera-Antolínez, A. and Baldi, M. and Bardelli, S. and Battaglia, P. and Bender, R. and Biviano, A. and Bodendorf, C. and Bozzo, E. and Branchini, E. and Brescia, M. and Burigana, C. and Cabanac, R. and Camera, S. and Capobianco, V. and Cappi, A. and Carbone, C. and Carretero, J. and Carvalho, C. S. and Casas, R. and Casas, S. and Castellano, M. and Castignani, G. and Cavuoti, S. and Cledassou, R. and Colodro-Conde, C. and Congedo, G. and Conselice, C. J. and Conversi, L. and Copin, Y. and Corcione, L. and Coupon, J. and Courtois, H. M. and Da Silva, A. and de la Torre, S. and Di Ferdinando, D. and Duncan, C. A. J. and Dupac, X. and Fabbian, G. and Farrens, S. and Ferreira, P. G. and Finelli, F. and Frailis, M. and Franceschi, E. and Galeotta, S. and Garilli, B. and Giocoli, C. and Gozaliasl, G. and Graciá-Carpio, J. and Grupp, F. and Guzzo, L. and Holmes, W. and Hormuth, F. and Israel, H. and Jahnke, K. and Keihanen, E. and Kermiche, S. and Kirkpatrick, C. C. and Kubik, B. and Kunz, M. and Kurki-Suonio, H. and Ligori, S. and Lilje, P. B. and Lloro, I. and Maino, D. and Marggraf, O. and Markovic, K. and Martinet, N. and Marulli, F. and Massey, R. and Mauri, N. and Maurogordato, S. and Medinaceli, E. and Meneghetti, M. and Metcalf, B. and Meylan, G. and Moresco, M. and Morin, B. and Moscardini, L. and Munari, E. and Neissner, C. and Niemi, S. M. and Padilla, C. and Paltani, S. and Pasian, F. and Patrizii, L. and Pettorino, V. and Pires, S. and Polenta, G. and Poncet, M. and Raison, F. and Renzi, A. and Rhodes, J. and Riccio, G. and Romelli, E. and Roncarelli, M. and Saglia, R. and Sánchez, A. G. and Sapone, D. and Schneider, P. and Scottez, V. and Secroun, A. and Serrano, S. and Sirignano, C. and Sirri, G. and Stanco, L. and Sureau, F. and Crespí, P. Tallada and Taylor, A. N. and Tenti, M. and Tereno, I. and Toledo-Moreo, R. and Torradeflot, F. and Valenziano, L. and Valiviita, J. and Vassallo, T. and Viel, M. and Wang, Y. and Welikala, N. and Whittaker, L. and Zacchei, A. and Zucca, E.},
	urldate = {2024-08-16},
	date = {2021-06-15},
	langid = {english},
	eprinttype = {arxiv},
	eprint = {2010.11288 [astro-ph]},
}

@article{ma_cosmological_1995,
	title = {Cosmological Perturbation Theory in the Synchronous and Conformal Newtonian Gauges},
	volume = {455},
	issn = {0004-637X, 1538-4357},
	url = {http://arxiv.org/abs/astro-ph/9506072},
	doi = {10.1086/176550},
	pages = {7},
	journaltitle = {The Astrophysical Journal},
	shortjournal = {{ApJ}},
	author = {Ma, Chung-Pei and Bertschinger, Edmund},
	urldate = {2024-08-16},
	date = {1995-12},
	langid = {english},
	eprinttype = {arxiv},
	eprint = {astro-ph/9506072},
}

@misc{lesgourgues_cosmic_2011,
	title = {The Cosmic Linear Anisotropy Solving System ({CLASS}) I: Overview},
	url = {http://arxiv.org/abs/1104.2932},
	shorttitle = {The Cosmic Linear Anisotropy Solving System ({CLASS}) I},
	number = {{arXiv}:1104.2932},
	publisher = {{arXiv}},
	author = {Lesgourgues, Julien},
	urldate = {2024-08-16},
	date = {2011-05-14},
	langid = {english},
	eprinttype = {arxiv},
	eprint = {1104.2932 [astro-ph]},
}

	\begin{annexes}
		
    \begingroup
        \renewcommand{\clearpage}{} 
        \renewcommand{\cleardoublepage}{}
        \chapter{Horndeski Gravity}
\label{annex: Horndeski}
\section{Computing the Equations of Motion}
In this chapter, we show how to derive the scalar and gravitational field equations for the reduced Horndeski action, given by:
\begin{equation}
    S[g_{\mu\nu},\phi] = \int \dd[4]x \mpls\sqrt{-g}\left[\sum_{i=2}^4\mathcal{L}_i(g_{\mu\nu},\phi) + \mathcal{L}_{\text{m}}(g_{\mu\nu},\psi_\text{m})\right].
\end{equation}
Before deriving the scalar field equations for the reduced Horndeski action, we recall that for a function $f$ of a single variable $x$, the action is defined by:
\begin{equation}
    S[f] = \int_{x_1}^{x_2}\L(x,f',f'',\ldots,f^{(k)})\dd{x},
\end{equation}
where $f'\equiv\dv*{f}{x}$, $f''\equiv\dv*[2]{f}{x}$, \ldots, $f^{(k)}\equiv\dv*[k]{f}{x}$. We note that in general, the Lagrangian could depend on any number of derivatives of $f$, and the Euler-Lagrange equations that it satisfies are:
\begin{equation}
    \pdv{\L}{f}-\dv{}{x}\qty(\pdv{\L}{f'}) + \dv[2]{}{x}\qty(\pdv{\L}{f''})-\ldots (-1)^k\dv[k]{}{x}\qty(\pdv{\L}{f^{(k)}})=0.
\end{equation}
In our case, the Lagrangian is a function of the scalar field $\phi$ and its derivatives up to second order. This is clearly seen in the following:
\begin{align}
    \mathcal{L}_2 &= G_2(\phi, X) = K(\phi, X),\\
    \mathcal{L}_3 &= -G_3(\phi, X)\Box\phi,\\
    \mathcal{L}_4 &= G_4(\phi) R,\\
    \mathcal{L}_5 &= 0,
\end{align}
where $X\equiv-\partial_\mu\phi\,\partial^\mu\phi/2$ and $\Box\phi\equiv g^{\mu\nu}\nabla_\mu\nabla_\nu\phi$. It is worth recalling that any derivative $\partial_\mu$ is interchangeable with the covariant derivative $\nabla_\mu$, when acting on a scalar. Given our definition of the d'Lambertian, we must take into account that the Lagrangians depend on the scalar field and its first and second comoving derivatives.

For instance, we can change the derivatives of the kinetic term for comoving derivatives, thus:
\begin{align}
    \pdv{X}{(\nabla_\rho\phi)} &= \pdv{}{(\nabla_\rho\phi)}\qty(-\frac{1}{2}g^{\mu\nu}\nabla_\mu\phi\nabla_\nu\phi),\nonumber\\
    &= -\frac{1}{2}g^{\mu\nu}\pdv{(\nabla_\mu\phi\nabla_\nu\phi)}{(\nabla_\rho\phi)},\nonumber\\
    & = -\frac{1}{2}g^{\mu\nu}\qty[\nabla_\nu\phi\,\tensor{\delta}{^\rho_\mu} + \nabla_\mu\phi\,\tensor{\delta}{^\rho_\nu}],\nonumber\\
    & = -\nabla^\rho\phi.
\end{align}

Now we turn to compute the scalar field equations, going step by step with each Lagrangian. We must remember that eventually we would be integrating by parts, where the boundary terms are negligible. Hence, we do it immediately a the following steps. For $\L_2$:
\begin{align}
    \delta\L_2(\phi,\nabla_\mu\phi) &= \pdv{\L_2}{\phi}\delta\phi + \pdv{\L_2}{(\nabla_\mu\phi)}\delta(\nabla_\mu\phi),\nonumber\\
    &= \pdv{\L_2}{\phi}\delta\phi - \nabla_\mu\qty(\pdv{\L_2}{(\nabla_\mu\phi)})\delta\phi,\nonumber\\
    &= K_\phi \delta\phi - \nabla_\mu\qty(\pdv{\L_2}{X}\pdv{X}{(\nabla_\mu\phi)})\delta\phi,\nonumber\\
    &= \qty[K_\phi +\nabla_\mu(K_X\nabla^\mu\phi)]\delta\phi.
\end{align}
Now, for $\L_3$:
\begin{align}
    \delta\L_3(\phi,\nabla_\mu\phi,\nabla_\mu\nabla_\nu\phi) &= \pdv{\L_3}{\phi}\delta\phi +\pdv{\L_3}{(\nabla_\mu\phi)}\delta(\nabla_\mu\phi) + \pdv{\L_3}{(\nabla_\mu\nabla_\nu\phi)} \delta(\nabla_\mu\nabla_\nu\phi),\nonumber \\
    &= \pdv{\L_3}{\phi}\delta\phi -\nabla_\mu\qty(\pdv{\L_3}{(\nabla_\mu\phi)}) \delta\phi + \nabla_\mu\nabla_\nu \qty(\pdv{\L_3}{(\nabla_\mu\nabla_\nu\phi)}) \delta\phi,\nonumber\\
    &= \qty[-G_{3\phi}\Box\phi + \nabla_\mu\qty(G_{3X}\pdv{X}{(\nabla_\mu\phi)}\Box\phi) -\nabla_\mu\nabla_\nu\qty(G_3g^{\mu\nu})]\delta\phi,\nonumber\\
    &= \qty[-G_{3\phi}\Box\phi - \nabla_\mu\qty(G_{3X}\nabla^\mu\phi\Box\phi) -\nabla_\mu\qty(G_{3\phi}\nabla^\mu\phi + G_{3X}\nabla^\mu X)]\delta\phi
\end{align}
Lastly, the term for $\L_4$:
\begin{align}
    \delta\L_4 &= \pdv{\L_4}{\phi}\delta\phi,\nonumber\\
                &= G_{4\phi}\,R\,\delta\phi.
\end{align}
Thus, for a fixed scalar field extreme, we find the scalar field equation:
\begin{equation}
    K_\phi +\nabla_\mu(K_X\nabla^\mu\phi) -G_{3\phi}\Box\phi - \nabla_\mu\qty(G_{3X}\nabla^\mu\phi\Box\phi) -\nabla_\mu\qty(G_{3\phi}\nabla^\mu\phi + G_{3X}\nabla^\mu X)+G_{4\phi}R = 0.\label{eq:annex-a scalar field}
\end{equation}
This may be further simplified by expanding the following terms:
\begin{align}
    \nabla_\mu\qty(K_X\nabla^\mu\phi) &= \nabla_\mu K_X\nabla^\mu\phi + K_X\Box\phi,\\
    \nabla_\mu\qty(G_{3\phi}\nabla^\mu\phi+G_{3X}\nabla^\mu X) &= \nabla_\mu G_{3\phi}\nabla^\mu\phi + G_{3\phi}\Box\phi +\nabla_\mu G_{3X}\nabla^\mu X + G_{3X}\Box X,\\
    \nabla_\mu(G_{3X}\nabla^\mu\phi\Box\phi) &= \nabla_\mu G_{3X}\nabla^\mu\phi\Box\phi + G_{3X}\qty[(\Box\phi)^2  + \nabla^\mu\phi\nabla_\mu\Box\phi],
\end{align}
where $\Box X$ and $\nabla_\mu\Box\phi$ may be expanded:
\begin{align}
\Box X &= -\frac{1}{2}\nabla_\mu\nabla^\mu
\qty(\nabla_\nu\phi\nabla^\nu\phi), \nonumber\\
       &= -\frac{1}{2}\nabla_\mu \qty(\nabla^\mu\nabla_\nu\phi\,\nabla^\nu\phi + \nabla_\nu\phi\,\nabla^\mu\nabla^\nu\phi),\nonumber\\
        &= -\frac{1}{2}\Big[ \nabla_\mu\nabla^\mu\nabla_\nu\phi\,\nabla^\nu\phi + \nabla^\mu\nabla_\nu\phi\,\nabla_\mu\nabla^\nu\phi \nonumber\\ 
        &\qquad\qquad + \nabla_\mu\nabla_\nu\phi\,\nabla^\mu\nabla^\nu\phi + \nabla_\nu\phi\,\nabla_\mu\nabla^\mu\nabla^\nu\phi \Big], \nonumber\\
        &= -\qty(\nabla_\mu\nabla_\nu\phi)^2 -\nabla_\nu\phi\,\nabla_\mu\nabla^\mu\nabla^\nu\phi,
\end{align}
\begin{align}
\nabla_\mu\Box\phi&= \nabla_\mu \nabla_\nu\nabla^\nu\phi,\nonumber\\
        & = \comm{\nabla_\mu}{\nabla_\nu}\nabla^\nu\phi +\nabla_\nu\nabla_\mu\nabla^\nu\phi,\nonumber\\
        &=\tensor{R}{^\nu_\rho_\mu_\nu}\nabla^\rho\phi +\nabla_\nu\nabla^\nu\nabla_\mu\phi\nonumber\\
        &= -\tensor{R}{_\mu_\nu}\nabla^\nu\phi + \nabla_\nu\nabla_\mu\nabla^\nu\phi.
\end{align}
Replacing everything into \cref{eq:annex-a scalar field} yields:
\begin{align}
    0 &= K_\phi + \nabla_\mu K_X\nabla^\mu\phi + K_X\Box\phi -2G_{3\phi}\Box\phi-\nabla_\mu G_{3\phi}\nabla^\mu\phi -\nabla_\mu G_{3X}\nabla^\mu X\nonumber\\
    &\quad+ G_{3X}\qty(\qty(\nabla_\mu\nabla_\nu\phi)^2 +\nabla_\nu\phi\,\nabla_\mu\nabla^\mu\nabla^\nu\phi) -\nabla_\mu G_{3X}\nabla^\mu\phi\Box\phi - G_{3X}\qty(\Box\phi)^2\nonumber\\
    &\quad- G_{3X}\nabla^\mu\phi\qty(-\tensor{R}{_\mu_\nu}\nabla^\nu\phi +\nabla_\nu\nabla^\nu\nabla_\mu\phi) + G_{4\phi}R,
\end{align}
and cancelling out the terms proportional to the three covariant derivatives of the scalar field, we have the final form of the scalar field equation:
\begin{align}
    0 &= K_\phi + \nabla_\mu K_X\nabla^\mu\phi + K_X\Box\phi -2G_{3\phi}\Box\phi-\nabla_\mu G_{3\phi}\nabla^\mu\phi -\nabla_\mu G_{3X}\nabla^\mu X\nonumber\\
    &\quad- G_{3X}\qty[\qty(\Box\phi)^2- \qty(\nabla_\mu\nabla_\nu\phi)^2 -\tensor{R}{_\mu_\nu}\nabla^\mu\phi\nabla^\nu\phi] -\nabla_\mu G_{3X}\nabla^\mu\phi\Box\phi+ G_{4\phi}R.
\end{align}

Now, we turn to the gravitational field equations. Similar to what we did for the scalar field equation, we go by each term of the action. By imposing that the variation of the action is null, we find:
\begin{align}
    \delta S = \delta S_2 + \delta S_3 + \delta S_4 + \delta(\sqrt{-g}\L_{\rm m}) = 0.
\end{align}
First  we study $\delta S_2$:
\begin{equation}
    \delta S_2 = \int\dd[4]x \mpls \qty(\delta\sqrt{-g}K + \sqrt{-g}\delta K).
\end{equation}
We use Jacobi's identity for the variation of $\sqrt{-g}$:
\begin{equation}
    \delta\sqrt{-g} = -\frac{1}{2}\sqrt{-g}g_{\mu\nu}\delta g^{\mu\nu}.
\end{equation}
On the other hand, the variation of the kinetic term is:
\begin{equation}
    \delta K = K_X \pdv{X}{g^{\mu\nu}} \delta g^{\mu\nu} = -\frac{1}{2}K_X\nabla_\mu\phi\nabla_\nu\phi\,\delta g^{\mu\nu},
\end{equation}
which allows us to write:
\begin{equation}
    \delta S_2 = \int\dd[4]x \mpls \sqrt{-g} \,\delta g^{\mu\nu} \qty[-\frac{1}{2}g_{\mu\nu} K -\frac{1}{2}K_X\nabla_\mu\phi\nabla_\nu\phi].
\end{equation}
For $\delta S_3$ we have:
\begin{equation}
    \delta S_3 = \int\dd[4]x \mpls\qty[-\delta\sqrt{-g}G_3\Box\phi - \sqrt{-g}\delta\qty(G_3\Box\phi)],
    \label{eq:annex deltas3}
\end{equation}
where the last term is:
\begin{align}
    \delta(G_3\Box\phi) &= \delta G_3\Box\phi + G_3\delta(\Box\phi)\nonumber\\
    &= -\frac{1}{2}G_{3X}\nabla_\mu\phi\nabla_\nu\phi\,\delta g^{\mu\nu}\Box\phi+ G_3\qty[\nabla_\mu\nabla_\nu\phi\,\delta g^{\mu\nu}+g^{\mu\nu}\delta(\nabla_\mu\nabla_\nu\phi)]\nonumber\\
    &= -\frac{1}{2}G_{3X}\nabla_\mu\phi\nabla_\nu\phi\,\delta g^{\mu\nu}\Box\phi+ G_3\qty[\nabla_\mu\nabla_\nu\phi\,\delta g^{\mu\nu}+g^{\mu\nu}(\nabla_\mu\nabla_\nu\,\delta\phi - \delta\tensor{\Gamma}{^\rho_{\mu\nu}}\nabla_\rho\phi)],
\end{align}
where in the last line we used the identity:
\begin{equation}
    \delta(\nabla_\mu\nabla_\nu\phi) = \nabla_\mu\nabla_\nu\,\delta\phi - \delta\tensor{\Gamma}{^\rho_{\mu\nu}}\nabla_\rho\phi.
\end{equation}
Now we plug our terms into \cref{eq:annex deltas3}, except for the one that is proportional to the variation of the scalar field:
\begin{align}
    \delta S_3 &= \int\dd[4]x\mpls\sqrt{-g}\Bigg[\frac{1}{2}G_3\Box\phi g_{\mu\nu}\delta g^{\mu\nu} +\frac{1}{2}G_{3X}\nabla_\mu\phi\nabla_\nu\phi\,\delta g^{\mu\nu}\Box\phi\nonumber\\
    &\qquad\qquad\qquad\qquad\qquad- G_3\qty(\nabla_\mu\nabla_\nu\phi\,\delta g^{\mu\nu} -g^{\mu\nu}\delta\tensor{\Gamma}{^\rho_{\mu\nu}}\nabla_\rho\phi)\Bigg].
\end{align}
Now, we group everything proportional to the variation of the inverse of the metric, and we use the identity:
\begin{equation}
    g^{\mu\nu} \delta\tensor{\Gamma}{^\rho_{\mu\nu}}=-\nabla_\sigma\delta g^{\rho\sigma} + \frac{1}{2}g_{\mu\nu}g^{\rho\sigma}\nabla_\sigma \delta g^{\mu\nu},
    \label{eq: annexa identity}
\end{equation}
such that we end up with:
\begin{align}
    \delta S_3 &= \int\dd[4]x\mpls\sqrt{-g}\,\delta g^{\mu\nu}\Bigg[\frac{1}{2}G_3\Box\phi g_{\mu\nu} +\frac{1}{2}G_{3X}\nabla_\mu\phi\nabla_\nu\phi\,\Box\phi\nonumber\\
    &\qquad\qquad\qquad\qquad\qquad- G_3\nabla_\mu\nabla_\nu\phi+\nabla_\nu\qty(G_3\nabla_\mu\phi)-\frac{1}{2}g_{\mu\nu}g^{\rho\sigma}\nabla_\sigma\qty(G_3\nabla_\rho\phi) \Bigg].
\end{align}
By expanding the derivative of the last term, we can exactly cancel out one of these terms with the first one. Lastly, since the metric is symmetric, we should ensure that everything inside the square brackets is symmetric. Then we arrive at:
\begin{align}
    \delta S_3 &= \int\dd[4]x\mpls\sqrt{-g}\,\delta g^{\mu\nu}\Bigg[\frac{1}{2}G_{3X}\nabla_\mu\phi\nabla_\nu\phi\,\Box\phi- G_3\nabla_\mu\nabla_\nu\phi\nonumber\\
    &\qquad\qquad\qquad\qquad\qquad\,\,+\nabla_{(\nu}\qty(G_3\nabla_{\mu)}\phi)-\frac{1}{2}g_{\mu\nu}\nabla^\rho G_3\nabla_\rho\phi \Bigg].
\end{align}
Now, we turn to the last term $\delta S_4$:
\begin{align}
    \delta S_4 &= \int\dd[4]x \mpls \delta(\sqrt{-g}\,G_4 R),\nonumber\\
    & = \int\dd[4]x \mpls (\delta\sqrt{-g}\,G_4 R + \sqrt{-g}\,\delta G_4 R + \sqrt{-g} \,G_4\delta R).
\end{align}
It is clear that $\delta G_4 = G_{4\phi}\delta\phi$, so we should ignore this term. In contrast, the variation of the Ricci scalar can be deduced by using the Palatini identity:
\begin{align}
    \delta R &= \delta(R_{\mu\nu}g^{\mu\nu}),\nonumber\\
    &= g^{\mu\nu}\delta R_{\mu\nu} + R_{\mu\nu}\delta g^{\mu\nu},\nonumber\\
    &= (\nabla_\rho\delta\tensor{\Gamma}{^\rho_{\mu\nu}} - \nabla_\nu\delta\tensor{\Gamma}{^\rho_{\rho\mu}})g^{\mu\nu} +R_{\mu\nu}\delta g^{\mu\nu},\nonumber\\
    &= \nabla_\rho\qty(g^{\mu\nu}\delta\tensor{\Gamma}{^\rho_{\mu\nu}}) - \nabla_\nu\qty(g^{\mu\nu}\delta\tensor{\Gamma}{^\rho_{\rho\mu}})+R_{\mu\nu}\delta g^{\mu\nu}.
\end{align}
Given that the difference of two connections is a tensor, we write:
\begin{equation}
    \delta\tensor{\Gamma}{^\rho_{\mu\nu}} = -\frac{1}{2}\qty(g_{\nu\sigma}\nabla_\mu\delta g^{\rho\sigma} + g_{\mu\sigma}\nabla_\nu\delta g^{\rho\sigma} - g_{\mu\alpha}g_{\nu\beta}\nabla^\rho\delta g^{\alpha\beta}).
\end{equation}
With the last equation, it is possible to show \cref{eq: annexa identity} and:
\begin{align}
    g^{\mu\nu}\delta\tensor{\Gamma}{^\rho_{\rho\mu}}
    &= -\frac{1}{2}\nabla^\nu\qty(g_{\mu\nu}\delta g^{\mu\nu}).
\end{align}
Substituting back:
\begin{align}
    \delta R &= \nabla_\rho\qty(-\nabla_\sigma\delta g^{\rho\sigma} +\frac{1}{2}g_{\mu\nu}\nabla^\rho\delta g^{\mu\nu})+\frac{1}{2}g_{\mu\nu}\Box\delta g^{\mu\nu} +R_{\mu\nu}\delta g^{\mu\nu}.\nonumber\\
    &= R_{\mu\nu}\delta g^{\mu\nu} + g_{\mu\nu}\Box\delta g^{\mu\nu} - \nabla_\mu\nabla_\nu\delta g^{\mu\nu}.
\end{align}

Thus, we get:
\begin{align}
    \delta S_4 &= \int\dd[4]x \mpls \sqrt{-g}\qty[-\frac{1}{2}G_4 R\,g_{\mu\nu}\delta g^{\mu\nu} + G_4 R_{\mu\nu}\delta g^{\mu\nu} + G_4 \,g_{\mu\nu}\Box\delta g^{\mu\nu} -G_4 \nabla_\mu\nabla_\nu\delta g^{\mu\nu}]\nonumber\\
    &= \int\dd[4]x \mpls \sqrt{-g}\delta g^{\mu\nu}\qty[G_{\mu\nu}G_4 +  \,g_{\mu\nu}\Box G_4 - \nabla_\mu\nabla_\nu G_4]\nonumber\\
    &= \int\dd[4]x \mpls \sqrt{-g}\delta g^{\mu\nu}\qty[G_{\mu\nu}G_4 +  \,g_{\mu\nu}\qty(G_{4\phi}\Box\phi -2XG_{4\phi\phi} ) - G_{4\phi} \nabla_\mu\nabla_\nu \phi - G_{4\phi\phi}\nabla_\mu\phi\nabla_\nu \phi].
\end{align}

By collecting all the terms that we calculated, we are in a position to write the gravitational field equations:
\begin{align}
    \frac{1}{\mpls}T_{\mu\nu}^{({\rm{m}})} &= -K_X\nabla_\mu\phi\nabla_\nu\phi - Kg_{\mu\nu} + G_{3X}\Box\phi\nabla_\mu\phi\nabla_\nu\phi + 2\nabla_{(\mu}G_3\nabla_{\nu)}\phi -g_{\mu\nu}\nabla_\lambda G_3\nabla^\lambda\phi \nonumber \\
    &\quad+2G_4G_{\mu\nu}+2g_{\mu\nu}(G_{4\phi}\Box\phi-2XG_{4\phi\phi})-2G_{4\phi}\nabla_\mu\nabla_\nu\phi - 2G_{4\phi\phi}\nabla_\mu\phi\nabla_\nu\phi.
\end{align}
\section{Coefficients of Horndeski}
\label{horndeski lin coeff}

The coefficients $A,B$ and $C$ present in the scalar field equation are:

\begin{align}
    \label{eqAannex}
    A&\equiv \frac{1}{2}a^3H^3(2-\alpha_{\rm{B}})\qty[\alpha_{\rm{K}}-3\alpha_{\rm{B}}(1+\alpha_{\rm{M}})] + aH\qty(\alpha_{\rm{K}}+\frac{3}{2}\alpha_{\rm{B}}^2)(2a^2H^2+aH')\nonumber\\
    &\quad+a^2H^2\qty[aH\alpha_{\rm{K}} + \frac{H'}{H}\alpha_{\rm{K}} + \alpha_{\rm{K}}' + 3\alpha_{\rm{B}}\qty(aH+\frac{H'}{H}+\frac{\alpha_{\rm{B}}'}{2})] + \frac{3(\rho+P)a^3H\alpha_{\rm{B}}}{2M^2_*\mpls},\\
    \label{eqBannex}
    B&\equiv a^2\frac{\rho+P}{M^2_*\mpls}\qty[a^2H^2+aH'- a^2H^2(1+\alpha_{\rm{M}})] + \frac{a^3H\alpha_{\rm{B}}P'}{2M^2_*\mpls} \nonumber\\
    &\quad + a^2H^2(2-\alpha_{\rm{B}})\qty[\frac{1}{6}a^2H^2\alpha_{\rm{K}} -aH'\alpha_{\rm{M}} - \frac{a^2H^2}{2}\alpha_{\rm{B}}\alpha_{\rm{M}}]+(2-\alpha_{\rm{B}})(aH')^2\nonumber\\
    &\quad +a^3H^3\qty(\frac{1}{3}\alpha_{\rm{K}}'+\alpha_{\rm{B}}') + \qty(\frac{aH}{2}\alpha_{\rm{B}} - aH-\frac{H'}{H})a^2H^2\alpha_{\rm{B}}' \nonumber\\
    &\quad+\qty(\alpha_{\rm{K}}+\frac{3}{2}\alpha_{\rm{B}}^2)a^3H^3\qty(aH+\frac{H'}{H}),\\
    \label{eqCannex}
    C&\equiv aH\alpha_{\rm{B}}'+(2-\alpha_{\rm{B}})\qty[a^2H^2(\alpha_{\rm{M}}-1)-aH'+\frac{a^2H^2}{2}(2+\alpha_{\rm{B}})]-a^2\frac{\rho+P}{M^2_*\mpls}
\end{align}

    \endgroup
    \setupsecondannexformat 

		\chapter{Basics of \class}
\label{annex: CLASS}
\section{Installing the code}
The easiest way to get the code is to clone the GitHub repository\footnote{\url{https://github.com/lesgourg/class_public}}. This allows a user to study, use, and modify the source code. Since the code is written in \texttt{C}, it is important to compile it when installing and modifying it. To do it, the user should navigate to the folder containing the code, and write these commands in a terminal:
\begin{verbatim}
    make clean
    make
\end{verbatim}

Alternatively, someone interested only in running the standard version of \class\ and using it through the \texttt{Python} wrapper \texttt{classy} might prefer to install it as a Python module with: \texttt{pip install classy}.

If the user plans to have multiple versions of {\class} on their machine, it is highly recommended to use virtual environments, and Anaconda or Python virtual environments are great options. For this work, we used \url{https://docs.astral.sh/uv/} to manage virtual environments. Then, it is possible to have multiple virtual environments with the different versions of {\class}, but the code must be compiled with the virtual environment enabled, or if there is a particular \texttt{Python} installation, through:
\begin{verbatim}
    make clean
    PYTHON=path/to/your/python/ make
\end{verbatim}
To find which instance of \texttt{Python} is currently being used:
\begin{verbatim}
    which python
\end{verbatim}
If the user is never going to use the \texttt{Python} wrapper, it is possible to just compile the \texttt{C} code by:
\begin{verbatim}
    make clean
    make class
\end{verbatim}

To check whether the code is properly installed, the user can run the \texttt{explanatory.ini} file, and no output means that everything is ready to use the code. To perform the test, the user must go to the \class\ directory and run \class\ with:
\begin{verbatim}
    ./class explanatory.ini
\end{verbatim}

\section{Settings and the \texttt{explanatory.ini} file}
The code has a large number of flags and settings that might be overwhelming at first sight. However, we strongly recommend exploring the \texttt{explanatory.ini} file. It has multiple comments, and virtually every single setting that can be fixed. For this reason, it is advisable to never touch this file and instead create new files with minimal settings and explore how the code works.

Some important aspects of the code concern how outputs are handled. To start, we have to specify, via the \texttt{output} flag, which output we want the code to compute. For instance, if we are interested in the matter power spectrum and the CMB temperature angular power spectrum, the output flag should look like:
\begin{verbatim}
    output = mPk, tCl
\end{verbatim}

In addition, it is possible to specify the wavenumber at which perturbations are to be computed, via:
\begin{verbatim}
    output_k_values = 0.01, 0.1, 1
\end{verbatim}
Note that this is just an example, and the units of the wave number are inverse megaparsecs.

There is a flag to output perturbations in the current gauge. It is important to note that {\class} by default computes everything in the synchronous gauge, but outputs the density contrasts in the Newtonian gauge. In newer versions of the code, it is possible

\section{The \texttt{Python} wrapper \texttt{classy}}
From now on, we assume that the user wants to work with classy. For this person, a minimal {\class} run looks like

\begin{verbatim}
    from classy import Class

    cosmo = Class()
    cosmo.set(params)
    cosmo.compute()
\end{verbatim}
where \texttt{params} should be a dictionary containing parameters. For example, we have run multiple instances of \class with:
\begin{verbatim}
kk = np.array([1e-3, 1e-2, 1e-1])
params = {
           'H0'      : 67,
           'Omega_b' : 0.049,
           'Omega_cdm': 0.27,
           'A_s'     : 2.1e-9,
           'n_s'     : 0.96,
           'output'  : 'tCl mPk dTk vTk',
           'k_output_values': ','.join([str(k) for k in kk]),
           'z_max_pk': 5,
           'evolver': 0,
           # Disable massless neutrinos
           'N_ur': 0,
           # Add 3 massive neutrinos of equal mass
           'N_ncdm'  : 1,
           'deg_ncdm': 3,
           'm_ncdm'  : 0.058,  # avoid exact value of 0.0
}
\end{verbatim}
Then, after running this set of parameters, we may study the background and perturbations. There are several predefined functions that can provide the user with specific functionality, such as density parameters. Instead of that, we usually worked with the \texttt{.get\_background()} and \texttt{.get\_perturbations()} functions, that gives us pretty much everything needed to study a cosmological model. They provide dictionaries containing the most relevant information to build observables, and for the perturbations, it is important to note that it gives an array containing dictionaries, where each entry corresponds to a specific wavenumber.

As a final note, we recommend that the user iterate over cosmological models using dictionaries. \class\ can be confused when iterating over models when not done properly, and in some cases, a model could yield completely unphysical results, like negative scale factors.
		\chapter[Gravitational Force in \concept]
        {Gravitational Force in\\\hspace*{5cm}\concept}
\label{annex:CONCEPT}
Here, we provide further details on the particle force in \concept. First, let us define some of the parameters chosen by the code and their default values.
\begin{itemize}
    \item \textbf{The number of particles $N$}: as the name suggests, correspond to the total number of particles that will be considered in the simulation.
    \item \textbf{Length of the box $L_{\rm box}$}: is the side of the toroidal cubic box where the $N$ particles reside. Its units are of Mpc$/h$.
    \item \textbf{Number of PM cells $n_\Phi$}: is the number of grids considered for the potential. Its default value is $n_\Phi = 2\sqrt[3]{N}$, meaning that there are 8 times PM cells as particles.
    \item \textbf{Grid size $L_\Phi$}: is the length of the cubic box, associated with the potential resolution. It is defined throught $L_\Phi=L_{\rm box}/n_\Phi$. Naturally, the condition $L_{\rm box}> L_\Phi$ must hold.
    \item \textbf{Splitting scale $x_s$}: is the scale that enters the exponential and complementary error functions in the short-range calculation. It depicts the scale, where for distances greater than $x_s$, the functions decay exponentially. By default, it is set to $x_s= 1.25L_\Phi$
    \item \textbf{Hard cut-off scale $x_r$}: this defines the hard cut-off where the sum of the short-range force is set to zero, and there is no longer the need to compute it up to infinity. Its default value is $x_r =4.5x_s$.
\end{itemize}

Now, let's turn to a detailed description of the force that \concept\ computes, given by:
\begin{equation}
    \begin{aligned}
    \vb{f}_i &= \frac{4\pi G m^2}{L^{4}_{\varphi}} \overbrace{
        \sum_{\mathclap{\phantom{\raisebox{1.9ex}{x}}}\mathclap{\phantom{\raisebox{-1.9ex}{x}}}\mathclap{
            \bigl\{ \, \vb{m} \, \bigm| \, |\vb{x}_i - \vb{x}_{\vb{m}}|_{\infty} < \frac{p_{\text{i}}L_{\varphi}}{2} \, \bigr\}
            }}
        W_{p_i}\biggl(\frac{\vb{x}_i - \vb{x}_{\vb{m}}}{L_{\varphi}}\biggr)
        }^{\smash{\text{particle $\leftarrow$ mesh}}}
        \vb{D}_{p_{\text{d}}}
        \,\underset{\smash{\hspace{-0.5em}\vb{m}\leftarrow\vb{h}}}{{\mathcal{F}_{\slashed{\mathbf{0}}}^{-1}}} \biggl\{ 
        \mathclap{\phantom{\raisebox{4.1ex}{x}}} \\
    &\qquad\quad \underbrace{\biggl[\frac{W_{p_{\text{i}}}(L_{\varphi}\vb{k}_{\vb{h}})}{L^3_{\varphi}}\biggr]^{-2}}_{\mathclap{\text{2 deconvolutions}}} 
        \smash{\overbrace{\frac{\exp\bigl(-x^2_{\text{s}}\vb{k}_{\vb{h}}^2\bigr)}{\vb{k}^2_{\vb{h}}}}^{\mathclap{\text{long-range kernel}}}}
        \underset{\smash{\mathclap{\vb{m}\rightarrow\vb{h}}}}{\mathcal{F}}\hspace{0.5em}
        \underbrace{\sum^N_{j=1} W_{p_{\text{i}}}\biggl(\frac{\vb{x}_{\vb{m}} - \vb{x}_j}{L_{\varphi}}\biggr)}_{\mathclap{\text{particles $\rightarrow$ mesh}}} \biggr\} \\
    &\qquad\quad - Gm^2\hspace{0.2em}
        \sum_{
            \mathclap{
                \substack{
                    \{ \, j \, \mid \, |\vb{x}_{ij\vb{n}'}| < x_{\text{r}} \, \} \\
                    j \neq i\mathclap{\phantom{\raisebox{0.8ex}{x}}}
                }
            }
        }
        \hspace{2.1em}
        \left[
        \begin{aligned}
        &\hspace{1.35em}|\vb{x}_{ij\vb{n}'}|^{-3} \text{erfc}\biggl(\frac{|\vb{x}_{ij\vb{n}'}|}{2x_{\text{s}}}\biggr) \\
        &+ \frac{|\vb{x}_{ij\vb{n}'}|^{-2}}{\sqrt{\pi}x_{\text{s}}}\exp\biggl(-\frac{\vb{x}_{ij\vb{n}'}^2}{4x^2_{\text{s}}}\biggr) \\
        &+ \phantom{\mathclap{|\vb{x}|^{-3}}} \smash{\underbrace{\bigl(|{\vb{x}}_{ij\vb{n}'}|^{-3}_S - |\vb{x}_{ij\vb{n}'}|^{-3}\bigr)}_{\smash{\text{softening}}}}
        \end{aligned}
        \right] \vb{x}_{ij\vb{n}'}
    \end{aligned}
    \label{eq:p3m_gravity}
    \vspace{2.1em}
\end{equation}
Although this expression is quite loaded, it is extremely important to understand it, as it is the core of the simulation code.

It is clearer to read the equation backwards, starting on the third line. This describes the force computed with the PP method, where direct particle-particle interactions are considered. This contribution must be read as the gradient of the gravitational potential between two particles, so in essence, we must have in mind the equation:
\begin{equation}
    \vb F = -\grad\phi = -\grad\qty(-\frac{G m^2}{ax}) = -\frac{Gm^2}{a}\frac{\vb x}{x^3},
\end{equation}
where $\vb x$ is the vector describing the separation between the two particles of mass $m$. In the third line of \cref{eq:p3m_gravity}, we have the same equation but in a more complicated form. The sum is done over all particles that are separated by a distance shorter than $x_r$, which is the hard cut-off of this sum. The distance between two particles is defined as:
\begin{equation}
    \abs{\vb x_{ij\vb n'}} \equiv \abs{\vb x_i - \vb x_j - L_{\rm box} \vb n'},
\end{equation}
where $L_{\rm box}$ is the length of the cubic box, and $\vb n'$ is a triplet integer, that is chosen such that:
\begin{equation}
    \abs{\vb x_{ij \vb n'}} = \min_{\vb n \in \Z^3}\abs{\vb x_{ij\vb n}}.
\end{equation}
This allows for the direct summation to be performed only on the nearest image $\vb n'$ of particle $j$, nearest to particle $i$, thus getting rid of the sum over $\vb n$. To this end, the splitting scale $x_s$ is chosen to be small enough that the hard cut-off $x_r$ is smaller than the box length. It is important to note that the self-interacting particle is neglected because it is unphysical and would lead to a mathematical divergence. In the square brackets, we find the reciprocal distance to the third power. This distance is split into short and long-range contributions, and softened to avoid unphysical velocities when two particles come close together. The complementary error and exponential function correspond to the short and long-range contributions, respectively. In the last term, related to the softening, we find the softened reciprocal distance $\abs{\vb x}_S$ for an arbitrary shape $S$. 

Moving on to the first and second lines, we have the computed force using the PM method. Starting with the sum over $j$, we find the deposition of the particles into a mesh, where $W_{p_i}$ is the window function\footnote{This is the mass assignment, for which schemes like Cloud-in-Cell are used.} of order $p_i$. In this term, a fast Fourier transform is performed, where the $\vb k=0$ mode is set to zero to remove the background density. Then, the long-range kernel is combined with two deconvolutions to construct the gravitational potential in Fourier space. The two deconvolutions are necessary, one for the mass assignment and one for the eventual interpolation of the force. It is important to use the same window function for both tasks, otherwise there would be a loss in resolution, and more importantly, it would lead to a mismatch between the exerted force and the corresponding particles. It might be the case that a particle could have self-interactions when using incorrect window functions. At this stage, the code can take two approaches: first, take the derivative in Fourier space and then perform the inverse Fourier transform; or, do the inverse Fourier transform first and take the derivative $\vb D_{p_{\rm d}}$ in real space. \concept\ takes the second approach, since the first one involves taking three inverse Fourier transforms (one per spatial component), which are computationally expensive. The last terms correspond to the interpolation of the force needed to evaluate it at the particle position.

	\end{annexes}

\end{document}